\documentclass[a4paper,12pt]{article}
\pdfoutput=1
\usepackage{epsfig}
\usepackage{amssymb}
\usepackage{amsfonts}
\usepackage{amsmath}
\usepackage{mathtools}
\usepackage{euscript}
\usepackage{verbatim}
\usepackage{latexsym}
\usepackage{graphicx}
\usepackage{caption}
\usepackage{float}
\usepackage{subcaption}
\usepackage{hyperref}

\usepackage{xcolor}

\usepackage{listings}

\usepackage{booktabs}

\newif\ifdtup

\catcode`\@=11

\@addtoreset{equation}{section}

\def\@normalsize{\@setsize\normalsize{15pt}\xiipt\@xiipt
\abovedisplayskip 14pt plus3pt minus3pt%
\belowdisplayskip \abovedisplayskip
\abovedisplayshortskip \z@ plus3pt%
\belowdisplayshortskip 7pt plus3.5pt minus0pt}

\def\small{\@setsize\small{13.6pt}\xipt\@xipt
\abovedisplayskip 13pt plus3pt minus3pt%
\belowdisplayskip \abovedisplayskip
\abovedisplayshortskip \z@ plus3pt%
\belowdisplayshortskip 7pt plus3.5pt minus0pt
\def\@listi{\parsep 4.5pt plus 2pt minus 1pt
     \itemsep \parsep
     \topsep 9pt plus 3pt minus 3pt}}

\relax

\catcode`@=12

\catcode`\@=11

\def\section{\@startsection{section}{1}{\z@}{3.5ex plus 1ex minus
   .2ex}{2.3ex plus .2ex}{\large\bf}}

\def\SymBoxes#1#2#3#4{\newdimen\un@t \un@t#3%
\raisebox{#1}{\rule{#2\un@t}{#4}\hskip-#2\un@t% lower horizontal
\@tempdimb\un@t \advance\@tempdimb by-#4\@tempcntb#2\relax%
\@whilenum{\@tempcntb>0}\do{%                         % #2 vertical lines
\rule{#4}{\un@t}\hskip\@tempdimb \advance\@tempcntb by\m@ne}%
\hskip-#2\un@t \rule[\un@t]{#2\un@t}{#4}%
\rule[\un@t]{#4}{#4}\hskip-#4%             % upper horizontal line
\rule{#4}{\un@t}}\hskip-#4}                % rightest vertical line
\begin{document}
%\begin{letter}{~}

%Define some new commands and  macros
\newcommand{\beq}{\begin{equation}}
\newcommand{\eeq}{\end{equation}}
\newcommand{\bea}{\begin{eqnarray}}
\newcommand{\eea}{\end{eqnarray}}
\newcommand{\beas}{\begin{eqnarray*}}
\newcommand{\eeas}{\end{eqnarray*}}
\newcommand{\defi}{\stackrel{\rm def}{=}}
\newcommand{\non}{\nonumber}
\newcommand{\bquo}{\begin{quote}}
\newcommand{\enqu}{\end{quote}}
\renewcommand{\(}{\begin{equation}}
\renewcommand{\)}{\end{equation}}
% definitions
\def \eqn#1#2{\begin{equation}#2\label{#1}\end{equation}}

\def\e{\epsilon}
\def\IZ{{\mathbb Z}}
\def\IR{{\mathbb R}}
\def\IC{{\mathbb C}}
\def\IQ{{\mathbb Q}}
\def\de{\partial}
\def\Tr{ \hbox{\rm Tr}}
\def\H{ \hbox{\rm H}}
\def\HE{ \hbox{$\rm H^{even}$}}
\def\HO{ \hbox{$\rm H^{odd}$}}
\def\K{ \hbox{\rm K}}
\def\Im{ \hbox{\rm Im}}
\def\Ker{ \hbox{\rm Ker}}
\def\const{\hbox {\rm const.}}
\def\o{\over}
\def\im{\hbox{\rm Im}}
\def\re{\hbox{\rm Re}}
\def\bra{\langle}\def\ket{\rangle}
\def\Arg{\hbox {\rm Arg}}
\def\Re{\hbox {\rm Re}}
\def\Im{\hbox {\rm Im}}
\def\exo{\hbox {\rm exp}}
\def\diag{\hbox{\rm diag}}
\def\longvert{{\rule[-2mm]{0.1mm}{7mm}}\,}
\def\a{\alpha}
\def\dag{{}^{\dagger}}
\def\tq{{\widetilde q}}
\def\p{{}^{\prime}}
\def\W{W}
\def\N{{\cal N}}
\def\hsp{,\hspace{.7cm}}

\def\br{\nonumber}
\def\IZ{{\mathbb Z}}
\def\IR{{\mathbb R}}
\def\IC{{\mathbb C}}
\def\IQ{{\mathbb Q}}
\def\IP{{\mathbb P}}
\def \eqn#1#2{\begin{equation}#2\label{#1}\end{equation}}

\newcommand{\C}{\ensuremath{\mathbb C}}
\newcommand{\Z}{\ensuremath{\mathbb Z}}
\newcommand{\R}{\ensuremath{\mathbb R}}
\newcommand{\rp}{\ensuremath{\mathbb {RP}}}
\newcommand{\cp}{\ensuremath{\mathbb {CP}}}
\newcommand{\vac}{\ensuremath{|0\rangle}}
\newcommand{\vact}{\ensuremath{|00\rangle}                    }
\newcommand{\oc}{\ensuremath{\overline{c}}}
\newcommand{\psizero}{\psi_{0}}
\newcommand{\phizero}{\phi_{0}}
\newcommand{\hzero}{h_{0}}
\newcommand{\psiin}{\psi_{\rh}}
\newcommand{\phiin}{\phi_{\rh}}
\newcommand{\hin}{h_{\rh}}
\newcommand{\rh}{r_{h}}
\newcommand{\rb}{r_{b}}
\newcommand{\psibnd}{\psi_{0}^{b}}
\newcommand{\psibndp}{\psi_{1}^{b}}
\newcommand{\phibnd}{\phi_{0}^{b}}
\newcommand{\phibndp}{\phi_{1}^{b}}
\newcommand{\gbnd}{g_{0}^{b}}
\newcommand{\hbnd}{h_{0}^{b}}
\newcommand{\zh}{z_{h}}
\newcommand{\zb}{z_{b}}
\newcommand{\man}{\mathcal{M}}
\newcommand{\hbr}{\bar{h}}
\newcommand{\tbr}{\bar{t}}

\begin{titlepage}
%\begin{flushright} CHEP XXXXX
%ULB-TH/09-10\\
%hep-th/yymmnnn\\ \end{flushright}
%\bigskip

\def\thefootnote{\fnsymbol{footnote}}

\begin{center}
{\bf {\large Holography with an Inner Boundary:} \\
\vspace{0.2cm}
{ A Smooth Horizon as a Sum over Horizonless States
}}
\end{center}

%\bigskip
\begin{center}
Chethan Krishnan$^a$\footnote{\texttt{chethan.krishnan.physics@gmail.com}} \quad
Pradipta S. Pathak$^a$\footnote{\texttt{pradiptap@iisc.ac.in}}
%\vspace{0.1in}
\end{center}

\renewcommand{\thefootnote}{\arabic{footnote}}

\begin{center}
%\vspace{0.2cm}

$^a$ {Center for High Energy Physics,\\
Indian Institute of Science, Bangalore 560012, India}\\

\end{center}
\vspace{-0.15in}
\noindent
\begin{center} {\bf Abstract} \end{center}
The (holomorphic) partition function of the Euclidean BTZ black hole with boundary modulus $\tau$, is the $S$-image of the Virasoro vacuum character, $\chi_{\rm vac}(-1/\tau)$. This object decomposes into primaries via the modular $S$-kernel: $\chi_{\rm vac}\left(-\frac{1}{\tau}\right)=\int_{0}^{\infty} dP S_{0P}(P,c)\chi_P(\tau)$. In this paper, we provide a bulk understanding of this spectral resolution using the Chern-Simons formulation of AdS$_3$ gravity with {\em two} boundaries: an asymptotic torus and an excised Wilson line at the origin (``stretched horizon"). At infinity, we impose standard AdS$_3$ Drinfel'd-Sokolov (DS) gauge to obtain the Alekseev-Shatashvili (AS) boundary action for a coadjoint orbit. At the inner boundary, removing the Wilson line prepares the state at the cut as a sum over orbits of the $spatial$ cycle. Re-inserting a spatial holonomy Wilson line acts as a delta-function projector onto the corresponding primary, which together with boundary gravitons, reproduces the Virasoro character (e.g., of a conical defect). But we can also consider projectors onto the {\em conjugate} basis $\tilde P$, of the dual cycle. A key observation is that this leads to $S$-kernels instead of delta functions, with the BTZ character arising when the dual cycle label is in the exceptional orbit. Our two-boundary construction provides a bulk understanding of BTZ entropy: holonomy zero modes at the horizon have an effective central charge $c_{\rm prim}=c-1$ from the kernel measure (primaries), while the universal Dedekind-$\eta$ in $\chi_P(\tau)$ contributes $c_{\rm desc}=1$ from boundary gravitons (descendants). Together, they reproduce the full Cardy entropy. While our methods are specific to AdS$_3$/CFT$_2$, they are an explicit illustration that smoothness of the (Euclidean) horizon may emerge from a $sum$ over bulk states which are manifestly unsmooth.

\vspace{1.6 cm}
\vfill

\end{titlepage}

\setcounter{footnote}{0}

\tableofcontents

\section{A Holographic ``Atomic" Hypothesis}

The successes of string theory with BPS black holes \cite{Sen, StromingerVafa, CassaniMurthyQuantumBH, ZaffaroniAdSBHLocalization, MurthyModularFormsBH} and their microstates might lead one to believe that explicitly constructing --or at least identifying-- individual UV complete microstates is a necessary step toward a satisfactory understanding of quantum black holes. If true, this would suggest that we are in a hopeless situation when confronted with generic, finite temperature black holes where supersymmetry is broken and the information paradox is in full force.

Yet this perception may be misleading. We do not possess a truly UV-complete description of the microstates of a system as ordinary as a gas in a box, and yet we regard our understanding of its micro- and macrophysics --in terms of atoms, molecules, and statistical principles-- quite satisfactory. A heuristic reason for this, is that the atomic hypothesis (or more broadly, the notion of particles and quanta) serves as a toy UV completion for thermal systems in generic $weakly$ coupled theories. With a ``little imagination and thinking" \cite{FeynmanLecturesVol1} applied, this lead to a revolution in our understanding of short distance physics in the last century. 

Given the expectation that gravity is a universal, holographic, low-energy manifestation of $strongly$ coupled large-$N$ theories, one could ask: is it conceivable that there exists an analogous, generic, toy UV completion for thermal systems (aka black holes) in gravity? In recent papers, we have argued \cite{Burman1, Burman2} that a suitably defined ``stretched horizon" may provide such a model. A finite Planck length in the bulk translates to finite-$N$ in the dual theory, and therefore a stretched horizon can also be viewed as a regulator for $N$. The discussions in \cite{Burman1, Burman2} demonstrate that black hole thermodynamics and smooth horizon correlators (as seen from the exterior\footnote{In fact, in the large-$N$ limit, these exterior correlators allow an analytic continuation into the interior as well \cite{Burman2}. It will be interesting to see if a satisfying physical interpretation for this emergent analytic continuation in terms of infall, exists.}) can be reproduced from fairly minimal variants of the stretched horizon prescription\footnote{The original paper that considers the stretched horizon/brick wall as a model for black hole thermodynamics is that of 't Hooft \cite{tHooft}. It managed to reproduce the area scaling of entropy. See also, \cite{SusskindThorlacius}. Recently there has been a revival of the stretched horizon, especially in the context of chaos and random matrices, see e.g., \cite{Adepu, FuzzRandom, Pradipta}.}. These are indications that a stretched horizon in the UV complete description may be compatible with low energy (smooth horizon!) correlators. In other words, degrees of freedom at the horizon in the UV complete description need not be in tension with the principle of equivalence, because the latter is an EFT statement. 

\subsection{Inner Boundary: On UV Completion of Bulk States}

Despite this promise, the nature of the stretched horizon remains mysterious. A particular manifestation of this is that while it is useful as a UV regulator in Lorentzian signature (as we discussed above), its role in Euclidean signature is quite unclear. The Euclidean horizon is not the setting in which  conventional forms of the information paradox are usually formulated\footnote{There do exist Euclidean versions of the paradox which are perhaps not as well known \cite{KapInfo,Datar}, and this was one of the motivations behind this paper.}. Indeed, to obtain black hole thermodynamics from gravity, what one uses are the heuristics first understood by Gibbons and Hawking \cite{Gibbons-Hawking}: statistical mechanics and models for microstates simply do not play a role. We can view this as the analogue of gas thermodynamics, which was well-developed long before any microscopic model for gases. 

However, in the context of thermodynamics of gases, it is a fact that the atomic hypothesis fits well with the Euclidean statistical physics of gases and the determinations of Euclidean partition functions. Motivated by this, we would like to have a Euclidean picture of the stretched horizon which would take our conventional understanding of black hole {\em thermodynamics}, closer towards {\em Euclidean statistical mechanics} from the bulk. If such a formulation is not possible, it should cause us worry that the stretched horizon is merely a stop-gap Lorentzian construct. 

In this paper, we will see that a suitably defined stretched horizon has a natural role to play in the {\em Euclidean} AdS$_3$/CFT$_2$ correspondence.  We work with AdS$_3$/CFT$_2$ because modular invariance is most powerful in controlling high-lying states, in the setting of 2-dimensional CFTs. An elementary (but far from exhaustive) manifestation of this, is the Cardy formula. While arguments that generalize modular invariance can be made in higher dimensions, and we expect some of them to be related to higher dimensional versions of the arguments we use here, in this paper we will stick to the basic AdS$_3$/CFT$_2$ setting. A second reason for our focus on AdS$_3$/CFT$_2$ is that Virasoro characters can be computed from both the bulk and the boundary, giving us confidence that our understanding of the bulk is sufficiently detailed. In the conclusions, we will argue that the broad lessons about the stretched horizon that we learn from AdS$_3$/CFT$_2$, may generalize to higher dimensions as well.

\subsubsection{A Key Identity and its Bulk Interpretation}

We will work with Euclidean AdS$_3$ gravity at finite temperature and angular momentum chemical potential. This is the context of the general rotating BTZ black hole (and the  spinning conical defect), and it implies that we are working with an asymptotic boundary that is a torus with complex modulus $\tau$ (and $\bar \tau$, which we will usually suppress). The holomorphic structure of the theory is most transparent in the Chern-Simons formulation \cite{Achucarro, Witten3Dgravity} which is what we will work with. 

A key fact that we will use as a jumping off point is that the (holomorphic) partition function of the BTZ black hole with modulus $\tau$ is known: 
\bea
Z^{(\mathrm{hol})}_{\mathrm{BTZ}}(\tau) = \chi_{\text{vac}}(-1/\tau), \label{BTZpart}
\eea
where $\chi_{\text{vac}}$ stands for the Virasoro character of the vacuum module. The fact that the BTZ partition function takes this form is easily argued by noticing that thermal AdS$_3$ and Euclidean BTZ are related by a switch of the contractible cycles in the solid torus, which can be implemented via $\tau \rightarrow -1/\tau$, a $PSL(2, \IZ)$ transform of the boundary torus.

A key fact for us in this paper will be the 2D CFT result \cite{SeibergLiouvilleNotes,TeschnerLiouvilleRevisited,PonsotTeschnerLiouvilleBootstrap} that 
\begin{equation}
\chi_{\text{vac}}\!\left(-\frac{1}{\tau}\right)
=\int_{0}^{\infty}\! dP\; S_{0P}(c)\;\frac{q^{P^2}}{\eta(\tau)}\,,\qquad q=e^{2\pi i\tau}\,,\label{key0}
\end{equation}
where the functional form of $S_{0P}$ (``vacuum row of the modular $S$-kernel") is explicitly known. We will explain the ingredients in this formula in greater detail later, we will not need them here. One of our goals in this paper will be to derive a variant/generalization of the above formula:
\begin{equation}
\chi_{\tilde P}\!\left(-\frac{1}{\tau}\right)
=\int_{0}^{\infty}\! dP\; S_{\tilde P P}(c)\;\frac{q^{P^2}}{\eta(\tau)}\,,\qquad q=e^{2\pi i\tau}\,,\label{key}
\end{equation}
from the bulk. Equation \eqref{key0} can be viewed as a special case (or more precisely an analytic continuation) of equation \eqref{key}. We emphasize that both $S_{0P}$ and $S_{\tilde P P}$ are explicitly known functions; we will write them down later. These integral equalities above are therefore mathematical {\em identities}, given that we know the explicit functional forms of the characters. But one can also view them as CFT facts, because they emerge naturally from modular properties of 2D CFTs.

\subsubsection{Outline}

In this paper, we will obtain these CFT results using a {\em bulk} AdS$_3$ gravity path integral in the Chern-Simons (CS) formulation, with {\em two} boundaries: one at the asymptotic torus and the other a small tube removed around origin (see figure \ref{torus}). We will think of this removed inner solid torus as an excised Wilson line. The main ingredients of the idea are best described after introducing Chern-Simons AdS$_3$ gravity with two boundaries, which we do in the next section. We will present an overview of the bulk calculation in Sections \ref{section:janus}. The details involve various side-quests, but the conclusion  is easy to state: the integral \eqref{key0} for the {\em smooth} Euclidean horizon character can be interpreted as a {\em sum} over black hole microstates which are {\em not} smooth. These ``universal" microstates are simply conical defect like states with scaling dimensions {\em above} the BTZ threshold. This and various related calculations are done in Section \ref{sec:partition} using the framework developed for Chern-Simons path integrals in Section \ref{sect:innerboundary}. The defects that source the microstates can be interpreted as Wilson line insertions, analogous to the more familiar conical defects -- but unlike conical defects, these microstate defects have hyperbolic as opposed to elliptic holonomy and they lie above the black hole threshold. Wilson line insertions as sources for geometries are discussed in Section \ref{sec:wilson}. 

Throughout the paper, we will present various arguments from the theory of Virasoro coadjoint orbits and their connections to CFT$_2$ primaries to strengthen our claims\footnote{We have made an effort to be self-contained in this discussion, by providing some introductory discussion in the early subsections of Section \ref{sec:orbits}.}. This includes putting some previous results in the literature in context, by refining and re-interpreting the relevant calculations. In the literature, the quantum corrected central charges of normal and exceptional orbits computed from AdS$_3$ Chern-Simons theory, are distinct. This is puzzling, if they have to live within the same boundary CFT. We will argue that there is a way to reconcile them that takes inspiration from our two-boundary description of primaries. Our refined central charge proposal and its implications are the content of the later subsections in Section  \ref{sec:orbits}, in particular \ref{sec:refinement}. We find strong evidence for our proposal (from both the bulk and the boundary) via simple extensions of existing calculations in the literature.

A well-known fact about ``pure" AdS$_3$ gravity is that it has no propagating gravitons (which might make one think that it is trivial), but it has black holes (which might make one think that it is not). Our point of view in this paper is that the Chern-Simons formulation manifests this quasi-topological feature, by associating degrees of freedom only to the boundaries. In the usual discussion of Euclidean Chern-Simons AdS$_3$ gravity, the only boundary is the asymptotic boundary. But for black holes, this captures only the descendants around a heavy primary state (and not the primaries themselves). In 2D CFT with central charge $c$, the Cardy growth requires a total central charge of $c$ \cite{Cardy86}, but the  descendants contribute only one degree of freedom. So the bulk origin of the remaining $c$$-$$1$ degrees of freedom (attributed to the primaries in the dual CFT \cite{Shouvik}) is mysterious. In other words, while modular arguments allow us to reproduce the partition function (and entropy) in \eqref{BTZpart}, it is an indirect argument  and the bulk origin of the count remains unclear. Our two boundary path integral instead leads directly to the RHS of \eqref{key}, providing a ``mechanical" understanding for the emergence of the modular $S$-image from the bulk: the primaries are to be associated to the stretched horizon and their ``density"
 is captured by the $S$-kernel. This together with the descendants at the boundary, give a bulk explanation for the Cardy growth and its primary-vs-descendant split. We elaborate on this using some simple calculations in Section \ref{sec:CardySplit}. 

Ours is a {\em bulk} radial quantization approach to defining the AdS$_3$/CFT$_2$ path integral. Unlike the usual picture where the asymptotic boundary conditions are given emphasis, we view the latter as part of the universal definition of {\em any} AdS$_3$/CFT$_2$ (the universal descendant structure). The precise specification of {\em which} 2D CFT we are working with, happens at the interior via the projection on to the primaries that define the CFT. (If one projects onto a specific primary instead of their sum, we get the corresponding character.) This perspective is discussed in the early subsections of Section \ref{sect:innerboundary} as well as Section \ref{sec:mod-bootstrap}. We believe a natural analogue of this structure exists even in the context of compact Chern-Simons theory. There the bulk CS theory path integral with a Wilson line insertion defines an extended chiral algebra character. But to define a full CFT, we again need discrete, spectrum data. This connection and its relationship to rational-vs-irrational CFTs is developed in Appendix \ref{sec:compact_vs_noncompact}.

This perspective suggests that pure (Chern-Simons) gravity, rather than being a standalone theory, is better viewed as the glue/toolkit for assembling holographic 2D CFTs from primaries/holonomies and descendants/gravitons. It knows about characters and modular invariance, but this does not fix the spectrum uniquely: modular bootstrap type constraints are still needed. This has connections to the recent proposal that AdS$_3$ pure gravity is most naturally compared to a Virasoro TQFT \cite{CollierEberhardtZhangVirasoroTQFT}. Some of our observations can be viewed as exploiting the ``functoriality" of a suitable TQFT on the thickened torus, to make claims about the bulk nature of states in AdS$_3$/CFT$_2$. 

In AdS$_3$/CFT$_2$, our two-boundary construction makes clean sense because the primaries (heavy microstates) and descendants (boundary gravitons) are naturally distinct at any $c$. But we believe that an inner boundary picture with an approximate differentiation (via large-$N$ factorization) between heavy and light states is valid even in higher dimensional AdS/CFT. While we do not attempt to make this picture explicit here, we do have some comments on various general aspects of holography with an inner boundary. This, and various other ideas that are not fully developed in this paper, as well as a selected summary of some of the main lessons, are the topics of discussion in the concluding section. 

\section{Chern-Simons with Two Boundaries}
\label{sec:twobdy-setup}

It is a famous fact that 2+1 dimensional Einstein gravity in AdS$_3$ has a re-writing as a Chern-Simons gauge theory with $SL(2,R) \times SL(2,\IR)$ gauge group \cite{Achucarro, Witten3Dgravity}. We will write the Chern-Simons action $S_{\rm CS}[A]$ as 
\beq \label{CScov}
S_{\rm CS}[A] = \frac{k \ell}{4\pi} \int_\mathcal{M} {\rm Tr}\left(A \wedge dA + \frac{2}{3} A \wedge A \wedge A\right),
\eeq
and take $A = A^a L_a$ and $\overline{A}^a = \overline{A}^a L_a$ as the $SL(2,\mathbb R)$-valued gauge potentials. We use the explicit generators in \cite{Ammon}:
\beq
\begin{aligned}
L_0 &= \frac12
\begin{pmatrix}
1 & 0\\
0 & -1
\end{pmatrix},
&
L_+ &= 
\begin{pmatrix}
0 & 0\\
-1 & 0
\end{pmatrix},
&
L_- &= 
\begin{pmatrix}
0 & 1\\
0 & 0
\end{pmatrix}.
\end{aligned}
\eeq
They satisfy 
\beq \label{SLgen}
[L_a,L_b] = f_{abc} L^c \ , \hspace{1cm} \, \quad \quad {\rm Tr}(L_a L_b) = \frac{1}{2} g_{ab} \eeq 
where the local tangent space metric is flat with mostly plus signature, but is of the form $g_{ab}= \Bigl(\begin{smallmatrix}
1 & 0 & 0\\
0 & 0 & -2\\
0 & -2 & 0
\end{smallmatrix}\Bigr)$. We use $g_{ab}$ to raise and lower the flat indices $a,b,c$. The $f_{abc}$ are fully anti-symmetric and are defined by $f_{0+-} = 2$. We will sometimes also use the notation $L_{+1}\equiv L_{+}$ and $L_{-1}\equiv L_{-}$, and write the commutation relations as $[L_i, L_j]=(i-j)L_{i+j}$, with $i,j \in (-1,0,1)$.

Usually the discussion of the translation from Chern-Simons to first order gravity is done in an $SL(2,\IR)$ basis where the $f_{abc}$ is replaced by $\epsilon_{abc}$ and $g_{ab}$ by $\delta_{ab}$. We work with the above forms to simplify our eventual gauge choices. With a general fully anti-symmetric $f_{abc}$ and symmetric $g_{ab}$ of mostly plus signature, the translation to Einstein action (on an asymptotically AdS spacetime $\mathcal{M}$ with $\ell$ being the AdS length scale), 
\beq
S_{\rm EH} = \frac{k}{4\pi} \int_\mathcal{M} d^3 x \sqrt{-g}\left(R+\frac{2}{\ell^2}\right), \quad {\rm where} \ k = \frac{1}{4 G_N},
\eeq
is accomplished via the first-order form
\beq \label{EHFirst}
S_{\rm EH} = \frac{k}{2\pi} \int_\mathcal{M} e^a \wedge \left(d\omega_a + \frac{1}{2} f_{abc} \omega^b \wedge \omega^c + \frac{1}{6\ell^2} f_{abc} e^b \wedge e^c\right),
\eeq
with the triad and spin connection one-forms, 
\beq
e^a = e^a_\mu dx^\mu \ , \hspace{1cm} \, \quad \quad \omega^a = \frac{1}{2} f^{abc}\omega_{bc\mu} dx^\mu.
\eeq
Since $f_{abc}$ is a fully anti-symmetric matrix of constants in three dimensions, it is proportional to $\epsilon^{abc}$, the invariant tensor of $(P)SL(2,\IR)$. These one-forms can be combined into two $SL(2,\mathbb R)$ connection one-forms, 
\beq
A^a = \omega^a + \frac{1}{\ell} e^a \ , \hspace{1cm} \, \quad \quad \overline{A}^a = \omega^a - \frac{1}{\ell} e^a
\eeq
in terms of which one can check that 
\beq \label{EHasCS}
   S_{\rm EH} = S_{\rm CS} [A] - S_{\rm CS}[\overline{A}] - \frac{k}{4 \pi} \int_{\partial \mathcal M} e^a \wedge \omega_a.
\eeq

\subsection{Canonical Bulk Action and Asymptotic Boundary}

We separate $A$ and $ \overline A$ into temporal and spatial parts, 
\beq
A = A_t dt + A_i dx^i \ , \hspace{0.5cm}  \overline A =  \overline A_t dt +  \overline A_i dx^i \hspace{1cm} \, \quad \quad {\rm where} \ i = r, \phi
\eeq
and plug it in \eqref{CScov} to get the canonical action: 
\beq \label{CScan}
S_{CS}[A] = \frac{k \ell}{4 \pi} \int_\mathcal{M} d^3 x \ {\rm Tr}\left(A_\phi \dot A_r - A_r \dot A_\phi + 2 A_t F_{r \phi}\right) - \frac{k \ell}{4 \pi} \int_{\partial\mathcal{M}} dt d\phi \ {\rm Tr}\left(A_t A_\phi\right) 
\eeq
where $d^3 x = dt \wedge dr \wedge d\phi$ as a result of choosing $\epsilon^{t\phi r} = 1/\sqrt{-g} $ orientation\footnote{This follows from $\epsilon^{\mu_1 .... \mu_n} \sqrt{-g} \ d^n x = (-1)^s  \ dx^{\mu_1} \wedge .... \wedge dx^{\mu_n}$ with $s=1$, where $s$ is the number of negative eigenvalues in the metric.}, the time-derivative is defined as $\dot A_i \equiv \partial_t A_i$ and the field strength $F_{\mu \nu} = \partial_\mu A_\nu - \partial_\nu A_\mu + [A_\mu , A_\nu]$.

We will define the bulk part of \eqref{CScan} as the canonical bulk action:
\beq
S[A] \equiv \frac{k \ell}{4 \pi} \int_\mathcal{M} d^3 x \ {\rm Tr}\left(A_\phi \dot A_r - A_r \dot A_\phi + 2 A_t F_{r \phi}\right). \label{canBulk}
\eeq
Since $S[A]-S[ \overline A]$ is the Einstein-Hilbert action up to boundary terms, the next immediate step will be to examine the variation of this action:
\beq \label{CScanvar}
\delta S[A] - \delta S[ \overline A] = \int_\mathcal{M} ({\rm EOM}) + \frac{k\ell}{2 \pi} \int_{\partial \mathcal{M}} dt d\phi \ {\rm Tr} (A_t \delta A_\phi -  \overline A_t \delta  \overline A_\phi) 
\eeq 
where EOM captures terms that vanish because of the classical equations of motion, 
\beq \label{bulkEOM}
F_{ti} = F_{ij} = 0 \hspace{0.5cm} \, {\rm and} \hspace{0.5cm}  \overline F_{ti} = \overline F_{ij} = 0.
\eeq
To impose asymptotically AdS$_3$ boundary conditions, we will add the following boundary term: 
\beq \label{Sbdry}
S_{\rm bdry}[A, \overline A] = -\frac{k \ell}{4 \pi} \int_{\partial \mathcal{M}} dt d\phi \ {\rm Tr} (A_\phi^2 +  \overline A_\phi^2) 
\eeq
to $S[A]-S[ \overline A]$. The utility of this is that with these added, the $chiral$ boundary condition $A_\phi - A_t =  \overline A_\phi + \overline A_t = 0$ at $\partial \mathcal M$, leads to a well-defined variational principle: the action
\beq \label{S1}
S_1 = S[A] - S[ \overline A] + S_{\rm bdry}[A, \overline A]
\eeq 
has vanishing on-shell variation when we demand chiral boundary conditions:
\beq \label{S1var}
\delta S_{1} = \underbrace{\int_\mathcal{M}({\rm EOM})}_{=0 , \  {\rm from} \ \eqref{bulkEOM}}-\frac{k \ell}{2 \pi} \int_{\partial \mathcal{M}} dt d\phi \ {\rm Tr}\left(A_- \delta A_\phi +  \overline A_+ \delta  \overline A_\phi\right) = 0
\eeq
where $A_\pm = A_\phi \pm A_t$. Chiral boundary conditions will eventually lead to the natural notion of ``asymptotically AdS$_3$" (or Brown-Henneaux \cite{BrHe}) spacetimes in the metric language, and this is the primary reason for interest in them in the holographic setting. 

We will not need the metric variables much in this paper. But we will crucially use the fact that in the Chern-Simons language, these same boundary conditions lead to a boundary reduction of the bulk Chern-Simons action to a chiral WZW model \cite{Coussaert}, and eventually to an Alekseev-Shatashvili action \cite{AS1, AS2, CJ1}. This is an action for the boundary gravitons and its path integral can be explicitly evaluated, and depending on the background (empty AdS$_3$, conical defect, BTZ) it leads to the appropriate Virasoro character \cite{CJ1}. We will present a variation of this calculation that applies universally to all cases of interest, later.

\subsection{Inner Boundary}

We will be interested in manifolds $\mathcal M$ having two boundaries, the outer one at the asymptotic infinity $\partial \mathcal M_+$ and the inner one near the ``origin" $\partial \mathcal M_-$. We will start the discussion here in Lorentzian signature, even though we will eventually move over to Euclidean. With this extra boundary, the covariant action in \eqref{CScan} and the canonical action \eqref{canBulk} are related by an extra term from the inner boundary: 
\beq \label{CScantwo}
S_{CS}[A] = S[A] - \frac{k \ell}{4 \pi} \int_{\partial\mathcal{M}_+} dt d\phi \ {\rm Tr}\left(A_t A_\phi\right) + \frac{k \ell}{4 \pi} \int_{\partial\mathcal{M}_-} dt d\phi \ {\rm Tr}\left(A_t A_\phi\right)
\eeq
To obtain a well-defined variational principle with chiral boundary conditions at the asymptotic region, in the previous subsection we removed the outer boundary term from the above expression and added \eqref{Sbdry}. This lead to well-defined chiral boundary conditions at the asymptotic boundary. 

It turns out that if we simply remove the boundary terms in \eqref{CScantwo} from either boundary, the resulting (bulk canonical) action $S[A]$ will lead to a well-defined variational principle with $A_\phi$ held fixed at that boundary. This ``Dirichlet" boundary condition is what we will demand at the inner boundary. In other words, even with the inner boundary, our action remains $S_1$ from \eqref{S1}. The variation of this action is now: 
\beq \label{S1vartwo}
\delta S_{1} = \int_\mathcal{M}({\rm EOM})-\frac{k \ell}{2 \pi} \int_{\partial \mathcal{M}_+} dt d\phi \ {\rm Tr}\left(A_- \delta A_\phi +  \overline A_+ \delta  \overline A_\phi\right) - \frac{k \ell}{2 \pi} \int_{\partial \mathcal{M}_-} dt d\phi \ {\rm Tr} (A_t \delta A_\phi -  \overline A_t \delta  \overline A_\phi) 
\eeq 
This defines chiral boundary conditions at the asymptotic boundary and Dirichlet ($\delta A_\phi \big|_{\partial \mathcal M _-}=0$) at the inner boundary, as the variational problem. 

The implementation and interpretation of this choice will become clear as we proceed, but for now we will simply make the following comments. (1) Dirichlet boundary conditions are natural from the perspective of removing a spatial holonomy Wilson line from the center of the spatial cycle of a solid cylinder, which we can think of as a particle or a state. (2) It turns out that in Chern-Simons theory (unlike in higher dimensional Yang-Mills theory) Dirichlet boundary conditions lead to no dynamical edge modes in the sense of (say) \cite{Donnelly}. We show this in Appendix \ref{app:symp}. (3) A key subtlety in this last comment is that on the annulus (which is our spatial section) the spatial holonomy mode is a degree of freedom that {\em is} dynamical. This will play a crucial role in our discussions. (4) Eventually, the only degrees of freedom that we will retain at the inner boundary will be holonomies. We will mod out by {\em all} periodic inner boundary gauge transformations. This is analogous to how one cuts and glues geometries in TQFT, where the states in the Hilbert space are labelled by holonomies. In this sense what our Dirichlet boundary term does is not so much as pick a privileged gauge field configuration at the inner boundary, but pick a {\em polarization} for the cut-open path integral in the holonomy Hilbert space. (5) It is also worth noting that Dirichlet boundary conditions in the metric language are natural from the perspective that ``black holes are D-branes". They also arise in the brick wall discussion \cite{tHooft}, even though in the philosophy adopted in \cite{Burman1, Burman2} where the brick-wall is viewed as a UV cut-off, the precise boundary condition is not particularly important as long as it is non-dissipative.

A final point to note is that in \cite{CJ1, Chua} Chern-Simons AdS$_3$ path integrals with two boundaries have been considered before. But the boundary conditions are chosen so that the two boundaries are both asymptotically AdS, in a manner adapted to constructing the two-sided/thermofield-double black hole. Our boundary conditions are different, and so are our results and conclusions. 

\subsection{The ``Janus" Cobordism Path Integral: An Overview}
\label{section:janus}

With this basic set up, we can now discuss the general picture. This subsection should be viewed as a semi-qualitative overview to orient the reader: the ideas and calculations will be elaborated on in subsequent sections, and the reader should not dwell on any detail that seems obscure.
 
We will be interested in Euclidean signature where we will consider the theory with a torus boundary, whose modulus $\tau$ (and $\bar \tau$) fixes the temperature and angular momentum chemical potential. These potentials are conjugate to scaling dimension and spin in the dual 2D CFT, which we will combine into the holomorphic scaling dimension $h$ (and $\bar h$). Often in Euclidean AdS$_3$/CFT$_2$, one considers bulk geometries that are solid tori: this means that one of the cycles is contractible. When the spatial cycle is smoothly contractible in the bulk this corresponds to the Euclidean AdS$_3$ saddle. Conical defects are geometries with contractible spatial cycles except for the Wilson loop placed at the origin of the (otherwise contractible) spatial disc. The holonomy around the Wilson line defines the conical defect geometry, via the scaling dimension of the primary. Conical defects have non-trivial spatial and thermal holonomies\footnote{Note that the temperature is a free parameter for them. There is no smoothness condition that relates the temperature to the charges.}.  The third class of geometries that are of interest are the BTZ black holes. Here the spatial holonomy is non-trivial, but the geometry has no singularities anywhere: the spatial circle of the solid torus is non-contractible, while the thermal cycle is smoothly contractible. The resulting vanishing holonomy condition acts as the glue relating ensemble parameters and charges of the geometry. 

Note that there is a small lesson already here. The fact that conical defects with arbitrary $h$ can be defined at {\em any} temperature and chemical potential, shows that they can naturally play the role of Hilbert space microstates (of course, below the BTZ threshold) in a way that the smooth horizon BTZ black hole cannot. The latter is a saddle -- a thermodynamic configuration, not a microstate.

\begin{figure}
    \centering
    \includegraphics[width=.55\textwidth]{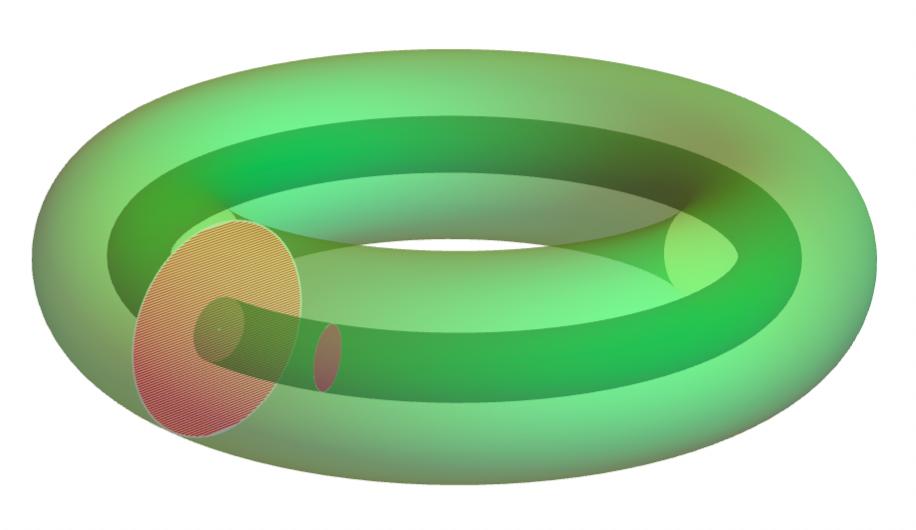}
   \caption{The thickened torus $T^2 \times I$, a ``solid torus with a solid torus removed".}
    \label{torus}
\end{figure}

Our goal will be to try and understand all of these solutions in terms of a single universal structure: a thickened torus, $M\simeq T^2\times I$, obtained by excising a thin tubular neighborhood sitting at the center of the spatial disc. See Figure \ref{torus}. This $M$ has two boundary tori: an \emph{outer} torus $\Sigma_{\rm out}$ at infinity and an \emph{inner} torus $\Sigma_{\rm in}$ that defines the (removed) tubular neighborhood. A useful starting point is to think of the conical defect as being sourced by a worldline along the thermal cycle (which acts as a source for the spatial holonomy). The key point is that the object we get when we remove this Wilson line is universal (it no longer has a holonomy label associated to it) and has the topology of the thickened torus. If you re-insert a new Wilson line with a different (elliptic)\footnote{We will discuss the connection between classes of holonomy labels and the corresponding states/geometries in the coming sections, in great detail. For the moment, all that is needed is that elliptic holonomy maps to sub-BTZ threshold primaries (conical defects) and hyperbolic holonomy maps to states above the BTZ threshold.} holonomy label, the geometry now turns into this new conical defect. If the holonomy of the inserted Wilson line is trivial, we get empty AdS$_3$ and if it is hyperbolic, we get a class of ``hyperbolic" defects corresponding to states above the BTZ threshold. These latter objects are usually ignored in discussions of AdS/CFT, but we will see that they are an important ingredient in our story. 

The path integral of (a chiral half of) the Chern-Simons theory on such a ``janus" cobordism, with conventional asymptotically AdS$_3$ boundary conditions at infinity and a cut in the interior adapted to the Dirichlet boundary condition (together with modding out by all periodic inner boundary gauge transformations), prepares a state that we will call an {\em annulus state}. Since the two-boundary path integral is Euclidean, we can view it as a propagator between the inner and outer boundaries with the bulk radial direction as time. As just mentioned, we demand that the boundary conditions at the outer torus are of the conventional asymptotically AdS$_3$ form. In effect, this means (loosely) that we declare that the state lives in a Virasoro representation, but do not specify {\em which} representation. In terms of Virasoro modules, this amounts to {\em not} specifying the primary while noticing that the descendant structure is essentially\footnote{There can be some special cases, like empty AdS$_3$, where some boundary graviton modes are part of the isometries of the geometry. We will have to remove these by hand, so the boundary graviton contribution can change slightly in these cases. We will discuss such issues in more detail later.} universal. Choosing a specific primary corresponds to choosing a particular asymptotically AdS$_3$ background (say, the vacuum, a conical defect, or a singular super-BTZ threshold state), and the descendants are the boundary graviton fluctuations on this geometry. As we mentioned previously, it is known that the CS gauge theory reduces on the boundary under these circumstances to a theory of the diffeomorphisms of the spatial circle \cite{Coussaert, CJ1} called the Alekseev-Shatashvili path integral \cite{AS1, AS2}. A noteworthy thing about this theory is that it is obtained from the bulk  \cite{CJ1}, and its path integral is one loop exact and reproduces the Virasoro character associated to the background. In other words, once we make the choice of the primary (i.e., background), the AS path integral reproduces the full character associated to that primary from the boundary gravitons. We will emphasize that this calculation is valid not just for conical defects as was discussed in \cite{CJ1}, but also for the super-threshold states sourced by hyperbolic Wilson lines that are usually ignored in AdS$_3$/CFT$_2$: they are both perfectly well-defined ``normal orbits". We will review the language of orbits in the next section.

We will elaborate this further, but the above discussion is equivalent to the statement that the annulus state produced by the path integral can be written schematically\footnote{We will write a more precise version of this, later.} as
\bea
|\Psi_{\rm ann}\rangle= \int dP |P\rangle \langle P | \Psi_{\rm out}\rangle,
\eea
where the sum is over the primary/holonomy label. The outer state $|\Psi_{\rm out}\rangle$ is defined by our asymptotically AdS$_3$ boundary conditions in bulk radial quantization. This means that from the timelike slicing of the path integral, we know that the $\langle P | \Psi_{\rm out}\rangle$ is simply the character associated to the primary/holonomy, $\chi_{P}(\tau)$.

\begin{figure}
    \hspace*{4cm}
    %\centering
    \includegraphics[width=.55\textwidth, angle=90]{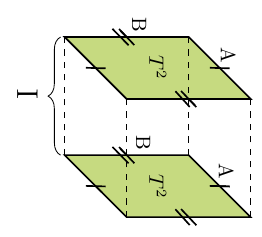}
   \caption{The topology of the janus $T^2 \times I$ geometry. The left end is the inner boundary, and projecting on to a primary corresponds to viewing the left torus as the boundary of an inner solid torus with a (not necessarily smoothly) contractible {\em  spatial} cycle.}
    \label{Spatial}
\end{figure}

Specifying the primary is equivalent to specifying the holonomy of the gauge field around the spatial cycle. It is precisely this data that is the degree of freedom at the inner boundary: as we discussed, the boundary conditions are such that all other inner boundary data are pure gauge. Specifying the holonomy at the inner boundary to be $P'$ is equivalent to projecting the above annulus state onto the primary specified by that holonomy. This is in effect a delta function projector, and the result is $\langle P' | \Psi_{\rm ann}\rangle = \chi_{P'}(\tau)$. Notably, this is precisely the result we get from the Alekseev-Shatashvili calculation on a conical defect background specified by $P'$. In short, specifying the holonomy at the inner boundary reproduces (in a natural way) the expectation from \cite{CJ1} regarding conical defects. 

But the two-boundary path integral does more. A key fact is that we have not made any assumption here that the holonomy should be elliptic\footnote{In the results in \cite{CJ1} only elliptic (i.e., conical defect) holonomies were considered. But as we will argue, only the fact that the orbit is normal is what affects the success of the calculation, and not the holonomy class.}. In particular, our observations go through even when the holonomy is hyperbolic (or parabolic). These defect states are usually not considered as physical and therefore ignored, in conventional discussions of AdS$_3$/CFT$_2$. But we will argue that it is extremely natural to consider these super-BTZ-threshold states (which also have  spatial cycles with defects) as part of the spectrum of a holographic CFT. These can be viewed as bulk microstates of the BTZ black hole. We will argue in the next sections, that the character is correctly reproduced universally from the annulus state, for super-threshold primaries, conical defects and the vacuum\footnote{The vacuum case requires handling extra zero modes. Our prescription will be to regulate all states the same way and {\em then} remove the extra zero modes for the vacuum. More on this in the next section.}.

\begin{figure}
     \hspace*{4cm}
    %\centering
    \includegraphics[width=.55\textwidth, angle=90]{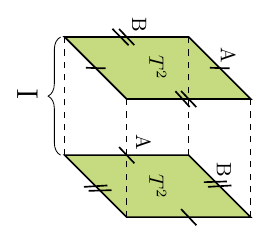}
   \caption{Projecting on to a conjugate cycle orbit corresponds to viewing the left torus as the boundary of an inner solid torus with a contractible conjugate cycle. The conjugate label in the exceptional orbit corresponds to the BTZ case: thermal cycle is {\em smoothly} contractible.}
    \label{Temporal}
\end{figure}

These observations by themselves would have been encouraging, but they come into full power when we consider BTZ black holes. In fact one of the main observations of this paper is that the BTZ partition function can be understood as a {\em sum} over these hyperbolic holonomy states, with a density of states that is  essentially\footnote{The states with hyperbolic spatial holonomy captures the primaries, which form the dominant contribution to the Cardy formula $\sim (c-1)$, in the large-$c$ limit. The boundary gravitons provide the remaining subleading contribution.} that predicted by the Cardy formula. Let us outline how this works, leaving details for later sections.

 Because the Chern-Simons action is first order, there are two canonically conjugate pieces of data: in radial quantization they correspond to holonomies around the spatial cycle and to holonomies around the thermal cycle \cite{WittenJones}. We will denote holonomies in the conjugate polarization by $\tilde P$ to distinguish them from $P$. If we specify a spatial cycle holonomy at the inner boundary, that is equivalent to specifying the state in the configuration polarization. As discussed: If this holonomy is elliptical, it is a state below the BTZ threshold and we get a conventional conical defect and the associated character, and if it is above the BTZ threshold, it denotes a black hole microstate and again reproduces the corresponding character. 

The crucial point is that we are not limited to considering projectors onto states labeled by $P$ at the inner boundary. We can also consider states labeled by the canonically conjugate variable $\tilde P$\footnote{This is a Hilbert space implementation of the re-insertion of the solid torus with a Wilson line into the tubular neighborhood, but with a contractible {\em thermal} cycle instead. See our figures and Appendix \ref{app:contractible-thermal}.}. This is analogous to working with the basis $\langle p |$ instead of  $\langle q |$ in non-relativistic quantum mechanics. The two are related by a Fourier kernel  $\langle p |= \int dq\, e^{i \,pq}\langle q |$. We will compute the analogous Fourier kernel in this paper for projectors onto the conjugate holonomy label, in the two-boundary Chern-Simons path integral. This can be done for general labels $\tilde P$, and one finds that the bulk Fourier transform is precisely the boundary modular $S$-kernel. In other words, our bulk path integral precisely reproduces \eqref{key}.

We can go one step further. One can consider the special case when the dual holonomy label is in the exceptional conjugacy class. This corresponds to putting the dual cycle in the smoothly contractible (``vacuum") configuration. The result is precisely \eqref{key0}, but now with a very transparent bulk physical interpretation: the BTZ character emerges from a sum over states with hyperbolic holonomy which are the generalizations of conical defects above the BTZ threshold (that we mentioned earlier as candidate black hole microstates). What is noteworthy about this result is that it shows explicitly that the BTZ character, which is usually obtained as a consequence of the smoothness of the geometry and the smooth contractibility of the thermal cycle, can explicitly be realized as a sum over ``unsmooth" bulk states. We will explicitly develop this and other related ideas in the following sections.

%We will outline the main storyline in this section. Some of the technical results that we will need here will be derived in later sections. The discussion here also has a review component, because (a) some ingredients may not be familiar to the average holographer, and (b) the way we are putting them together is new.

\section{Orbits and Holonomies}\label{sec:orbits}

The connection between holonomies of the Chern-Simons gauge field, coadjoint orbits of the Virasoro group, CFT states, and their bulk duals will be an essential part of our story. So we will start with a brief overview on the nature of states in our holographic system.

The standard perspective in AdS$_3$/CFT$_2$ is that the state is specified by a boundary stress tensor. The constant mode of this boundary stress tensor specifies the mass and angular momentum of the background (say $M$ and $J$ of BTZ, or equivalently, its holomorphic scaling dimension $h$) and the non-zero modes describe the boundary fluctuations (i.e., boundary gravitons). In the CFT language, $M$ and $J$ specify  the primary and the  non-constant modes correspond to the descendants. In the bulk language, specifying the stress tensor fixes a unique bulk solution of the Einstein equations called the Banados solution \cite{Banados}. In Chern-Simons language, this translates to a specific form for the gauge field, which is also uniquely specified by the boundary stress tensor, usually presented in the so-called Drinfel'd-Sokolov gauge. 

For $c > 1$ unitary 2D CFTs it turns out that there is a one-to-one correspondence between modules ${\cal V}_{c,h}$ labeled by $c$ and $h$ and the co-adjoint orbits of the Virasoro group that contain a constant representative. We explain this statement below, and use it as an opportunity to review, organize, and comment on these topics in a way that orients us for the rest of the paper. Some standard references on these topics are \cite{OrbitRefs}. We will also present some new ideas that refine existing statements in the literature in a way significant for our story, in the final two subsections.

\subsection{Virasoro Coadjoint Orbits}

Coadjoint orbits are a more general entity, but it will suffice for our purposes to start with the ``Virasoro coadjoint action" on the stress tensor. This is simply the usual transformation law of the 2D CFT stress tensor, viewed as a statement about quadratic differentials $T(\theta)\,d\theta^{2}$ on $S^{1}$:
\begin{equation}
  T \;\mapsto\; (f')^{2}\,T\!\circ f \;-\; \frac{C}{12}\,\{f,\theta\}, 
  \qquad 
  \{f,\theta\}=\frac{f'''}{f'}-\frac{3}{2}\Big(\frac{f''}{f'}\Big)^{2},
  \quad f\in \mathrm{Diff}^{+}(S^{1}). \label{coadact}
\end{equation}
$\mathrm{Diff}^{+}(S^{1})$ stands for orientation preserving diffeomorphism of the circle: in terms of $e^{i \theta} \mapsto e^{i f(\theta)}$ with a smooth lift $f : \IR \rightarrow \IR$ such that $f(\theta +2 \pi) = f(\theta) + 2\pi$, the map preserves the orientation on $S^1$ iff
\bea
f'(\theta) > 0,\ \ {\rm for\ all}\ \theta.
\eea 
A \emph{coadjoint orbit} $\mathcal O_{T}$ is the equivalence class of a given $T(\theta)$ under the co-adjoint action 
\eqref{coadact}. In other words, it is the full trajectory of a given stress tensor under the Schwarzian transformation \eqref{coadact}. The configuration variable on an orbit is the diffeomorphism $f(\theta)$. It turns out that an orbit is a symplectic manifold (with a symplectic form called the Kirilov-Kostant form) and therefore it can be (geometrically) quantized. The central charge $C$ of the orbit here should be viewed as a classical central charge, and can get finite quantum corrections (which we will discuss). The CFT central charge $c$ therefore should be distinguished from $C$.

Loosely, we can think of the quantization of an orbit as something like a Verma module. In fact, for unitary theories when there exists a stress tensor representative $T(\theta)$ on the orbit that is a constant independent of $\theta$, we can find one-to-one maps between orbits and Virasoro modules that we will describe below. This is our case of interest, and it will be convenient for us to think about orbits and modules somewhat interchangeably, even though the former is a classical entity. We will think of $\mathcal O_{T}$ as a
``classical phase space'' for the CFT stress tensor as described by the boundary gravitons on a primary state. 

\subsection{Drinfel'd-Sokolov Gauge} 

Chern-Simons gauge fields that are asymptotically AdS$_3$ naturally give rise to a coadjoint orbit structure. Let us see this in some detail; is a crucial part of our story.

Unlike in higher dimensions, the general asymptotically AdS solution of Einstein's equations in 2+1 dimensions can be explicitly written down in closed form. The Fefferman-Graham form of the asymptotically AdS$_3$ metric is therefore fully fixed, and this is the so-called Banados solution \cite{Banados}. We will not need the details in the metric language, but in terms of the Chern-Simons gauge field, we will write it as a further gauge fixing of the Banados gauge:
\[
A=b^{-1}ab + b^{-1} db,\qquad b=e^{r L_0},\qquad a=a_t\,dt+a_\phi\,d\phi,
\]
where $A_r=L_0$ and $a_t,a_\phi$ are $r$-independent\footnote{See Appendix \ref{AppConventions} for our conventions. Some aspects of DS and related gauges in the context of holonomy phase space will be further discussed in Section \ref{InnerEdge}.}. Drinfel'd-Sokolov (highest-weight) gauge further fixes
\bea
a_\phi(\phi,t)=L_{+1}-\frac{2\pi}{k}\,\mathcal L(\phi,t)\,L_{-1}
=\begin{pmatrix}
0 & -\dfrac{2\pi}{k}\,\mathcal L(\phi,t)\\[6pt]
-1 & 0
\end{pmatrix}.
\eea
The key point is that the coefficient of $L_{+1}$ is unity (highest weight condition) and that of $L_0$ is zero. This is what we will mean by an asymptotically AdS$_3$ solution in this paper. It may seem surprising that a choice of gauge can be used to define a physically meaningful idea such as asymptotic-AdS-ness. The point here is that gauge transformations are more than mere redundancies in the asymptotic region, and therefore fixing the gauge has physical content.

The gauge transformations that preserve the DS form of the gauge field can be described by the gauge parameter $\epsilon(t,\phi)$ packaged as
\begin{equation}
\lambda[\epsilon]=\epsilon\,L_{+1}-\epsilon'\,L_0
+\Big(\tfrac12\,\epsilon''-\tfrac{2\pi}{k}\,\epsilon\,\mathcal L\Big)L_{-1}
=
\begin{pmatrix}
-\tfrac12\,\epsilon' & \ \tfrac12\,\epsilon''-\tfrac{2\pi}{k}\,\epsilon\,\mathcal L\\[6pt]
-\epsilon & \ \tfrac12\,\epsilon' \label{lambda}
\end{pmatrix},
\end{equation}
where primes denote angular derivatives.
The transformation on the bare gauge field can be shown to be $\delta_\epsilon a_\phi=\partial_\phi\lambda+[a_\phi,\lambda]$, since the general gauge transformation is $\delta A = d \Lambda + [A,\Lambda]$ with $\Lambda=b^{-1} \lambda b$. One can check by direct calculation from $\delta a_\phi$ that $\lambda[\epsilon]$ keeps the DS {\em form} of $a_\phi$ unchanged, but changes the function $\mathcal L$. 

The key point is that $\mathcal L$ changes precisely by a Virasoro coadjoint action. Explicitly, on $\mathcal L$, $\epsilon$ acts as:
\begin{equation}
\delta_\epsilon\mathcal L=\epsilon\,\mathcal L'+2\,\epsilon'\,\mathcal L-\frac{k}{4 \pi}\,\epsilon'''. \label{coad-DS}
\end{equation}
Equivalently, for $T=2\pi\mathcal L$ and $c=6k$: \(\delta_\epsilon T=\epsilon T'+2\epsilon' T-\tfrac{c}{12}\epsilon'''\) 
which is the infinitesimal form of the coadjoint action on the stress tensor.

\subsection{Digression on Chemical Potential}

Even though eventually not too important for us, the discussion becomes a lot more transparent if we digress briefly to discuss the other component of the gauge field as well: $a_t$. Demanding flatness $F_{t\phi}=0$ together with the form of $a_\phi$ in DS gauge, fixes $a_t$ to be
\begin{equation}
a_t=\lambda[\mu]=\mu\,L_{+1}-\mu'\,L_0+\Big(\tfrac12\,\mu''-\tfrac{2\pi}{k}\,\mu\,\mathcal L\Big)L_{-1}
=
\begin{pmatrix}
-\tfrac12\,\mu' & \ \tfrac12\,\mu''-\tfrac{2\pi}{k}\,\mu\,\mathcal L\\[6pt]
-\mu & \ \tfrac12\,\mu'
\end{pmatrix}\label{DS-at}
\end{equation}
and also specifies the coadjoint flow of $\mathcal L$:
\begin{equation}
\dot{\mathcal L}=\delta_\mu\mathcal L
=\mu\,\mathcal L'+2\,\mu'\,\mathcal L-\frac{k}{4\pi}\,\mu'''. \label{coadflow}
\end{equation}
The quantity $\mu(t,\phi)$ is sometimes called the chemical potential and can be viewed as a source. 

Note that if we have no particular requirements on $a_t$, then the gauge parameter $\epsilon(t,\phi)$ can have arbitrary time-dependence. The coadjoint action that arises from the requirement that the $a_\phi$ should retain DS form, only depends on the angular derivatives. 

However, if we demand that the {\em form} of $a_\phi$ and $a_t$ are fixed to be DS-compatible, while also requiring that the source $\mu$ is held fixed as a $function$, then the time-dependence of the gauge parameter is constrained to be
\bea
\partial_t \epsilon = \mu \epsilon' - \epsilon \mu'
\eea
We show this explicitly, in Appendix \ref{ResidualDS}.

Fixing $\mu$ can often be viewed as a choice of boundary condition: the Brown-Henneaux (asymptotically AdS$_3$) boundary condition that we are interested in, sets $\mu=1$ in DS gauge. From the forms of $a_t$ and $a_\phi$ that we have presented above, it follows that this forces the ``chirality" condition $a_t =a_\phi$. In radial gauge, $a_t$ and $a_\phi$ are radially constant, therefore this condition can be viewed as emerging from a boundary condition. In fact, the boundary term that we discussed in the previous section guarantees a good variational principle when the chirality condition holds, 
\begin{equation}
a_t\big|_{\partial}=a_\phi\big|_{\partial}\,
\end{equation}
and is the underlying reason behind the emergence of chiral WZW (and eventually Alekseev-Shatashvili) theories from boundary reduction. This can also be seen by noting from \eqref{coadflow} that the gauge transformation leads to flows that are chiral when $\mu=1$:
\begin{equation}
\dot{\mathcal L}=\mathcal L'\,.
\end{equation}

\subsection{Hill's Equation}

Consider the linear problem for a 2-component column vector $\Psi=(\psi_1,\psi_2)^{\!\top}$: 
\((\partial_\phi+a_\phi)\Psi=0.\) In components, with $a_\phi$ being DS gauge field, we get
\[
\psi_1'-\frac{2\pi}{k}\,\mathcal L\,\psi_2=0,\qquad \psi_2'-\psi_1=0.
\]
Eliminating $\psi_1$ yields Hill's equation
\begin{equation}
\psi''(\phi) - q(\phi)\,\psi(\phi)=0,
\qquad
q(\phi)\equiv \frac{2\pi}{k}\,\mathcal L(\phi) = \frac{6}{C} T(\phi).
\label{Hill}
\end{equation}
The notation $(q, \mathcal{L}, T)$ is quite redundant, but it is also quite convenient: $q$ is simplest to describe the holonomies, $T$ is natural to connect with the CFT stress tensor, and $\mathcal{L}$ is the standard in connecting with the gravitational side via the Banados metric.
In any event, in DS gauge the boundary stress-tensor density plays the role of the potential in Hill's equation. We are emphasizing the $\phi$-dependence of the quantities here, but as discussed before, the true dependence is chiral with Brown-Henneaux boundary conditions. This does not affect the following holonomy/monodromy discussion.

Let $u(\phi)$ and $v(\phi)$ be two independent solutions of \eqref{Hill}. Their Wronskian $W:=u'v-uv'$ is constant, and by rescaling the basis one can set $W=1$ without loss of generality. If we define the projective coordinate $f(\phi):=u(\phi)/v(\phi)$,
\begin{equation}
f'=\frac{1}{v^2},
\qquad
\frac{f''}{f'}=-2\,\frac{v'}{v},
\end{equation}
and 
\begin{equation}
S(f,\phi):=\frac{f'''}{f'}-\frac{3}{2}\Big(\frac{f''}{f'}\Big)^2
= -\,2\,q(\phi).
\label{eq:Schwarzian}
\end{equation}
Equation \eqref{eq:Schwarzian} is the standard identification of the projective connection\footnote{Connection, because $q \sim \mathcal L \sim T$ fixes the gauge field, and projective, because it only depends on the ratio of $u$ and $v$ and therefore the transport around the circle (as we will see) lives in $PSL(2,\IR)$ and not $SL(2,\IR)$.} with the Schwarzian of $f$. Coadjoint action of the stress tensor can be obtained via compositions of Schwarzians. This shows that Virasoro coadjoint orbits are the same as the orbits of projective connections under circle diffeomorphisms.

\subsection{Orbit Invariants and Monodromy as Holonomy}

Assemble $u,v$ into the fundamental matrix
\begin{equation}
Y(\phi):=
\begin{pmatrix}
u'(\phi) & v'(\phi)\\[2pt]
u(\phi)  & v(\phi)
\end{pmatrix},
\qquad
Y'(\phi) = -\,a_\phi(\phi)\,Y(\phi),
\qquad
\det Y(\phi)=W=1.
\label{eq:Ydef}
\end{equation}
The monodromy of the fundamental matrix is (as we will argue) an orbit invariant and therefore we will use it to label orbits, and eventually connect with primaries and Virasoro modules. The monodromy of $Y$ as it goes around the circle is simply the matrix $M$ defined by $Y(2 \pi) = M\, Y(0)$. Therefore, 
\begin{equation}
M := Y(2\pi)\,Y(0)^{-1}
= \mathcal P\exp\!\Big(-\!\int_0^{2\pi} a_\phi\,d\phi\Big)\in SL(2,\mathbb R),
\label{eq:monodromy}
\end{equation}
which is exactly the $SL(2,\mathbb R)$ Chern-Simons holonomy along the boundary loop. In other words, the holonomy of the Chern-Simons gauge field can be used to label orbits.

Writing $M=\bigl(\begin{smallmatrix}a&b\\ c&d\end{smallmatrix}\bigr)$ and transporting once around the circle gives
\begin{equation}
\begin{pmatrix} u(2\pi)\\ v(2\pi)\end{pmatrix}
= M \begin{pmatrix} u(0)\\ v(0)\end{pmatrix}
\quad\Longrightarrow\quad
f(2\pi)=\frac{u(2\pi)}{v(2\pi)}
=\frac{a\,u(0)+b\,v(0)}{c\,u(0)+d\,v(0)}
=\frac{a\,f(0)+b}{c\,f(0)+d}.
\end{equation}
Thus the analytic continuation of $f$ around $S^1$ is the $PSL(2,\mathbb R)$ Möbius action of $M$. The $PSL(2,\mathbb R)$ monodromy of Hill's equation (equivalently, of $f$) is the $SL(2,\mathbb R)$ holonomy \eqref{eq:monodromy} modulo the center $\{\pm \mathbf 1\}$. This projective nature of the solutions has significance for (asymptotically) AdS$_3$ solutions of Chern-Simons theory, as emphasized in \cite{CJ1}.

Let us prove that the monodromy is an orbit invariant. Since we have already shown the map from monodromy to holonomy, we will work with holonomies instead. Under any finite gauge transformation $G$,
\begin{equation}
a_\phi \;\longrightarrow\; a_\phi^{G} = G^{-1}a_\phi G + G^{-1}\partial_\phi G,
\qquad
Y(\phi)\;\longrightarrow\; Y^{G}(\phi)=G(\phi)^{-1}\,Y(\phi)\,G(0).
\end{equation}
To demonstrate the latter relation, we need to show that $Y^{G}(\phi)=G(\phi)^{-1}\,Y(\phi)\,G(0)$ satisfies $Y^{G}{}'=-\,a_\phi^{G}\,Y^{G}$ and $Y^{G}(0)=\mathbf 1$, given that $Y(\phi)$ solves $Y'(\phi)=-\,a_\phi(\phi)\,Y(\phi)$ with $Y(0)=\mathbf 1$. To shows this, use $\frac{d}{d\phi}\big(G^{-1}Y\,G(0)\big)=(G^{-1})'\,Y\,G(0)+G^{-1}\,Y'\,G(0)$ and $
(G^{-1})'=-\,G^{-1}G'\,G^{-1}$ in:
\[
\begin{aligned}
(Y^{G})'
&=(G^{-1})'Y\,G(0)+G^{-1}Y'\,G(0) \\
&=\big(-G^{-1}G'G^{-1}\big)Y\,G(0)+G^{-1}(-a_\phi Y)\,G(0) \\
&=-\big(G^{-1}G'+G^{-1}a_\phi G\big)\,\underbrace{G^{-1}Y\,G(0)}_{Y^{G}} \\
&=-\big(G^{-1}a_\phi G+G^{-1}G'\big)\,Y^{G} =-\,a_\phi^{G}\,Y^{G}.
\end{aligned}
\]
Together with $Y^{G}(0)=G(0)^{-1}Y(0)G(0)=\mathbf 1$, this shows that $Y^{G}$ satisfies the transformed equation with the same initial condition:
\[
Y^{G}{}'=-\,a_\phi^{G}\,Y^{G},\qquad Y^{G}(0)=\mathbf 1.
\]
This means that the monodromy transforms by endpoint conjugation
\begin{equation}
M \;\longrightarrow\; M^{G}=G(2\pi)^{-1}\,M\,G(0).
\label{eq:Mendpoint}
\end{equation}
For our DS-preserving, periodic $G$ one has $G(2\pi)=G(0)$, hence
\begin{equation}
M^{G}=G(0)^{-1}\,M\,G(0),
\end{equation}
i.e. the \emph{conjugacy class} $[M]$ is unchanged\footnote{Because we will only be interested in its conjugacy class, we will often use the word ``holonomy" to stand for ``the conjugacy class of the holonomy".}. Passing to $PSL(2,\mathbb R)$ removes the central ambiguity $\pm \mathbf 1$, so the $PSL(2,\mathbb R)$ monodromy of $f$ (equivalently, of the first-order system) is invariant along the entire Virasoro coadjoint orbit through $q$.

\subsection{Conjugacy Classes for Orbits} 

Because $[M]$ is an orbit invariant, it is useful to specify an orbit by describing its conjugacy-class. For $M\in SL(2,\mathbb R)$ the characteristic polynomial is
\begin{equation}
\chi_M(\lambda)=\lambda^2-(\mathrm{Tr}\,M)\lambda+1,
\qquad
\Delta:=(\mathrm{Tr}\,M)^2-4.
\label{eq:charpoly}
\end{equation}
Standard facts about matrices can be used to argue that conjugacy classes in $SL(2,\mathbb R)$ are determined by the sign of $\Delta$ together with whether $M$ is central ($\pm I_2$) or non\-central at $\Delta=0$:
\begin{itemize}
  \item \emph{Hyperbolic} ($\Delta>0$, equivalently $|\mathrm{Tr}\,M|>2$): two distinct real eigenvalues $\lambda,\lambda^{-1}$ with $\lambda>1$. One may conjugate to a diagonal form
  \begin{equation}
  M\sim \begin{pmatrix}\lambda&0\\[2pt]0&\lambda^{-1}\end{pmatrix},\qquad \lambda\in\mathbb R\setminus\{0,\pm1\}.
  \end{equation}
\item \emph{Parabolic} ($\Delta=0$, equivalently $|\mathrm{Tr}\,M|=2$): here we reserve “parabolic” for the \emph{noncentral} case (non\-diagonalizable with a single Jordan block). Up to $SL(2,\mathbb R)$-conjugacy,
\[
M \sim \begin{pmatrix}1&1\\[2pt]0&1\end{pmatrix}
\quad\text{or}\quad
M \sim \begin{pmatrix}-1&1\\[2pt]0&-1\end{pmatrix}.
\]
The central matrices $M=\pm I_2$ at the same trace are usually \emph{not} treated as parabolic. They form their own conjugacy classes in $SL(2,\mathbb R)$ and become the identity in $PSL(2,\mathbb R)$. We will mostly be concerned with elliptic and hyperbolic cases, so will often not discuss the parabolic case in detail.
  \item \emph{Elliptic} ($\Delta<0$, equivalently $|\mathrm{Tr}\,M|<2$): complex-conjugate eigenvalues $e^{\pm i\theta}$ with $\theta\in(0,\pi)$. One may conjugate to a rotation
  \begin{equation}
  M\sim R(\theta):=\begin{pmatrix}\cos\theta&\sin\theta\\[2pt]-\sin\theta&\cos\theta\end{pmatrix},
  \qquad \mathrm{Tr}\,M=2\cos\theta.
  \end{equation}
\end{itemize}
Within hyperbolic and elliptic types there is a unique class for each value of the continuous parameter ($\lambda>1$ or $\theta\in(0,\pi)$).

\subsection{Specialization to Constant Stress Tensor: $\mathcal L=b_0$} 

It should be emphasized that while the holonomy/monodromy conjugacy class is an invariant of the orbit, it does not uniquely fix the orbit: it is a coarse invariant. Things improve, when we consider orbits with constant representatives for the stress tensor. In AdS$_3$/CFT$_2$ language, these are the cases that map to Virasoro modules. Loosely: the scaling dimension of the primary in the module is what maps to the constant representative in the orbit, and the descendants furnish the orbit itself (after quantization). 

In DS gauge, as we saw, the angular component is $a_\phi=L_{+1}-q\,L_{-1}$ with $q=\frac{2\pi}{k}\,\mathcal L(\phi)$.
If $\mathcal L(\phi)=b_0$ is constant, then $a_\phi$ is constant and the holonomy along the spatial circle is
\begin{equation}
M=\exp\!\big(-2\pi\,a_\phi\big).
\end{equation}
The eigenvalues of $a_\phi$ are $\pm\sqrt{q_0}$ (with $q_0 \equiv \frac{2\pi}{k}\, b_0 $), so the eigenvalues of $M$ are $e^{\mp 2\pi\sqrt{q_0}}$. Hence
\begin{equation}
\frac{1}{2}\,\mathrm{Tr}\,M=
\begin{cases}
\cosh\!\big(2\pi\sqrt{q_0}\big), & q_0>0,\\[4pt]
1, & q_0=0,\\[4pt]
\cos\!\big(2\pi\sqrt{|q_0|}\big), & q_0<0.
\end{cases}
\label{eq:halftrace-piecewise}
\end{equation}
Therefore:
\begin{itemize}
  \item $q_0>0$ : $M$ is hyperbolic $\Rightarrow$ BTZ black hole (general BTZ when combined with the right\-moving sector) or super-BTZ defects,
  \item $q_0=0$ : $M$ is parabolic/unipotent $\Rightarrow$ massless BTZ/defect,
  \item $q_0<0$ : $M$ is elliptic $\Rightarrow$ conical defect geometry. For elliptic orbits it is convenient to introduce the orbit label $\mu > 0$ via $q_0 = -\mu^2$, in terms of which $M$ is a rotation by an angle $2 \pi \mu$. This gives an explicit realization of our earlier observation that holonomy class (or ${\rm Tr}M$) does not uniquely fix the orbit: both $\mu^2$ and $(\mu+n)^2$ with $n \in \IZ$ lead to the same ${\rm Tr}M$, but it is a fact (even though we will not prove it) that they are distinct Virasoro orbits. We will restrict our attention to one fundamental domain of $\mu$, that is $\mu \in(0,1)$.
\end{itemize}
The central holonomy cases also play a role, and includes the vacuum (empty AdS$_3$). But it is useful to discuss an alternate classification of orbits in order to discuss them.

\subsection{Normal and Exceptional Orbits}
\label{subsec:normal-exceptional}

So far we have discussed the classification of orbits in terms of holonomies. But it is also useful to classify them in terms of their stabilizer groups.

Roughly, the stabilizer is the subgroup of diffeomorphisms that leaves a given stress-tensor configuration unchanged. Concretely, start from some stress tensor $T(\theta)$ on the circle and act on it with a diffeomorphism $f(\theta)$ using the Virasoro coadjoint action (the Schwarzian transformation law in \eqref{coadact}). The \emph{stabilizer} $\mathrm{Stab}(T)\subset \mathrm{Diff}^+(S^1)$ is defined as the set of $f$ for which the transformed stress tensor is exactly the same function:
\begin{equation}
  T(\theta)\ \longrightarrow\ T^f(\theta) = T(\theta)
  \quad\Longleftrightarrow\quad
  f\in \mathrm{Stab}(T)\,.
\end{equation}
The coadjoint orbit $\mathcal{O}_T$ can then be thought of schematically as
\begin{equation}
  \mathcal{O}_T \;\simeq\; \mathrm{Diff}^+(S^1)\big/\mathrm{Stab}(T)\,,
\end{equation}
in direct analogy with the Wigner classification of particles as ``group/little group''. Physically, $\mathrm{Stab}(T)$ is the group of global symmetries (rigid reparametrizations) that the configuration has: in the CFT it is the subgroup of the Virasoro group that leaves the background stress tensor invariant, and in the bulk it matches onto the isometry group of the corresponding Bañados geometry.

For the orbits we care about, the representative is a \emph{constant} stress tensor. On the cylinder we write $T(\theta)=2\pi b_0$ with classical central charge $C$, so that $b_0$ is the constant Virasoro coadjoint element, and the orbit is generated by acting with arbitrary diffeomorphisms. The size of the stabilizer then tells us how ``symmetric'' that constant background is.

\subsubsection{Normal Orbits: $U(1)$ Stabilizer}

For a generic constant value $b_0$ (equivalently generic holonomy parameter $q$), the only circle diffeomorphisms that leave $T(\theta)=2\pi b_0$ invariant are rigid rotations
\begin{equation}
  \theta \;\to\; \theta + \text{constant}\,,
\end{equation}
generated by the zero-mode of the stress tensor. In other words, the stabilizer is just the circle of rigid translations:
\begin{equation}
  \mathrm{Stab}_{\text{normal}}(b_0) \;\simeq\; U(1)\,.
\end{equation}
We will refer to the corresponding coadjoint orbits as \emph{normal orbits}. They are the ``generic'' constant base point orbits: the Virasoro group acts freely except for the obvious $U(1)$ of going around the cylinder. In orbit language one can write
\begin{equation}
  \mathcal{O}_{\text{normal}} \;\simeq\; \mathrm{Diff}^+(S^1)\big/ U(1)\,,
\end{equation}
and (it turns out) after quantization these orbits give the usual highest-weight Virasoro modules with no extra null states.

In the gravitational dictionary, constant normal orbits include both:
\begin{itemize}
  \item elliptic monodromy with $b_0<0$: conical defects below the BTZ threshold;
  \item hyperbolic monodromy with $b_0>0$: BTZ-type primaries (ie., defects above threshold);
\end{itemize}
as well as the marginal parabolic case. We will relate $b_0$ to the scaling dimension in the next subsection.

\subsubsection{Exceptional Orbits: $PSL(2,\mathbb{R})$  Stabilizer}

There is, however, a discrete set of constant stress tensors for which the stabilizer is \emph{larger} than $U(1)$. These are the \emph{exceptional orbits}. It turns out that these are all orbits with central holonomy, i.e, $q_0 = - m^2/4$ with $m \in \IZ^+$ in our discussion in the previous subsection (equivalently, $b_0 = -k\, m^2/8 \pi= -C\, m^2/48 \pi$). The most important one for us is the vacuum orbit, corresponding to $m=1$:
\begin{itemize}
  \item For the particular value
  \begin{equation}
    b_0^{\text{vac}} = -\frac{C}{48\pi}\,,
  \end{equation}
  corresponding to the vacuum stress tensor $T_{\text{vac}} = -C/24$ on the cylinder, the stabilizer is the full global conformal group
  \begin{equation}
    \mathrm{Stab}_{\text{vac}} \;\simeq\; PSL(2,\mathbb{R})\,,
  \end{equation}
  generated by the modes $L_{-1}, L_0, L_{+1}$. The orbit is
  \begin{equation}
    \mathcal{O}_{\text{vac}} \;\simeq\; \mathrm{Diff}^+(S^1)\big/ PSL(2,\mathbb{R})\,,
  \end{equation}
  and its quantization gives precisely the Virasoro vacuum representation. In the bulk this is the orbit corresponding to smooth global $\mathrm{AdS}_3$: its extra stabilizer modes are just the three $SL(2,\mathbb{R})$ isometries of the $\mathrm{AdS}_3$ ground state.
\end{itemize}

More generally, the Virasoro classification allows a discrete family of ``higher exceptional'' constant orbits where the stabilizer is again three-dimensional but embedded using modes $(L_{-m},L_0,L_{+m})$ instead of $(L_{-1},L_0,L_{+1})$. They occur at special constant values
\begin{equation}
  b_0 = -\frac{C\,m^2}{48\pi}\,, \qquad m=1,2,3,\dots\, 
\end{equation}
For $m=1$ this is the vacuum orbit above; for $m>1$ one gets orbits whose quantization yields degenerate Virasoro representations of type $(1,m)$. These higher exceptional representations are non-unitary for large central charge, so we will not discuss them further. From a holographic point of view these are states with energy ``below the vacuum".

In the next subsection, we will now relate these orbit labels and their stabilizers to the usual CFT data $(c,h,\bar h)$ and to the bulk parameters. Note that normal orbits can have holonomies that lie above or below the BTZ threshold (ie., the orbit can be elliptic or hyperbolic).

\subsection{Quantization, Orbit-Primary Map, and Virasoro Characters}\label{sec:refinement}

Consider an orbit on the cylinder with representative left/right stress tensors that are constants: $\mathcal L=b_0$ and $\bar{\mathcal L}=\bar b_0$. We take $C=6k$ and keep the AdS radius $\ell$ (sometimes set to 1) and Newton's constant $G$. %Consolidating our previous observations, in this subsection we write down the semi-classical connection between orbits, primaries and bulk states. We present the holomorphic and anti-holomorphic pieces explicitly for completeness, elsewhere in the paper we will generally suppress the anti-holomorphic piece.

From standard semi-classical AdS$_3$/CFT$_2$ dictionary, as well as  
$T=2\pi\,\mathcal L=2\pi\,b_0$ and $ \bar T=2\pi\,\bar{\mathcal L}=2\pi\,\bar b_0$, it follows that
\begin{equation}
h_0 - \frac{C}{24} \;=\;  2\pi\,b_0, \qquad \bar h_0 -\frac{C}{24} \;=\;  2\pi\,\bar b_0.
\label{eq:h-b0-map}
\end{equation}
The subscript on (say) $h_0$ is to emphasize that these quantities are classical  and may get quantum corrections. The classical central charge $C$ can also get $\mathcal{O}(1)$ quantum corrections, and the full CFT central charge we will write as $c$. Because they are quantum effects, we expect the corrections in $h_0$ and $C$ to {\em not} scale with $C \sim c$ in a large-$c$ holographic theory, as we will see momentarily.

The quantization of orbits in a form that we will directly use is given in the illuminating paper of Cotler and Jensen \cite{CJ1}. We will state their results as follows:
\begin{itemize}
\item For the exceptional orbit, which they interpret as the vacuum, 
\bea
h=h_0=0, \ \ \ c=C+13.
\eea
\item For normal orbits, which they view as conical defects \footnote{The result is in fact valid for any normal orbit -- above or below the BTZ-threshold.}, 
\bea
h=h_0=2 \pi b_0+\frac{C}{24}, \ \ \ c=C+1.
\eea
\end{itemize}
The basic strategy of \cite{CJ1} was to write down the boundary Alekseev-Shatashvili actions associated to chiral boundary gravitons for bulk backgrounds (vacuum AdS$_3$ and conical defects\footnote{Using modular-S on the vacuum, they also obtained the BTZ character.}). This leads to one-loop exact quantizations of exceptional and normal orbits, reproducing the respective Virasoro characters of these bulk geometries. By matching the bulk/orbit parameters $2\pi b_0$ and $C$ in the results obtained this way, with the expected forms of the characters
\bea
\chi_{vac}(\tau)=q^{-\frac{c}{24}} \prod_{n=2}\frac{1}{1-q^n}, \ \ \ \chi_{h}(\tau)=q^{\,h-\frac{c}{24}} \prod_{n=1}\frac{1}{1-q^n},
\eea
we can fix the central charge and the quantum scaling dimension in the two cases. This is what we have presented above\footnote{For the normal orbit, one also uses the stress tensor two point function in the primary, to break the degeneracy between $h$ and $c$ in the character. More on this, later.}.

Note that in order to view the set of orbits/primaries as belonging to a {\em single} holographic CFT, it is crucial for us that there is a {\em single} CFT central charge. This point was not important in the discussions in \cite{CJ1} where the quantizations of each of the orbits were largely treated as independent. A related point is that the theories in \cite{CJ1} are not modular invariant, because they are viewed as independent theories of boundary gravitons. Since our goal is to try and understand features of the heavy bulk spectrum by exploiting modular invariance, we will need to be more ambitious: we want these quantized orbits to all fit inside a single CFT, with a single central charge $c$. 

This is {\em not} possible as it stands with the results we have reported above\footnote{We thank Kristan Jensen for confirming that this is indeed the case.}. The reason is that there is immediate tension between the $c$'s for the two kinds of orbits. It is clear from our formulas above, that the quantum corrections in the classical Chern-Simons level $k$ (or equivalently, $C$) are background dependent. This is not believed to be natural. For example, in the case of compact Chern-Simons theories, if we start with level $k$ in the classical theory, the quantum corrections do not depend on the background \cite{WittenJones, Seiberg}. This is in fact a necessary condition for the ``functoriality" of Chern-Simons TQFT. More directly (and problematically for the holographer), it is clear that if we have the same value of $C$ to start with, our $c$'s cannot both be the same. This is very much a deal-breaker if we want to work with a single holographic CFT\footnote{Here is a candidate (perhaps not so unnatural) way to bypass this problem, which we do {\em not} pursue: Maybe we could allow the possibility that we are starting with distinct classical levels $k$ in different (ie., normal vs exceptional) backgrounds, so that the quantum central charges match up. It is not clear to us that this is an {\em a priori} bad idea, because at large-$c$, $k$ is also large and therefore $\mathcal{O}(1)$ corrections in the classical couplings are as minuscule as quantum gravity effects. It is not clear that we have precise enough control on classical couplings to truly call this a problem. But it turns out that this possibility runs into trouble in later sections, with our Fourier kernel (which depends on the spatial holonomy, and through it the scaling dimension). The (technical) culprit is the fact that the orbit scaling dimension gets no quantum corrections \cite{CJ1}: $h=h_0$. We will later run into requirements like $h-\frac{(c-1)}{24} =2 \pi b_0 \equiv h_0-\frac{C}{24}$ from normal orbits, which are impossible to satisfy with $c=C+13$ arising from the exceptional orbit, because $h=h_0$.}. 

Our proposal that holographic spacetimes should be understood as configurations with inner boundaries (that are projected on to the relevant primary/Wilson line), suggests a resolution to this problem. In particular, it suggests that empty AdS$_3$ should also be viewed as a special case of such a geometry, where the state in the interior {\em happens} to be projected on to the vacuum. This happens when the re-inserted Wilson line has vanishing spatial holonomy. From the orbit point of view, this implies that we should treat {\em all} solutions as normal orbits, and view the vacuum case as a suitable {\em limiting} normal orbit. Operationally, this means (as we will see) that we should ``regulate all orbits like they were normal orbits first, and {\em then} quotient by the appropriate stabilizer". \\

\noindent
Let us discuss how and why this works:
\begin{itemize}
    \item Firstly and tautologically, it bypasses the clash between the quantizations of the different orbits we mentioned previously. Now all orbits satisfy the general rule:
    \bea
    h=h_0=2 \pi b_0 +\frac{C}{24}, \ \ \ \ c=C+1.
    \eea
    So we at least have the {\em possibility} that all quantized orbits can live within a single holographic CFT.
    \item For this proposal to be acceptable, we need that the  limiting versions of the normal orbit results should be consistent with our expectations about the vacuum state of the CFT on the cylinder, with $c=C+1$. The first check we can do is eqn. (5.57) in \cite{CJ1}:
    \begin{equation}
\langle T(w) T(0) \rangle_{b_0}
= \frac{C+1}{32 \sin^{4}\!\left(\frac{w}{2}\right)}
  - \frac{2\pi\left( b_{0} + \frac{C}{48\pi} \right)}{2 \sin^{2}\!\left(\frac{w}{2}\right)} \, .
\label{TT-primary-CJ}
\end{equation}
    This is the result of a normal orbit calculation, exploiting the fact that the corresponding AS action is secretly a free theory. By comparison with the standard stress tensor two-point function on a cylinder in a primary of weight $h$, it was argued that the coefficient of the second term should in fact equal $h$. It immediately follows from our definition of the exceptional orbit earlier that this coefficient vanishes on it. In other words, even though we have moved the central charge from $c=C+13$ to $c=C+1$ (which is a consequence of the coefficient of the first term in the TT correlator above), the scaling dimension remains $h=0$. Note that this need not have been the case. The fact that $h=h_0=0$ here, is the result of a calculation for {\em normal} orbits \cite{CJ1}. Stated differently, the fact that the coefficient of the second term in \eqref{TT-primary-CJ} has the form it does in a {\em normal} orbit calculation, is the reason why we are able to straightforwardly interpret it as a valid result for the exceptional orbit\footnote{The authors of \cite{CJ1} do not present the 2-point correlator calculation, directly in the exceptional orbit. The reason seems to be that the field redefinition that makes the action free, still does not quite result in a free theory because of the $PSL(2,\IR)$ quotienting. Instead they make the (plausible) assumption that $h=0$ after quantization, to avoid relying on the TT calculation. It is conceivable that there is a way around this, that exploits some clever heuristic to get a ``manifestly exceptional orbit" version of \eqref{TT-primary-CJ}. For this calculation to be consistent, the $C+1$ in the first term must be  replaced by $C+13$ and the second term should remain unaffected. That will establish that $h=h_0=0$ directly in the exceptional orbit. In any event, for our purposes a direct exceptional orbit calculation is of not much interest, because we view it as a regulated normal orbit.}.  
    \item Let us also point out a calculation in \cite{Hijano} where the boundary stress tensor 2-point function in the vacuum is computed from bulk gravity, and again results in the $c=C+1$ shift and {\em not} $c=C+13$. This is loosely because the asymptotic behavior of the bulk fields is what is crucial for their calculation, and therefore their regulator is not sensitive to global isometries of the background. While it is not entirely clear to us whether this calculation is truly an independent piece of evidence (in light of the various known facts about AdS$_3$/CFT$_2$), we believe it comes close in spirit to our perspective.
    \item But the stress tensor 2-point function is not enough. We need to also make sure that there is a natural way to obtain the vacuum character as a limit of our normal orbit. Interestingly, we can make such a proposal, both from the orbit perspective as well as the bulk gravity perspective. Let us first consider the former. It is straightforward to see that the crucial difference between the normal and exceptional orbits in the (one loop exact) calculation, is in the range of modes that get summed over in the fluctuation determinant. Let us phrase this difference at the level of (say) eqn. (5.21) in \cite{CJ1} by adapting it to a normal orbit: 
    \begin{equation}
\sum_{n=1}^{\infty} n \cot(\pi n \tau)
\;\to\;
\sum_{n=1}^{\infty} n\bigl(\cot(\pi n \tau) + i\bigr)
\;-\;
i \sum_{n=1}^{\infty} n \, .
\label{eq:CJ-5.21}
\end{equation}
Note that the sum here is from $n=1$, to allow for the fact that we are {\em not} quotienting by the enlarged stabilizer $PSL(2,\IR)$. This leads to a $-\frac{1}{12}$ shift instead of $-\frac{13}{12}$ in zeta function regularization. Together with the classical contribution, it leads to eqn. (5.24) of \cite{CJ1}:
\bea
Z_{1\text{-loop}}
= q^{\,h - \frac{c}{24}} \prod_{n=1}^{\infty} \frac{1}{1 - q^{n}},
\qquad
h - \frac{c - 1}{24} = 2\pi b_{0} \, .
\label{eq:CJ-5.24}
\eea
How do we adapt this result to the exceptional orbit? By our previous bullet points, the vacuum corresponds to $h=0$: the one-loop partition function is {\em almost} that of the vacuum, except that we have not removed the $n=1$ modes. But removing it is a natural thing to do because in the limit when the normal orbit tends to the vacuum, there is an enhanced symmetry. In effect, what we are proposing is to regulate first and then quotient by the extra modes corresponding to $PSL(2,\IR)$ (normal orbit philosophy), instead of removing the modes first and then regulating (exceptional orbit philosophy) which was what was done in \cite{CJ1}. We will see more evidence for the validity of our philosophy, below.
\item There is also a gravity side to this story, implicit in the work of Giombi, Maloney and Yin \cite{Giombi} and made explicit in \cite{CJ1}. The $+13$ arises here from a short distance divergence in the regulated one loop determinant evaluated by heat kernel methods. The details are in fact not important for us: what we need for our proposal to work, are four things. Firstly, we need to compute the analogous (regulated) one loop gravity partition function on a conical defect. Secondly, we need to take the limit where the conical defect tends to the vacuum AdS$_3$: that is, in the oft-used notation for the $\IZ_N$ conical defect (see e.g., \cite{Benjamin}), we want to consider $N\rightarrow 1$. Thirdly, when we take this limit, the geometry will have extra isometries, and we should remove contributions coming from those modes from the product. Fourth, when we are done, we need the (chiral half of the) partition function to be precisely,
\bea
Z_{1\text{-loop}}
= q^{\, - \frac{C+1}{24}} \prod_{n=2}^{\infty} \frac{1}{1 - q^{n}}.
\label{BHY}
\eea
Note that this involves multiple moving parts and is a fairly non-trivial test of the viability of our proposal. In particular, it means that there has to be a sort of ``cancellation" between the $+13$ from the empty AdS$_3$ geometry and the divergences intrinsic to the conical defect. Remarkably, it turns out that this is precisely true: the conical defect calculation has been done in \cite{Benjamin}. Taking a look at their eqn.(4.23), we see that the chiral half is simply the $h=0$ version of the $c=C+1$ Virasoro character. And modding out by the extra isometries in the empty AdS$_3$ limit leads us to remove exactly the $n=1$ modes. The result is the vacuum character above. Note that again, the basic idea is the same: regulate-and-quotient rather than quotient-and-regulate.
    \item A somewhat circumstantial reason\footnote{It is also worth pointing out here, that for chiral CFTs, it is known that the $+13$ shift (or for that matter any shift that is not a multiple of $24$) is inconsistent \cite{Eberhardt}. It is argued in \cite{Eberhardt} that the $+13$ can be attributed to a ``metaplectic correction" term in geometric quantization which is not present in the chiral CFT setting. While this discussion is for chiral/extremal CFTs \cite{Witten3Dnew} and not directly applicable to our setting, it does illustrate that these quantum corrections are not beyond question. See also \cite{Hijano}. } to think that our proposal is reasonable, is that (as we will emphasize later) the space of Virasoro characters can be viewed as a basis for a Hilbert space. The natural path to obtain the vacuum character in this language is by versions of analytic continuation (and demands of unitarity) from the non-degenerate characters. In effect, this is a consequence of the fact that all orbits are now contained within a single CFT, and we can adapt 2D CFT methods. We will use a related approach in section \ref{conjugBTZ} to show the emergence of the BTZ character as a sum over bulk states. 
\end{itemize}

We have presented some of the immediate technical reasons that support our proposal, and prevent it from arriving dead in the water. But it should be emphasized that this entire paper implicitly uses this perspective. In what follows, whenever we talk about the exceptional orbit, it will be in this regulated normal orbit way, inspired by the janus cobordism path integral of the previous section.

\subsection{A Notational Interlude}

We will often write Virasoro characters for normal orbits as
\[
  \chi_{P}(\tau) \;=\; \frac{q^{\,P^{2}}}{\eta(\tau)},
\]
with $P$ serving as an orbit label. For a general (non-vacuum) primary, the character is usually labeled using its scaling dimension and has a few useful explicit forms:
\[
  \chi_{h}(\tau) \;=\; q^{\,h-\frac{c}{24}} \prod_{n\ge 1}\frac{1}{1-q^n}
  \;=\; q^{\,h-\frac{c}{24}} \frac{q^{\,1/24}}{\eta(\tau)}
  \;=\; \frac{q^{\,h-\frac{c-1}{24}}}{\eta(\tau)}\,.
\]
This means that after factoring the universal descendant tower $1/\eta(\tau)$ (which carries a $q^{+1/24}$) we have a map between orbit labels and scaling dimensions: 
\bea
P^{2}=h-\frac{c-1}{24}.
\eea
As noted in the previous subsection, one can quantize a normal co-adjoint orbit via the AS action associated to the classical CS gauge field/orbit \cite{CJ1} and we find the relation:
\bea
2 \pi b_0=h-\frac{c-1}{24}.
\eea
Note that the LHS is a classical quantity and the RHS involves the difference between quantum quantities. This expression means that the $P^2$ that we have introduced above, is in fact the same quantity as the classical object $2 \pi b_0$: it controls the spatial holonomy of the gauge field. We can view it as a quantity that is ``protected" under quantization.

We can also view the above relation as a quantum corrected expression that demarcates the boundary between hyperbolic and elliptic holonomy. In other words, just as the classical quantity $h_0 -C/24$ controls the character of the holonomy in classical settings, in the quantum theory, we should be considering the sign of $h-(c-1)/24$. When it is positive, we will view it as describing black hole states and when it is negative, we will view it as describing conical defects. This means that in the quantum theory, $P^2 \ge 0$ is the black hole regime. In the elliptic holonomy regime, sometimes it is convenient to introduce an alternate label $\mu=\sqrt{-P^2}$, which we can use to label sub-BTZ threshold states. In such cases, we can write $
  \chi_{\mu}(\tau) \;=\; \frac{q^{\,-\mu^{2}}}{\eta(\tau)}$, to denote the same Virasoro character.

\section{Bulk States from Wilson Lines}\label{sec:wilson}

In this paper, one of our goals is to get some intuition for bulk representations of CFT states. Below the BTZ threshold and above the vacuum, one often does this in terms of Wilson line insertions \cite{DeserJackiwtHooft,DeserJackiw,CarlipBook,AFS,Castro} at the origin of the contractible spatial cycle. These are the conical defect states with elliptic holonomy that we discussed in the previous section. The key idea here is that this state is completely captured by its spatial holonomy label because the primary scaling dimension $h$ fixes it. We will discuss this below, but we will also generalize it to states above the BTZ threshold. Eventually, our perspective will be that these super-threshold states capture microstates of the BTZ black hole.

\subsection{Worldline Sources for Holonomy}

In this subsection we make precise what we mean by ``inserting a Wilson line that sources a conical defect/BTZ microstate geometry''. The technically relevant object is a worldline action for a pointlike excitation coupled to the Chern-Simons gauge field. For clarity we work with a single (holomorphic/left) $SL(2,\mathbb{R})$ sector. Our discussion is loosely based on \cite{Castro} but with an eye towards generalizing for super-threshold states. Super-threshold states mean, at the level of representations of $SL(2,\IR)$, that we are working with the principal continuous series and not the discrete series \cite{CK}. But the representational label (which can be viewed roughly as the holonomy) is all that will matter for our bulk discussion, not the details of the representation itself.

To describe a line defect we add to $S_{\text{CS}}$ a worldline action for an $SL(2,\mathbb{R})$-valued field
$g(\tau)$ living on a curve $\gamma \subset M$, parameterised by $\tau$:
\begin{equation}
  S_{\text{WL}}[A,g;\lambda]
  \;=\;
  \int_\gamma d\tau\;
  \big\langle \lambda,\,
    g^{-1} D_\tau g
  \big\rangle
  \;=\;
  \int_\gamma d\tau\;
  \mathrm{Tr}\Big(
    \lambda\,
    g^{-1}\big( \partial_\tau + \dot x^\mu A_\mu(x(\tau)) \big)g
  \Big).
  \label{eq:WL-action}
\end{equation}
Here $x^\mu(\tau)$ is the embedding of the worldline, $\dot x^\mu = d x^\mu / d\tau$, and $D_\tau g = (\partial_\tau + \dot x^\mu A_\mu) g$ is the pullback of the gauge-covariant derivative. The fixed Lie-algebra element
\[
  \lambda = \lambda^a L_a \in \mathfrak{sl}(2,\mathbb{R})
\]
labels the coadjoint orbit of the worldline degrees of freedom -- it is the ``charge'' of the line and will become the precise holonomy label below. The bracket $\langle \cdot,\cdot\rangle$ is proportional to the matrix trace and can be chosen so that all numerical factors are absorbed into $\lambda^a$.

Gauge transformations act as
\[
  A \;\mapsto\; A^h = h^{-1} A h + h^{-1} dh,
  \qquad
  g(\tau) \;\mapsto\; h^{-1}(x(\tau))\,g(\tau),
\]
under which $S_{\text{WL}}$ is invariant provided $\lambda$ transforms in the coadjoint representation
$\lambda \mapsto h^{-1}(x_\gamma)\lambda h(x_\gamma)$. Thus only the conjugacy class of $\lambda$ (equivalently, of $e^\lambda$) is gauge invariant -- this is exactly the same data as the conjugacy class of the holonomy around the line.

In the Euclidean torus setup we will eventually take $\gamma$ to run along the thermal $B$-cycle at the origin of the spatial disc, with the bulk connection written in Banados gauge as
\[
  A
  \;=\;
  b^{-1}\big( a_\phi\, d\phi + a_t\, dt \big) b + b^{-1} db,
  \qquad
  b = e^{r L_0},
\]
so that the spatial and thermal holonomies are controlled by the constant matrices $a_\phi$ and $a_t$, respectively. The worldline along the thermal cycle will source the holonomy of $a_\phi$ around the spatial $A$-cycle.

\subsubsection{Spacetimes from Sourced Flat Connections}

The total action is\footnote{We suppress the boundary terms in the CS action because it is not important for our discussions here.}
\begin{equation}
  S_{\text{tot}}[A,g;\lambda]
  \;=\;
  S_{\text{CS}}[A] + S_{\text{WL}}[A,g;\lambda].
  \label{eq:total-action}
\end{equation}
The variation of $S_{\text{WL}}$ with respect to the gauge field is
\begin{equation}
  \delta S_{\text{WL}}
  \;=\;
  \int_\gamma d\tau\;
  \mathrm{Tr}\!\left(
    \lambda\, g^{-1} \delta A_\mu(x(\tau)) g
  \right)\dot x^\mu(\tau).
  \label{eq:WL-variation-1d}
\end{equation}
It is convenient to rewrite this as a three-dimensional integral by inserting a delta-function localising the current on the worldline:
\begin{equation}
  \delta S_{\text{WL}}
  \;=\;
  \int_M d^3x\;
  \mathrm{Tr}\big(
    \delta A_\mu(x) J^\mu(x)
  \big),
  \qquad
  J^\mu(x)
  =
  \int_\gamma d\tau\;
  \dot x^\mu(\tau)\,
  g(\tau)\lambda g(\tau)^{-1}\,
  \delta^{(3)}\big(x - x(\tau)\big).
  \label{eq:WL-current}
\end{equation}
Combining the bulk variation and \eqref{eq:WL-current}, the equation of motion obtained from $\delta S_{\text{tot}}=0$ is (in differential-form notation),
\begin{equation}
  \frac{k}{2\pi} F \;+\; J \;=\; 0,
  \qquad
  J = J_\mu dx^\mu \;\;\text{(a distributional Lie-algebra valued 2-form)}.
  \label{eq:sourced-flatness-form}
\end{equation}
Thus the connection is flat ($F=0$) everywhere away from the worldline, but there is a delta-function source supported on $\gamma$. This is the precise sense in which the worldline \emph{sources} the geometry: it appears as a source in the Chern-Simons equations of motion.

To see how this fixes the holonomy, consider a small two-disc $D$ that intersects the worldline once transversely, with boundary $C=\partial D$ linking the line. Integrating \eqref{eq:sourced-flatness-form} over $D$ gives
\begin{equation}
  \frac{k}{2\pi} \int_D F
  \;+\;
  \int_D J
  \;=\; 0.
  \label{eq:integrated-EOM}
\end{equation}
We define the Lie-algebra valued ``holonomy charge''
\begin{equation}
  Q \;\equiv\; \int_D J
  \;=\;
  \int_\gamma d\tau\;
  g(\tau)\lambda g(\tau)^{-1}\,
  \delta^{(2)}\big(x_\perp - x_\perp(\tau)\big),
  \label{eq:holonomy-charge}
\end{equation}
where $x_\perp$ are coordinates transverse to $\gamma$. Because $J$ is supported only at the intersection point, $Q$ is simply $g_*\lambda g_*^{-1}$ for some $g_*$ evaluated at that point, and transforms by conjugation under bulk gauge transformations. Equation \eqref{eq:integrated-EOM} can then be written as
\begin{equation}
  \int_D F \;=\; -\,\frac{2\pi}{k}\,Q.
  \label{eq:F-flux-vs-Q}
\end{equation}

Since $F=0$ away from the line, we can write $A =-dU U^{-1}$ on $D\setminus\{0\}$ for some $U(x) \in SL(2,\mathbb{R})$ defined up to right-multiplication by a constant group element. The nonzero flux \eqref{eq:F-flux-vs-Q} implies that $U$ cannot be globally single-valued around the puncture: going once around $C$ one finds
\[
  U(\phi+2\pi) = M\,U(\phi),
\]
for some constant $M\in SL(2,\mathbb{R})$. Using the non-Abelian Stokes theorem together with \eqref{eq:F-flux-vs-Q}, one finds that $M$ is conjugate to the exponential of $Q$:
\begin{equation}
  M \sim \exp\!\Big(-\frac{2\pi}{k} Q\Big).
  \label{eq:monodromy-expQ}
\end{equation}
The holonomy of $A$ around $C$ is
\begin{equation}
  H_C \;\equiv\; P\exp\!\Big(-\oint_C A\Big)
  \;\sim\;
   U(0)^{-1}U(2\pi)
  \;\sim\; M
  \;\sim\; \exp\!\Big(-\frac{2\pi}{k} Q\Big),
  \label{eq:holonomy-vs-Q}
\end{equation}
where $\sim$ denotes equality up to conjugation in $SL(2,\mathbb{R})$. Thus the conjugacy class of the holonomy $H_C$ is fixed by the coadjoint-orbit label $\lambda$ via $Q=g_*\lambda g_*^{-1}$. This is what we mean by calling $\lambda$ (or, equivalently, the eigenvalues of $Q$) the \emph{holonomy label} of the worldline.

In particular, when $H_C$ is elliptic (hyperbolic), the eigenvalues of $Q$ determine the deficit angle (boost parameter) of the corresponding conical-defect (BTZ-microstate) geometry. In usual discussions of AdS$_3$/CFT$_2$, the super-threshold ``defect" geometries  are ignored. But we will argue that the Chern-Simons formulation and the modular S-transform can be used to interpret the BTZ character as a sum over such states. We discuss super-threshold states in some detail below.

We will have more to say about these Wilson line constructions in an upcoming paper, where we will use them to construct candidate (quantum) BTZ microstates in Lorentzian signature \cite{CK}. We will argue that the Chern-Simons path integral on a Euclidean semi-infinite cylinder geometry with a (quantum) Wilson line insertion at the center of the spatial disc (ie., along the cylinder Euclidean time direction), when cut open at some Euclidean time, constructs a heavy Lorentzian bulk Virasoro primary. The $SL(2,\IR)$ representation of the Wilson line has to be above the BTZ threshold, and therefore in the principal continuous series, for this to have the interpretation as a BTZ microstate. Below the threshold, it prepares a quantum conical defect primary state.

\subsection{Hyperbolic Holonomy Without a Smooth Horizon}
\label{subsec:hyperbolic-wilson-geometry}

In this subsection we spell out the bulk geometry sourced by a Wilson line
with \emph{hyperbolic} holonomy, inserted at the origin of the spatial
disc and running along the Euclidean thermal circle. The point of this
construction is to keep the spatial cycle holonomy in the hyperbolic
conjugacy class, while \emph{not} demanding that the Euclidean thermal
cycle be smoothly contractible. This distinguishes the geometry sharply
from the standard Euclidean BTZ handlebody.

\paragraph{Manifold and Cycles.}
We start from a Euclidean ``solid cylinder''
\begin{equation}
  M \;=\; \mathbb{D}_\phi \times S^1_\tau,
\end{equation}
where $\mathbb{D}_\phi$ is a spatial disc and $S^1_\tau$ is the Euclidean
thermal circle. The conformal boundary of $M$ is a torus
\begin{equation}
  \partial M \;\simeq\; S^1_\phi \times S^1_\tau,
\end{equation}
with the usual interpretation of $S^1_\phi$ as the \emph{spatial} cycle
and $S^1_\tau$ as the \emph{thermal} or Euclidean time cycle. On
$\partial M$ we therefore have two distinguished homology classes, which
we denote by
\begin{equation}
  \gamma_\phi \in H_1(\partial M,\mathbb{Z}),\qquad
  \gamma_\tau \in H_1(\partial M,\mathbb{Z}),
\end{equation}
corresponding to the $S^1_\phi$ and $S^1_\tau$ directions respectively.

\paragraph{Wilson Line Insertion.}
We insert a bulk Wilson line along the Euclidean thermal direction at
the center of the spatial disc,
\begin{equation}
  L \;=\; \{0\} \times S^1_\tau \;\subset\; M,
\end{equation}
in some representation of the (chiral) $SL(2,\mathbb{R})$ Chern-Simons
theory. We choose the representation such that the holonomy of the
connection around $\gamma_\phi$ is in a \emph{hyperbolic} conjugacy
class. Concretely, if $A$ denotes the $SL(2,\mathbb{R})$ connection and ${\cal P}\exp (-\oint A$) is the corresponding Wilson loop,
we impose that
\begin{equation}
  \mathrm{Hol}(\gamma_\phi) \;=\;
  \mathcal{P}\exp \left(-\oint_{\gamma_\phi} A \right)
\end{equation}
be conjugate to a diagonal matrix with real eigenvalues
$e^{\pm \lambda/2}$ with $\lambda>0$. This fixes the ``mass'' (and, in
the rotating case, the spin) parameters of the geometry in the usual
BTZ sense. We do \emph{not} impose any constraint on the holonomy along
$\gamma_\tau$ at this stage; in particular, the Euclidean period
$\beta$ of $\tau$ is treated as a free parameter and is \emph{not} fixed
by any smoothness requirement in the bulk. 

Specifying a super-threshold state corresponds to specifying the
hyperbolic holonomy along $\gamma_\phi$, as discussed above. 

\paragraph{Complement of the Wilson Line.}
In our janus path integral we remove the worldline
$L$ itself and work on the complement
\begin{equation}
  M' \;=\; M \setminus L 
      \;\cong\; (\mathbb{D}_\phi \setminus\{0\}) \times S^1_\tau
      \;\cong\; \text{annulus} \times S^1
      \;\cong\; T^2 \times I.
\end{equation}
Topologically, $M'$ is a \emph{thickened torus}. Its fundamental group
is
\begin{equation}
  \pi_1(M') \;\cong\; \mathbb{Z} \oplus \mathbb{Z},
\end{equation}
generated by (for example)
\begin{itemize}
  \item a loop $\gamma_\phi$ going once around the missing point in the
        disc (this is homotopic to the spatial $S^1_\phi$ at the
        boundary), and
  \item a loop $\gamma_\tau$ running along the thermal circle $S^1_\tau$.
\end{itemize}
On $M'$ there is no relation between these generators: neither
$\gamma_\phi$ nor $\gamma_\tau$ bounds a disc in $M'$. In particular,
both $\gamma_\phi$ and $\gamma_\tau$ are non-contractible.

\paragraph{Distinction from the BTZ Handlebody.}
It is useful to emphasize how this discussion differs from the more conventional
Euclidean BTZ geometry. The Euclidean BTZ handlebody is obtained by
taking a solid torus and declaring some primitive cycle on the boundary
torus to be contractible in the bulk; for the standard BTZ black hole
this is the thermal cycle $\gamma_\tau$. Smoothness at the ``horizon''
then imposes that the holonomy around the contractible cycle is central
(trivial in $PSL(2,\mathbb{C})$), which fixes $\beta$ to the Hawking
value $\beta_H$.

In our janus path integral, we
work with the thickened torus $M' = T^2 \times I$ obtained by
removing the Wilson line from the solid cylinder, and we do not require
any cycle to be contractible. To define a super-threshold states, we re-insert a Wilson line which in turn fixes the (hyperbolic) spatial holonomy, but do not demand {\em smooth} contractibility of either cycle. The thermal cycle $\gamma_\tau$ remains
non-contractible, and $\beta$ remains a free parameter; the only
holonomy we fix is the hyperbolic one along the spatial cycle
$\gamma_\phi$.

\paragraph{Metric Interpretation.}
From the metric point of view, we may choose a locally AdS$_3$ metric on
$M'$ which, in a convenient radial coordinate, looks exactly like the
Euclidean BTZ metric in the exterior region,
\begin{equation}
  ds^2 \;=\;
  f(r)\,d\tau^2 + f(r)^{-1}\,dr^2 + r^2\,d\phi^2,\qquad
  f(r) = \frac{r^2 - r_+^2}{\ell^2}, \label{non-rotBTZ}
\end{equation}
with $r\ge r_++\epsilon$ and with $(\tau,\phi)$
identified as
\begin{equation}
  \tau \sim \tau + \beta,\qquad
  \phi \sim \phi + 2\pi.
\end{equation}
We allow generic $\beta$ but there is a Wilson line source at the core 
$r=r_+$. The parameter $r_+$ (and its rotating generalization $(r_+,r_-)$)
is determined by the hyperbolic holonomy along $\gamma_\phi$ in the usual
way. The
metric on $M'$ is therefore locally identical to the  BTZ
exterior, with the Wilson line sitting at the core and carrying
the appropriate representation data.

The geometry sourced by a hyperbolic holonomy Wilson line at the
origin of the spatial disc fixes the BTZ parameters, while the thermal cycle remains
non-contractible and its period $\beta$ is an unfixed modulus rather
than being tied to a smoothness condition at a horizon. In this sense, these are ``off-shell" geometries in classical gravity, but we will see that they allow a natural microstate interpretation. Note that an unfixed $\beta$ is perfectly natural (and in fact necessary) when one is working with microstates. These states will get summed over, and it is only on the {\em saddle} configuration in the partition function, that we expect relationships between $r_+$ and $\beta$. 

A direct generalization exists in terms of $\beta$ and $\Omega$ (or $\beta_+$ and $\beta_-$) when we have both $r_+$ and $r_-$. The fact that even in the rotating case, the core of the geometry is at $r=r_+$ in Euclidean signature (and that there are no regions beyond it), is useful to us: it is at $r=r_+$ that we insert Wilson lines in the off-shell geometries.  

Finally: it should be clear that by taking a cylinder geometry instead of a torus with some thermal periodicity, we can also discuss zero temperature Euclidean configurations in this set up. Cutting such states open will result in Lorentzian microstates \cite{CK}.

\section{Specifying States at the Inner Boundary}\label{sect:innerboundary}

In AdS$_3$/CFT$_2$, states can be organized in terms of primaries and their (essentially universal) descendants. One of our goals in this paper is to emphasize that primaries allow the bulk perspective that they are being specified at the core of the geometry, because they are essentially holonomies. The descendants live at the asymptotic boundary as boundary gravitons. To recapitulate what we have already said: the theory for these boundary gravitons is controlled by the Alekseev-Shatashvili action \cite{AS1, AS2, CJ1}, the path integral for them is one-loop exact \cite{CJ1}, it can be evaluated following localization methods similar to those of \cite{StanfordWitten}, and the result reproduces the Virasoro character once one chooses the background  \cite{CJ1}. One goal of this section is to interpret these statements in a bulk radial path integral language, developing and extracting its consequences.

\subsection{Temporal vs Radial Slicing of the Path Integral}

Our main tool is the Euclidean path integral in bulk radial quantization. This gives some new perspectives compared to the more familiar temporal quantization, which was the language in which our discussion of outer boundary conditions and boundary terms was phrased. Both approaches are fundamentally equivalent because we will be working in Euclidean signature in this section (specifically, on the Euclidean thickened torus $T^2 \times I$).

Time slicing language is suitable for some of our discussions, with the fields on the ``spatial" slice being $(A_r,A_\phi)$, with $A_t$ playing the role of a Lagrange multiplier enforcing $F_{r\phi}=0$. The chiral/AdS$_3$ boundary conditions at the asymptotic boundary (such as $a_t=a_\phi$ in Drinfel'd-Sokolov gauge) lead to the Alekseev--Shatashvili boundary action in this language. This is the conventional Brown-Henneaux set up. In the temporal quantization language, evaluating the path integral with a fixed holonomy (or equivalently a Wilson line insertion of the kind we discussed) gives us a Virasoro character. This was shown for conical defects\footnote{And for the vacuum with a different quantized central charge.} in \cite{CJ1}. We will show in Section \ref{ASaction} that a universal normal orbit action that is meaningful for all orbits can be written down. 

The radial slicing approach is complementary, and we have already implicitly alluded to it in some of the previous discussions. We develop it in the next two subsections. It is the appropriate
perspective when interpreting the thickened-torus Chern-Simons path integral where specifying the boundary conditions and fall-offs at the asymptotic boundary is equivalent to specifying a state $|\Psi_{\rm out}\rangle$ there. This means that the same result obtained in \cite{CJ1} would be viewed here as computing the wave function $\langle P | \Psi_{\rm out}\rangle$, once one specifies the primary/holonomy/defect: $\langle P|\Psi_{\rm out}\rangle = \chi_P(\tau)$. 
This perspective of viewing characters as wave functions is familiar in compact $SU(2)$ Chern-Simons theory \cite{WittenJones, Seiberg, Axelrod}, where the analogue of the AS action is simply the chiral WZW action, and the characters that result from it are those of the $SU(2)_k$ current algebra. The Wilson line label is the spin label in that case. We contrast some aspects of compact Chern-Simons theory in Appendix \ref{sec:compact_vs_noncompact}. The reader may find it useful to consult it after Section \ref{sec:CFT-bra} and \ref{sec:mod-bootstrap}.

\subsection{The Annulus State}

In this paper, our interest is less on the descendants and more on the primaries (backgrounds). We will view the background as being specified by a Wilson line inserted at the origin. As we saw, this data is completely specified by the holonomy label of the Wilson line. Once one specifies this holonomy (thereby choosing a primary), the bulk Chern-Simons path integral in effect reduces to the above-mentioned AS path integral for gravitons at the boundary. By the standard rules for cutting open path integrals, this means that removing the Wilson line at the origin of the spatial cycle in a Euclidean path integral can be viewed as defining a wave functional in the holonomy basis. We will write this as the \emph{annulus state}:
\bea
\big|\Psi_{\rm ann}; C\big\rangle
&=&\int_{0}^{\infty}\! dP\ \, |P;C\rangle \,\langle P;C\,|\,\Psi_{\rm out}\rangle + \sum_{\mu}|\mu;C\rangle \langle \mu;C\,|\,\Psi_{\rm out}\rangle \\
&=&\int_{0}^{\infty}\! dP\ \, |P;C\rangle \,\chi_P(\tau) +\sum_{\mu}|\mu;C\rangle \, \chi_{\mu}(\tau).
\label{annulus-state}
\eea
In other words, this is the state that a bulk path integral with chiral Brown-Henneaux boundary conditions at the asymptotic region and a cut (a removed Wilson line) at the origin, prepares. The fact that Chern-Simons theory is topological is implicit: only the two boundaries are really playing a role. We will outline these claims here and discuss the details in some of  the subsequent subsections.

In words: the two-boundary Chern-Simons path integral on the janus cobordism in radial quantization prepares the annulus state. We take the bulk manifold to be a thickened torus $T^2 \times I$, with $r$ the radial coordinate and $(t,\phi)$ coordinates on each $T^2$ slice. The inner boundary at $r=\epsilon$ is the cut where the Wilson line has been removed, and the outer boundary at $r \rightarrow \infty$ carries the usual Brown-Henneaux/Drinfel'd-Sokolov boundary conditions. In this language, the two-boundary path integral is a Euclidean propagator in “time” $r$, which prepares a state in the Hilbert space $\mathcal{H}_{T^2}$ living at the inner boundary.

It is useful to write the Chern-Simons action in a ``radial" form\footnote{In this re-writing, the covariant CS action goes over directly to this form without any extra boundary terms. This means that when working with this form instead of the canonical bulk action \eqref{canBulk}, we will need to add extra boundary terms on both ends so that the total actions are the same (and the variational principles remain intact).}:
\bea
S_{\text{CS}}[A] \;=\; \frac{k\ell}{4\pi} \int_{r_{\rm in}}^{r_{\rm out}} dr \int_{T^2} \mathrm{Tr}\!\left( A_\phi \,\partial_r A_t \;-\; A_t \,\partial_r A_\phi \;+\; 2 A_r F_{t\phi} \right),
\eea
which makes it clear which components play a role in the radial Hamiltonian description. In particular, the radial component $A_r$ enters the action only linearly, multiplying the field strength $F_{t\phi} = \partial_t A_\phi - \partial_\phi A_t + [A_t,A_\phi]$. Integrating out $A_r$ therefore produces a functional delta-constraint $\delta[F_{t\phi}]$
and enforces $F_{t\phi}(r,t,\phi)=0$ everywhere in the bulk. Equivalently, for each fixed value of $r$, the components $(A_t(r,\cdot),A_\phi(r,\cdot))$ must define a \emph{flat} $SL(2,\mathbb{R})$ connection on the two-dimensional slice $T^2_r$. (Note that flatness here means flatness on the torus slice at fixed $r$.)

On a torus, a flat connection is completely characterized by its holonomies around the two non-contractible cycles $C$ and $\tilde C$. Equivalently, in each chiral sector we will label spatial flat connections by a continuous hyperbolic parameter $P\ge 0$ (above the BTZ threshold) and a discrete set of elliptic labels $\mu$ (conical defects and exceptional orbits), which encode the eigenvalues of the holonomy matrix\footnote{We will be fairly agnostic about the exact set of labels here, other than that there is a continuum of hyperbolic holonomies. We comment on this point more, later.}. As we will argue in detail later, the only genuine degree of freedom at the cut is this holonomy label. It is natural to choose a basis in $\mathcal{H}_{T^2}$ adapted to this data,
\bea
|P;C\rangle,\qquad |\mu;C\rangle,
\eea
where the label $C$ reminds us that the holonomy is taken along the spatial cycle. At the inner-boundary in Dirichlet polarization, we specify the gauge field configuration $A_\phi$. This is superficially more data than the holonomies. A natural way to get rid of these extra gauge data is to quotient by it, and view the cut Hilbert space as a holonomy Hilbert space. This is natural from a TQFT perspective (as well as from our holographic intuition), but is in sharp contrast with the outer boundary, where the boundary term and boundary conditions lead to dynamical degrees of freedom in the form of boundary gravitons. This is of course physically reasonable. 

In any event, we define our system after modding out by these ``inner boundary gauge transformations". What remains is pure holonomy data at the inner boundary. We will see that holonomy data is constant in $t, \phi$ (as well as $r$). Since all coordinate dependence is gauge, it is natural to take the representative inner boundary gauge field configuration $A_\phi$ in a given holonomy class, to be constant along the spatial cycle. This naturally ties in with the fact that a normal orbit has a $U(1)$ circle isometry as the stabilizer. 

The outer boundary at $r\rightarrow \infty$ is treated differently: there we impose asymptotically $\text{AdS}_3$ (Drinfel'd-Sokolov) boundary conditions. As we have outlined more than once \cite{CJ1}, if we fix a particular holonomy (equivalently, primary) label, the outer-boundary path integral computes
\bea
\langle P;C|\Psi_{\text{out}}\rangle = \chi_P(\tau),\qquad
\langle \mu;C|\Psi_{\text{out}}\rangle = \chi_\mu(\tau), \label{outer-proj}
\eea
where $|\Psi_{\text{out}}\rangle$ is the state defined by our $\text{AdS}_3$ boundary conditions and $\chi_P(\tau), \chi_\mu(\tau)$ are the corresponding characters. The dependence on the modulus $\tau$ resides entirely in these boundary graviton characters. This calculation is easiest done in temporal slicing, and we will discuss it in Section \ref{ASaction}.

\subsection{Holonomy as the Physical Phase Space on the Cobordism}

What remains of the bulk path integral\footnote{The CS cylinder cobordism discussion in this subsection is standard, but we include it for readability.}, after imposing the bulk constraint $F_{t\phi}=0$ and enforcing the boundary conditions at both ends, is most cleanly described in terms of the gauge-invariant variables of a flat connection on a torus slice. We first review why holonomies are the correct bulk variables, and why they are
$t,\phi$-independent (up to conjugation) on the support of $\delta[F_{t\phi}]$.

On the support of $\delta[F_{t\phi}]$, the restriction of $A$ to each torus slice\footnote{By a torus slice, we mean the torus at each radial location on the cobordism $T^2 \times I$, where $I$ is the radial direction.} is flat. Pick primitive cycles
$C$ (spatial/$\phi$-cycle) and $\tilde C$ (dual/$t$-cycle) on $T^2$ and a basepoint $x_0$ on the slice\footnote{All statements in this paragraph work within a given radial slice $T^2_r$ (e.g, the movements of the base point). The goal is to show that on any given torus slice, the dynamical data are the conjugacy classes of the two commuting (before quantization) holonomies arising from solving the flat ($F_{t\phi}=0$) connections on that slice.}. Define the
based holonomies
$H_C(x_0)[A]\equiv{\cal P}\exp(-\oint_{C}A)$ and $H_{\tilde C}(x_0)[A]\equiv{\cal P}\exp(-\oint_{\tilde C}A)$.
Under a gauge transformation $g(x)$, $H_C(x_0)\mapsto g(x_0)^{-1}H_C(x_0)g(x_0)$ and similarly for
$H_{\tilde C}(x_0)$, so only their conjugacy classes (or class functions such as ${\rm Tr}_R H$) are gauge invariant.
Flatness implies parallel transport is homotopy-invariant: if we slide the cycle $C$ around the torus
(e.g. change the ``time'' at which we take the $\phi$-loop), the resulting loop is homotopic to the original,
so the corresponding based holonomy is unchanged as an element of $\pi_1(T^2,x_0)$. If we instead use
a moving basepoint, the holonomy changes only by conjugation by the transport between basepoints.
Therefore the conjugacy class $[H_C]$ is independent of $t$ and $\phi$ (and likewise for $[H_{\tilde C}]$).
This is a genuine path-integral statement here because it holds configuration-by-configuration on the
support of $\delta[F_{t\phi}]$. In short: on the support of $\delta[F_{t\phi}]$, the gauge-invariant bulk data on a torus slice is the pair of commuting conjugacy classes $([H_C],[H_{\widetilde C}])$ (reflecting $\pi_1(T^2)=\mathbb Z\times\mathbb Z$). We will present closely related, but slightly more explicit, discussion of the holonomy in a later subsection in the context of temporal quantization. 

In the rest of this subsection we assemble the bulk radial path integral. Some of the claims we make here without proof  are justified in Appendix  \ref{app:cylinder_identity}.

\subsubsection{The Torus Hilbert Space}

The Dirichlet polarization at the inner boundary is adapted to the spatial cycle $C$. So we label sectors by the conjugacy class $[H_C]$. This motivates our introduction earlier of the holonomy-adapted basis of $\mathcal H_{T^2}$,
$|P;C\rangle$ and $|\mu;C\rangle$. We choose the normalization so that these states form an orthonormal and complete basis of the physical torus Hilbert space,
\bea
\langle P;C|P';C\rangle=\delta(P-P'),\qquad \langle \mu;C|\mu';C\rangle=\delta_{\mu\mu'}, \\ 
\mathbf 1_{\mathcal H_{T^2}}
=\int_0^\infty\! dP\,|P;C\rangle\langle P;C|+\sum_\mu |\mu;C\rangle\langle \mu;C|. \hspace{0.3cm}
\eea

\subsubsection{Holonomy Propagator on the Cylinder}

With $r$ viewed as Euclidean ``time'', Chern-Simons theory is a constrained first-order system.
After imposing $\delta[F_{t\phi}]$, the reduced phase space on a torus slice is the moduli space of flat connections,
equivalently the space of commuting holonomies $(H_C,H_{\widetilde C})$ modulo conjugation.
Crucially, there is no local Hamiltonian in the bulk: a standard fact (that we review Appendix  \ref{app:cylinder_identity}) is that on the thickened torus $T^2\times I$,
the bulk path integral defines the identity operator on $\mathcal H_{T^2}$:
\begin{equation}\label{eq:cylinder_identity}
U_{\rm bulk}[T^2\times I] \;=\; \,\mathbf 1_{\mathcal H_{T^2}}.
\end{equation}
A normalization $\mathcal N_{\rm cyl}$ that is independent of the $P$-label has been absorbed, in writing this. 
In the holonomy basis this means that bulk propagation is diagonal and simply identifies the holonomy labels at the
two ends,
\begin{equation}\label{eq:holonomy_propagator}
\langle P;C|U_{\rm bulk}|P';C\rangle=\delta(P-P'),
\qquad
\langle \mu;C|U_{\rm bulk}|\mu';C\rangle=\delta_{\mu\mu'}.
\end{equation}
This is the gauge-invariant sense in which ``the bulk dynamics is trivial''.

\subsubsection{Gluing the Outer Boundary and the Annulus State}

Using all these facts and inserting the holonomy completeness relation, we obtain the annulus state in the form we wrote earlier:
\begin{align}
|\Psi_{\rm ann};C\rangle
&= \int_0^\infty\! dP\,|P;C\rangle\langle P;C|U_{\rm bulk}|\Psi_{\rm out}\rangle
   +\sum_\mu |\mu;C\rangle\langle \mu;C|U_{\rm bulk}|\Psi_{\rm out}\rangle \nonumber\\
&= 
\int_0^\infty\! dP\,|P;C\rangle\langle P;C|\Psi_{\rm out}\rangle
+\sum_\mu |\mu;C\rangle\langle \mu;C|\Psi_{\rm out}\rangle,
\end{align}
If we re-insert a Wilson line with a given holonomy at the cut, it is equivalent to projecting the annulus state onto the state with that holonomy. But the same configuration has another interpretation: this is now simply a solid torus with a Wilson loop insertion along the longitude. This can be evaluated in temporal quantization \cite{CJ1}. The result is the Virasoro character corresponding to the holonomy label as we wrote in equations \eqref{outer-proj}. We will do a version of the \cite{CJ1,Chua} calculation adapted to super-threshold states and reproduce the AS action responsible for these characters,  in Section \ref{ASaction}. 

Putting these all together, we get the annulus state expanded as a sum over holonomy eigenstates as we wrote earlier. 
The topological nature of Chern-Simons theory was essential to the discussion. This is what lead $A_r$ to enforce flatness of the connection on each torus slice even off-shell. The bulk dynamics is trivial away from the boundaries. Note also that the propagator on $T^2\times I$ is independent of the holonomy and its normalization can therefore be absorbed into the definition of the holonomy basis states at the cut.

\subsubsection{Aside: A Note on Ray-Singer Torsion}\label{Ray-Singer}

The demonstration that the propagator is the identity is reviewed in Appendix \ref{app:cylinder_identity}, where a Hamiltonian-like language is used. Here we pause to make some comments which are more path integral-based in flavor. 

A convenient gauge for many purposes is the $A_r=0$ radial gauge. For the gauge-fixing functional $F[A]\equiv A_r$ corresponding to radial gauge, and infinitesimal gauge transformation $\delta_\epsilon A_r=D_r\epsilon$, the Faddeev-Popov operator is $D_r$. Evaluated on the gauge slice $A_r=0$ this reduces to $\partial_r$, so 
that  $\Delta_{\rm FP}=\det(\partial_r)$.
This determinant is independent of the holonomy labels on the $T^2$ slices and therefore this state-independent constant can be absorbed into the path integral normalization. 

A crucial point is that {\em reaching} the $A_r=0$ gauge is possible, because we are working on $T^2 \times I$ and the radial direction is a segment\footnote{We discuss radial gauge more explicitly, but in a different context, in Section \ref{temporalQ}.}. If the quantization direction has non-trivial topology, this need not be possible. A good example is the Chern-Simons path integral on $\Sigma \times S^1$ with the $S^1$ taken as the quantization direction
. This is a well-studied object: we refer the reader to the beautiful paper by Blau and Thompson \cite{Blau} for a very explicit discussion. The analogue of our $A_r=0$ gauge would be the $A_0=0$ gauge on $\Sigma \times S^1$. But this gauge is not reachable if there is non-vanishing holonomy around the circle. Therefore \cite{Blau} works with the $\partial_0 A_0=0$ gauge. The resulting non-trivial FP/ghost determinant contribution is an example of Ray-Singer torsion. 

It is interesting to compare these observations to temporal quantization on the same janus geometry. The time circle of the $T^2$ in $T^2\times I$ is the natural quantization direction. In this case the spatial slice is an annulus: the full geometry can be viewed as $\Sigma \times S^1$, with $\Sigma={\rm annulus}$. The results of \cite{Blau} can therefore be adapted here. In particular, the Ray-Singer torsion has the form $({\rm stuff})^{\chi(\Sigma)/2}$, with $\chi(\Sigma)$ the Euler characteristic, which is zero for the annulus, leading again to the conclusion that the Ray-Singer torsion contribution there is also trivial.

\subsubsection{Comments}

Let us say a few words about the nature of the basis expansion in the annulus state. Firstly, as already alluded to, the janus geometry after the removal of the Wilson line has no information about any specific holonomy: it is a sum over all holonomies. So if we re-insert a new Wilson line with some other holonomy, the result should yield the character of the new primary. The above structure is naturally compatible with that because it is a sum over primaries and simply projects on to the appropriate primary. 

Based on holographic intuition, we expect the basis to consist of black hole microstate-like primaries (labeled by hyperbolic $P$) above the BTZ threshold and light primaries (labeled by ellipic $\mu$) below\footnote{We will include the vacuum (exceptional) orbit also in the sum over $\mu$, and will not write it separately.}. The canonical quantization of pure Chern-Simons theory on $T^2 \times I$ leads to a Hilbert space associated to the moduli space of flat connections. The quantization is well-understood for the compact $SU(2)$ case and leads to a discrete spectrum \cite{WittenJones, Seiberg, Axelrod}, but for the $SL(2,R)$ case the non-compactness of the gauge group leads to various ambiguities and divergences which lead to further choices. The full story we believe is not fully understood: at any rate, it is beyond our expertise. But it is expected that quantizing the hyperbolic sector will lead to a Liouville-like continuum \cite{Teschner, Killingback}, and this is the part that we will be mostly concerned with. (We do not have strong statements about the sub-threshold basis over which we sum.) The basis over which we sum in the definition of the annulus state, we will view as having been obtained from a suitable definition of the Hilbert space of the quantized moduli space of flat connections of $SL(2,R)$ Chern-Simons theory on the torus. Fortunately, we will only need some generalities. See Appendix \ref{sec:compact_vs_noncompact} for more related discussions, and also comments on the connection to the recent discussion on Virasoro TQFT \cite{CollierEberhardtZhangVirasoroTQFT}. 

The true nature of the holographic CFT that we are working with (e.g., the discreteness of its spectrum) enters not through the basis, but via the bra on to which we project the annulus state. This is what we turn to next.

\subsection{CFT as a State in Bulk Radial Quantization}
\label{sec:CFT-bra}

Eventually, we will be interested in projecting the annulus state onto a specific primary or a general state constructed from the basis of primaries. This will define partition functions of various kinds, starting from Virasoro characters, to the full partition function of the holographic CFT itself. Loosely, we expect to write the full CFT partition function $Z_{\rm CFT}(\tau, \bar \tau)$ as a projection of the annulus state (and its anti-chiral counterpart) onto an inner bra of the form\footnote{In this integral, we have only kept the $P$ labels for notational simplicity -- if there are discrete $\mu$ labels, they should also be included in the obvious way.}:
\bea
\langle\!\langle  \Psi_{\rm CFT}|\!| \equiv \int dP\, d{\bar P} \, \rho(P, \bar P)\, \langle P; C|  \otimes \langle {\bar P}; C|. \label{CFTstate}
\eea
Here the $\rho(P, \bar P)$ captures the distribution of primaries in the CFT. On the cylinder we expect it to be a sum of delta functions. In the black hole regime, this delta function comb will be exponentially dense (a ``discretuum"). The $\bar P$ (the anti-chiral spatial holonomy variable, analogous to $\bar h$) here should not be confused with $\tilde P$, the chiral holonomy on the conjugate cycle of the torus. This is one of the formulas in this paper where we will need $\bar P$, because we are describing the full partition function of the CFT which includes both halves. With this understanding the CFT partition function is simply the projection
\bea
Z_{\rm CFT}(\tau, \bar \tau)=\langle\!\langle  \Psi_{\rm CFT}|\! |\Big( \big|\Psi_{\rm ann}; C\big\rangle \otimes \big|\bar{\Psi}_{\rm ann}; \bar{C}\big\rangle\Big).
\eea
Explicitly, it takes the form
\bea
Z_{\rm CFT}(\tau, \bar \tau)= \int dP\, d{\bar P} \, \rho(P, \bar P)\, \chi_P(\tau) \chi_{\bar P}(\bar \tau), \label{distrib-Z}
\eea
which is exactly the expression for a CFT partition function written in terms of the characters of the primaries in its spectrum. Note that all the usual expectations and consistency conditions on the CFT spectrum act as constraints on the  choice of $\rho(P,\bar P)$. 

The ideas presented above seem to give an alternative perspective on AdS/CFT, where instead of viewing the CFT partition function as a partition function with certain boundary values viewed as sources \cite{WittenAdSCFT}, we view it as preparing a wave-function in bulk radial quantization with asymptotically AdS (and if hard-wired, ignorable) boundary conditions. This perspective is especially transparent in AdS$_3$ where the split between boundary gravitons and holonomies makes the dynamics in the bulk, trivial everywhere except at the cut. 

\subsubsection{Chern-Simons Hilbert Space and Chiral Characters}

It is worth emphasizing that we are {\em not} expanding the annulus state in a basis of holographic CFT states\footnote{Such an interpretation would be in tension with the fact that $SL(2,\IR)$ holonomies seem to contain the hyperbolic continuum, whereas holographic CFTs on the cylinder are expected to have a discrete spectrum.}. In the bulk path integral language, we are working with a Chern-Simons theory. What the CS path integral gives us, is a potential quantization of the holonomies which parameterize the moduli space of flat connections. It is a suitably defined Hilbert space of {\em this} system (tensored with the anti-chiral half) that we are working with at the inner boundary. It is only after the projection onto a specific entangled state in this tensor product Hilbert space (as defined by \eqref{CFTstate}), that we get a full CFT partition function. 

This state of affairs warrants the following speculation:  we can view a suitable quantization for $SL(2,\IR)$ Chern-Simons theory\footnote{Roughly, the sectors of holonomy that one allows.} as leading to the {\em universal space} of all (chiral) characters from which a holographic CFT can be assembled. Note in particular, that this gives us a heuristic understanding of why the modular transforms of individual characters are over a continuum of primaries, even though in a given CFT (despite the fact that it is modular invariant!) we only have a discrete set. The former is controlled by this universal Chern-Simons quantization, while the choice of primaries requires specific UV complete data in the form of the $\rho(P,\bar P)$. We further discuss closely related matters in Appendix \ref{sec:compact_vs_noncompact}.

\subsubsection{Conjugate Holonomy and Smoothness}

Eventually, we will make contact with the ``modular image" picture of BTZ black holes, and for that it is useful to consider a continuum above the BTZ threshold. The threshold is labeled by $P=0$ in this notation.  In the definition of the basis states above (e.g., $|P; C\rangle$) we have specified the cycle on which the holonomy label is defined. The label $C$ denotes the spatial cycle. But it will turn out that it is also useful (especially to connect with black holes) to consider states with holonomy labels associated to the dual ``thermal" cycle $\tilde C$. These are canonically conjugate to the holonomy labels on the spatial cycle: for compact Chern-Simons theory, this is a basic fact known from the celebrated work of \cite{WittenJones}. We will discuss an analogous story for the canonically conjugate nature of holonomies on the torus in AdS$_3$ Chern-Simons theory, in some of what follows. (A word on notation: if a state is written without a label, it should be clear from the context which cycle is being referred to, but it is usually the spatial cycle.)

The fact that we are expanding $| \Psi_{\rm ann}\rangle$ in  the spatial cycle holonomy basis is correlated with the fact that the canonical bulk Chern-Simons action \eqref{S1} (without any extra boundary terms at the inner boundary), is a well-defined Dirichlet problem in $a_\phi$. When we allow a slight generalization by allowing an angular momentum chemical potential as well, this leads to working naturally with the canonical spatial and temporal ($\theta$-$y$) cycles on the torus. In this first-order radial formulation, one cannot fix
both $A_\theta$ and $A_y$ at the inner boundary; the choice of action and canonical cycles selects $A_\theta$ as the configuration variable. So
the raw annulus state $\Psi_{\rm ann}$ is naturally expressed in the corresponding
polarization. Any state in the conjugate holonomy basis (as required in understanding black holes) requires a
polarization-changing Fourier transform. We will discuss this later in detail.

We can now describe Virasoro characters (of thermal AdS$_3$, conical defects, super-threshold states and BTZ black holes) as wave functions from our annulus state. As mentioned, the conical defect and empty AdS$_3$ cases were derived via single-boundary Chern-Simons partition functions in \cite{CJ1}, while the BTZ case was obtained from the vacuum case (and the associated single-boundary path integral) via the modular $S$-transform. We will instead argue that these results can all be viewed as different cases of the same two-boundary path integral, where the inner boundary ``bra" is chosen appropriately. There are two key integral identities that naturally emerge:
\begin{align}
&\hspace{2cm} \chi_{P}(\tau)=\int dP'\ \delta(P-P')\,\chi_{P'}(\tau), \label{inner-delta} &\\
\chi_{\widetilde{P}}\!\left(-\frac{1}{\tau}\right)
&=\int_0^\infty dP\ S(\widetilde{P},P)\,\chi_P(\tau),
\ \ \ {\rm with} \ \ \ S(\widetilde{P},P)=2\sqrt{2}\cos\!\big(4\pi\,P\,\widetilde{P}\big). & \label{inner-Fourier}
\end{align}
(In the first line, the ``integral and delta function" combination should be understood to also include ``sums and Kronecker deltas" when the $P$ label is elliptic and/or exceptional.) In all cases, the outer boundary gives rise to the AS action contribution that captures the descendant piece of the character. When the inner boundary projection is on to a state carrying a holonomy class of the spatial cycle, we get the delta function projector and Virasoro characters. But when the state carries a conjugate label (corresponding to the dual cycle) we get the cosine $S$-kernel. Our discussions so far effectively derive the former case. In later sections we derive the cosine kernel case by explicitly evaluating the Fourier transform associated to the changeover to the conjugate basis. Together, these two classes  correspond to the spatial and dual cycles being in normal orbits\footnote{We will use the language of orbit labels to discuss even the conjugate cycle holonomies, even though strictly speaking the orbits are associated to the spatial cycle. Note that they span the same Hilbert space after quantization.}. 

Once we have the cosine kernel, we have a basis of characters of primaries to work with and this allows us to define characters of degenerate states, by analytically continuing in the orbit label. This allows us to ask the question: what if we demand that the inner boundary is projected onto the exceptional orbit (vacuum) of the conjugate cycle? It turns out that this construction immediately leads us to the RHS of \eqref{key0}, showing that $\chi_{vac}(-1/\tau)$ is obtained manifestly as a sum over super-threshold primaries in this approach: a smoothly contractible thermal cycle corresponds to a {\em sum} over spatial cycle holonomies corresponding to (unsmooth) microstates. Thanks to modular invariance, we are able to show this {\em without} access to the explicit spectrum of a holographic CFT: the continuum approximation to the density of states as seen by the modular kernel can already see this\footnote{Also suggesting that with a discrete but dense spectrum, smoothness may be an approximation.}. The density of states in the sum is fixed by the calculation -- it is the 
vacuum row of the modular $S$-kernel.

\subsection{Holonomy Zero Mode at the Inner Boundary}\label{InnerEdge}

In this subsection we discuss what it means to be working
with the canonical bulk action $S[A]-S[\tilde A]$ of eq. \eqref{canBulk} together with the outer
boundary term \eqref{Sbdry}, but \emph{no} extra boundary term at the inner boundary. One
key point is that with this choice the inner torus does not support dynamical edge
modes in the sense of a chiral WZW (or AS) theory: all putative inner-boundary modes compatible with Dirichlet
are pure gauge. We demonstrate this in Appendix \ref{app:symp}. The only dynamical mode is a global zero-mode that captures the spatial holonomy. 

But we need more. We want to view the data at the inner boundary as being comprised {\em only} of holonomy. The Dirichlet problem does not quite accomplish this, because the profile of the gauge field at the inner boundary (even though non-dynamical) is data that we need to specify there. In order to avoid this, we will simply declare that we need to mod out by all (periodic) gauge transformations at the inner boundary. 

A more direct way to do this is perhaps to formulate a ``holonomy-fixed variational problem" for Chern-Simons theory at the inner boundary, by taking inspiration from the worldine action we discussed earlier. This seems possible \cite{Seiberg}, but we will not develop this approach in this paper\footnote{But we do take closely related perspectives in the follow-up work \cite{CK}.}. Here we will stick to Dirichlet, together with the prescription for quotienting by the inner boundary gauge transformations.

Taking only holonomy as the data at the inner boundary is natural from the usual point of view of cutting and gluing path integrals in TQFT: we are viewing the inner boundary as a cut. As already mentioned, this makes two things natural --
\begin{itemize}
\item From standard AdS$_3$/CFT$_2$ and the picture of primaries and descendants, we know that primaries are in one-to-one correspondence with holonomy data. There cannot be any extra degrees of freedom at the inner boundary, because the outer boundary already takes care of the descendants.
\item Our perspective that primaries should be viewed as normal orbits suggests that we should have a $U(1)$ isometry in our description arising from the stabilizer. Once we demand that all profile dependence is pure gauge at the inner boundary, this is automatic\footnote{When we work in Drinfel'd-Sokolov gauge with descendants, the $A_\phi$ need not have a manifest $U(1)$ symmetry. But it is important to remember that this is due to {\em boundary} diffeomorphisms and that while the isometry of the primary is no longer a stabilizer, the isometry of the primary conjugated by the boundary diffeomorphism, is still a $U(1)$ stabilizer. With this implicit understanding, we will still refer to the stabilizer as an isometry.  We can accommodate this at the inner boundary {\em while} staying in DS gauge by working with a holonomy representative that is also dressed by the same boundary graviton. At the inner boundary, this dressing is pure gauge and the holonomy is unaffected.}.
  
\end{itemize}

The results of this section will be useful for us in the next section to demonstrate the emergence of the Alekseev-Shatashvili action at the outer boundary. Apart from the standard references we have mentioned elsewhere, we have also benefited from \cite{AtiyahBott, AlekseevMalkin, Sengupta, Meusburger}.

\subsubsection{Temporal Quantization: The Spatial Holonomy Modulus} \label{temporalQ}

As shown in Appendix \ref{app:symp}, with inner
Dirichlet boundary conditions, the inner torus does not carry a nontrivial pre-symplectic
structure: local would-be edge degrees of freedom at $\partial M_-$ remain pure gauge.
The remaining question is whether there can nevertheless be \emph{global} configuration-space
data localized at $\partial M_-$. The answer is yes: because the spatial slice in temporal
quantization is an annulus, there is a holonomy zero-mode around the non-contractible
spatial circle. In this subsection we isolate this mode systematically. This will be 
our \emph{only} gauge-invariant inner datum in this polarization. This calculation will also help us set up the gauge field in a form suitable for boundary reduction.

In temporal slicing, the canonical bulk action \eqref{canBulk} is first order in time and $A_t$
appears as a Lagrange multiplier. In the path integral, integrating over $A_t$ enforces the
\emph{spatial} flatness constraint on each time slice,
$F_{r\phi}=0$, as an exact delta-functional constraint. We will see explicitly below that ``time evolution'' is
pure gauge: it preserves the conjugacy class of the spatial holonomy. For clarity we phrase the discussion with a single $SL(2,\mathbb{R})$ copy $A$ on a manifold
$M\simeq [r_{\rm in},r_{\rm out}\rightarrow \infty]\times S^1_\phi\times\mathbb{R}_t$, with coordinates $(t,r,\phi)$
and an inner boundary at $r=r_{\rm in}$. 

We first impose the constraint $F_{r\phi}=0$ on the annulus.
Because $r$ is an interval, we can fix \emph{radial gauge} on the annulus by a gauge
transformation that is trivial at $r=r_{\rm in}$:
choose $G(t,r,\phi)$ solving the linear ODE
\begin{equation}
\partial_r G(t,r,\phi)=-A_r(t,r,\phi)\,G(t,r,\phi),\qquad G(t,r_{\rm in},\phi)=\mathbf{1}, \label{rad-ini}
\end{equation}
and gauge-transform by $G$. This sets $A_r=0$ globally on the annulus.
In this gauge, the spatial flatness constraint becomes
\begin{equation}
F_{r\phi}=0\qquad\implies\qquad \partial_r A_\phi(t,r,\phi)=0,\qquad {\rm so} \quad
A_\phi(t,r,\phi)=A_\phi(t,\phi)\,,
\end{equation}
i.e.\ the $\phi$-component is $r$-independent off-shell and in the constrained path integral.

In close parallel to some of our discussions of orbit monodromy and holonomy, now we compute the spatial holonomy of the gauge field and show that its conjugacy class is a gauge invariant.
At fixed $t$, we are left with a one-dimensional gauge field $A_\phi(t,\phi)\,d\phi$ on a
circle. Define the parallel-transport operator $U(t,\phi)$ by
\begin{equation}
\partial_\phi U(t,\phi)=-A_\phi(t,\phi)\,U(t,\phi),\qquad U(t,0)=\mathbf{1},
\end{equation}
so $U(t,\phi)=\mathcal{P}\exp\!\big(-\int_0^\phi A_\phi(t,\phi')\,d\phi'\big)$.
The holonomy around the spatial circle is by definition
\begin{equation}
H(t)\;\equiv\;U(t,2\pi)\;=\;\mathcal{P}\exp\!\Big(-\oint_{S^1_\phi} A_\phi\,d\phi\Big)\in SL(2,\mathbb{R}).
\end{equation}
Under a gauge transformation $u(t,\phi)$ that is single-valued on $S^1_\phi$,
\begin{equation}
A_\phi\ \mapsto\ A_\phi^u=u^{-1}A_\phi u+u^{-1}\partial_\phi u,
\end{equation}
one finds
\begin{equation}
U(t,\phi)\ \mapsto\ U^u(t,\phi)=u(t,\phi)^{-1}U(t,\phi)\,u(t,0),
\qquad
H(t)\ \mapsto\ H^u(t)=u(t,0)^{-1}H(t)\,u(t,0).
\end{equation}
Note that we used single-valuedness in the last step, and it shows that holonomy $H$ transforms by conjugation. Thus the gauge-invariant content of the spatial holonomy is its conjugacy class $[H]$.
Equivalently, any class function of $H$ (e.g.\ $\Tr H$ in a chosen representation) is gauge
invariant.

Next we write the gauge field $A_{\phi} (t, \phi)$ in a standard form where its holonomy class is manifest.
The previous step already identifies $[H]$ as the natural global datum. We now show 
directly that \emph{all} $\phi$-dependence of $A_\phi(t,\phi)$ is pure gauge once $[H]$ is fixed.

Choose a Lie-algebra element $K(t)\in\mathfrak{sl}(2,\mathbb{R})$ and a (time-dependent)
 group element $h_0(t)$ such that
\begin{equation}
H(t)=h_0(t)^{-1}\,e^{-2\pi K(t)}\,h_0(t),
\end{equation}
i.e.\ $e^{-2\pi K(t)}$ is a chosen representative of the conjugacy class $[H(t)]$.
Define
\begin{equation}
g(t,\phi)\ \equiv\ U(t,\phi)\,h_0(t)^{-1}\,e^{\phi K(t)}.
\end{equation}
Using $U(t,2\pi)=H(t)$ and the defining relation above, we get $g(t,2\pi)=g(t,0)$, so
$g(t,\phi)$ is single-valued on the circle. A short computation then shows that $g$ gauges
$A_\phi$ to the constant\footnote{By constant here, we mean $\phi$-independent.} connection $K$:
\begin{equation}
A_\phi(t,\phi)\ =\ g(t,\phi)\big(\partial_\phi+K(t)\big)g(t,\phi)^{-1}.
\end{equation}
This makes the structure transparent:
all non-constant $\phi$-dependence is pure gauge because we can write any $A_\phi$ (with arbitrary $\phi$-dependence) in the above form, with a periodic $g$, with the $\phi$-independent datum $K(t)$. The latter is equivalent to the holonomy conjugacy class $[H]$ as we already discussed. Note that this form is specific to radial gauge.

% Holonomy conservation explicit calculation.

Finally, we argue that the time evolution is pure gauge and $[H]$ is conserved. We incorporate the remaining bulk equation $F_{t\phi}=0$ in radial gauge $A_r=0$:
\begin{equation}
F_{t\phi}=0\qquad\Longleftrightarrow\qquad \partial_t A_\phi = D_\phi A_t,
\qquad D_\phi A_t\equiv \partial_\phi A_t+[A_\phi,A_t]. \label{Ftphi}
\end{equation}
Note that $\partial_t A_\phi = D_\phi A_t$ is just the gauge transformation $\delta_\lambda A_\phi= D_\phi \lambda$ with the gauge parameter $\lambda = A_t dt$. In other words, $F_{t \phi}=0$ is the statement that time evolution of $A_\phi$ is pure gauge. We turn this into a statement about the holonomy below\footnote{Note that the equations like $F_{t \phi}=0$ should be viewed as Schwinger-Dyson type equations or operator equations in these discussions. They are more than classical flatness conditions.}.

Differentiating our parallel transport equation above with respect to $t$ gives
\begin{align*}
\partial_\phi(\partial_t U)
&=-(\partial_t A_\phi)\,U-A_\phi\,(\partial_t U)
=-(\partial_\phi A_t+[A_\phi,A_t])\,U-A_\phi\,(\partial_t U),
\end{align*}
where in the second equality we used \eqref{Ftphi}.
Define $V(t,\phi):=\partial_t U(t,\phi)+A_t(t,\phi)\,U(t,\phi)$. Using $\partial_\phi U=-A_\phi U$ and the previous expression gives,
\begin{align*}
\partial_\phi V
&=\partial_\phi(\partial_t U)+(\partial_\phi A_t)U+A_t(\partial_\phi U) \\
&=\big[-(\partial_\phi A_t+[A_\phi,A_t])U-A_\phi(\partial_t U)\big]+(\partial_\phi A_t)U+A_t(-A_\phi U)
=-A_\phi\,V.
\end{align*}
Thus $V$ obeys the same homogeneous $\phi$-equation as $U$, so $V(t,\phi)=U(t,\phi)\,V(t,0)$.\footnote{For fixed $t$, view $\phi$ as the evolution variable and suppose $U$ and $V$ both solve the same homogeneous matrix ODE
$\partial_\phi X(\phi)=-A_\phi(\phi)\,X(\phi)$, with $U(0)=\mathbf{1}$. Since $U(\phi)$ is invertible, define $W(\phi):=U(\phi)^{-1}V(\phi)$. Then
\[
\partial_\phi W=(\partial_\phi U^{-1})V+U^{-1}\partial_\phi V.
\]
Using $\partial_\phi U=-A_\phi U$ gives $\partial_\phi U^{-1}=U^{-1}A_\phi$, and using $\partial_\phi V=-A_\phi V$ yields
$\partial_\phi W=U^{-1}A_\phi V+U^{-1}(-A_\phi V)=0$, so $W(\phi)=W(0)$ is $\phi$-independent. Hence
$V(\phi)=U(\phi)\,W(0)=U(\phi)\,U(0)^{-1}V(0)=U(\phi)\,V(0)$.}
But since $U(t,0)=\mathbf{1}$ for all $t$, we have $\partial_t U(t,0)=0$ and hence from the definition of $V$ it follows that $V(t,0)=A_t(t,0)$. Again using the definition of $V$, we have
$\partial_t U(t,\phi)+A_t(t,\phi)U(t,\phi)=U(t,\phi)A_t(t,0)$, i.e.
$\partial_t U(t,\phi)=-A_t(t,\phi)U(t,\phi)+U(t,\phi)A_t(t,0)$.
Evaluating at $\phi=2\pi$ yields
\[
\frac{d}{dt}H(t)=-A_t(t,2\pi)\,H(t)+H(t)\,A_t(t,0),
\]
and for periodic $A_t$ (so $A_t(t,2\pi)=A_t(t,0)$) this becomes 
\begin{equation}
\frac{d}{dt}H(t)=\big[H(t),\,A_t(t,0)\big].
\end{equation}
which is a Lax equation: $\dot H=[H,X(t)]$ with $X(t)=A_t(t,0)$, whose evolution is by conjugation. Indeed if $G(t)$ solves $\dot G(t)=-X(t)\,G(t)$, with $G(0)=\mathbf{1}$, then 
\bea
\frac{d}{dt}\!\big(G(t)^{-1}H(t)G(t)\big)
=G^{-1}\dot H\,G+(\dot G^{-1})HG+G^{-1}H\dot G
=G^{-1}\big(\dot H-[H,X]\big)G=0, \nonumber
\eea
where we used $\dot G^{-1}=-G^{-1}\dot G\,G^{-1}=-G^{-1}X$.
Hence $G(t)^{-1}H(t)G(t)$ is $t$-independent and therefore
\[
H(t)=G(t)\,H(0)\,G(t)^{-1}.
\]
In particular, the conjugacy class $[H]$ is conserved along $t$. Equivalently, we can take $K(t)$ to be a time-independent constant $K$. This is as expected for a topological theory with no local propagating degrees of freedom.

We now use these results to write down the general Chern-Simons gauge field in temporal quantization. This will be useful in Section \ref{ASaction} when we derive the AS action from boundary reduction.

\subsubsection{General Gauge Field in Temporal Quantization}
\label{gen-gauge}

The above discussion shows (among other things) that a {\em general} gauge field configuration in temporal quantization on the janus cobordism can be written as:
\beq \label{spatialgauge-first}
\begin{aligned}
A_r = G_0 \partial_r G_0^{-1} \ ,  \quad A_\phi = G_0 \left(\partial_\phi + K(t)\right) G_0^{-1} 
\\[6pt]
\overline A_r = \overline G_0 \partial_r \overline G_0^{-1} \ ,  \quad \overline A_\phi = \overline G_0 \left(\partial_\phi + K(t)\right) \overline G_0^{-1} 
\end{aligned}
\eeq
where $G_0$ and $\overline G_0$ are elements of the two $SL(2,\mathbb R)$'s, and functions of $(t,r,\phi)$. Note that $A_t$ does not affect the calculation because it only shows up as a Lagrange multiplier multiplying $F_{r \phi}$, and the latter is zero even off-shell. In the notation $g$ and $G$ used in section \ref{temporalQ}, we can think of $G_0$ as
\beq
G_0(t,r,\phi) \equiv g(t,\phi) G(t,r,\phi) \label{gG}
\eeq
with  $G(t,r_{in},\phi)=1$. Note that in the discussion in section \ref{temporalQ}, we were specifically interested in identifying a $G(t,r,\phi)$ that trivialized a given $A_r$. Here we are going the other way, and writing the gauge field configuration in a form compatible with holonomy data and boundary conditions. (In other words, here, if we simply set $G(t,r,\phi)=1$ everywhere, we will be in radial gauge.)

Since the conjugacy class of the holonomy is time independent, we can write: 
\beq
K(t) = \frac{2 P}{\sqrt{k}} L_0 = \frac{P}{\sqrt{k}} \begin{pmatrix}
    1 & 0 \\
    0 & -1
\end{pmatrix} \label{K-matrix}
\eeq
where the normal orbit label $P$ is chosen to be consistent with our previous and upcoming discussions\footnote{We will see that $P^2 = 2 \pi b_0$ in the notation used in our discussion of orbits.}. This is the form that is useful in discussions of the Fourier kernel later, but to reach Banados/DS gauge one can work with a different form as we discuss presently.

Undoing the gauge transform in \eqref{spatialgauge-first} we can first reach
\bea
A_r=0,
\qquad
A_\phi = K ,
\qquad
K=\frac{2P}{\sqrt{k}}\,L_0.
\eea
Now we can perform a \emph{constant} gauge transformation
\[
A \;\longrightarrow\; A^{(2)} = s_P^{-1} A\, s_P ,
\qquad s_P\in SL(2,\mathbb R),
\]
which preserves flatness and does not affect the holonomy conjugacy class. One convenient choice is
\bea
s_P=
\begin{pmatrix}
1 & -\dfrac{P}{\sqrt{k}}\\[6pt]
\dfrac{\sqrt{k}}{2P} & \dfrac12
\end{pmatrix},
\qquad
\det s_P=1 .
\eea
A direct computation gives
\bea
s_P^{-1}\Big(\frac{2P}{\sqrt{k}}\,L_0\Big)s_P
\;=\;
L_1-\frac{P^2}{k}\,L_{-1},
\eea
and the connection takes the Drinfel'd--Sokolov
(highest-weight) form
\bea
A_r^{(2)}=0,
\qquad
A_\phi^{(2)}=L_1-\frac{P^2}{k}\,L_{-1},
\eea
which is the standard DS representative for a flat connection with $\phi$-cycle holonomy labeled by $P$. We can do a further gauge transformation by choosing $b(r)=e^{r L_0}$ to reach the forms we presented in Section \ref{sec:orbits} with $A_r=L_0$.   To go to more general DS-configurations with boundary gravitons turned on (i.e., $\mathcal L(t,\phi)$ instead of constant $b_0 \sim P^2$) we can rely on coadjoint actions as we discussed in section \ref{sec:orbits}. This essentially corresponds to turning on non-trivial $g(t,\phi)$ in our notations above, and as expected, it cannot change the holonomy. 

Writing the gauge field configuration in terms of $G_0$ is useful for boundary reduction and connecting with the work of \cite{CJ1} as we will see in Section \ref{ASaction}. 

\subsubsection{Inner Boundary Gauge Transformations}

The initial condition $G(t,r_{\rm in},\phi)=\mathbf{1}$ that we have taken in \eqref{rad-ini} for reaching the radial gauge, is worth commenting on. Indeed, for any $f(t,\phi)\in SL(2,\mathbb{R})$ the solution
with $G(t,r_{\rm in},\phi)=f(t,\phi)$ is related by right-multiplication:
$G_f(t,r,\phi)=G_1(t,r,\phi)\,f(t,\phi)$, where $G_1$ denotes the solution with $f=\mathbf{1}$.
Both choices set $A_r\to 0$, but they differ by the residual $r$-independent transformation $f(t,\phi)$
which preserves $A_r=0$ and acts on the remaining component as
$A_\phi\to f^{-1}A_\phi f+f^{-1}\partial_\phi f$. Since the full gauge transformation is the $G_0$ that we have defined above in \eqref{gG}, we could also view this freedom as a redefinition of $g$ while keeping the $G$ intact.

But at the inner boundary, our action principle leads to a well-defined Dirichlet problem with $A_\phi|_{\partial M_-}$ fixed. So one needs a separate statement about what to do with single-valued transformations that can change $A_\phi$. We argued earlier that the correct thing to do is to quotient by these inner boundary gauge transformations as well. This is natural from the TQFT perspective where the Hilbert space only contains holonomy labels, but it is also reasonable from a holographic perspective: the only labels for states (other than those for boundary gravitons) comes from holonomies in AdS$_3$/CFT$_2$. Once one adopts this philosophy, arbitrary gauge field configurations at the inner boundary within the same holonomy class, are treated as the same physical configuration. The fact that angle-dependence is pure gauge also means that the configurations have circular isometry: a virtue, since we wish to interpret them as normal orbits with a $U(1)$ stabilizer. These observations suggest that $G(t, r_{\rm in}, \phi)=\mathbf{1}$ is a natural choice at the inner boundary, but along with it we can also choose $f(t,\phi)$ so that it cancels the $g(t,\phi)$. The net gauge field configuration that we will be working with then, will be of the same form as written in \eqref{spatialgauge-first}, but with the understanding that $g(t,\phi)=1$ in \eqref{gG} and $G(t,r_{\rm in},\phi)=1$. In such a gauge, $G_0(t,r,\phi) = G(t,r,\phi)$ will capture the asymptotic boundary gravitons as $r \rightarrow \infty$. Note the crucial fact however, that $A_r$ is neither zero nor constant in such gauges.

In the context of conventional Banados/DS gauge, the above discussion changes slightly. Banados gauge has $A_r =L_0$, which we can reach by taking $G(t,r,\phi)=e^{r L_0}$. But if we want to consider a generic DS configuration with non-trivial gravitons, we also have to do a further ``gauge" transformation, which (as discussed at the end of section \ref{gen-gauge}) can be achieved by turning on a $g(t,\phi)$.  This is the analogue of the co-adjoint action circle diffeomorphism $\epsilon(t,\phi)$ in the context of our discussions on orbits. One subtlety here is that such a $g(t,\phi)$ will lead to a profile for $A_\phi$ at the inner boundary. We can think of it as giving rise to a non-constant representative of the holonomy conjugacy class at the inner boundary. 

Let us make one more comment about $f(t,\phi)$. As we just discussed, the generic scenario is where an arbitrary periodic $f$ changes the gauge field value at the inner boundary. But there is also the non-generic case where some choices of $f$ do not change the $A_\phi|_{\partial M_-}$: these are the stabilizers of $A_\phi$. This is an example of the case considered in the Appendix \ref{app:symp}, where we show that gauge transformations that leave the $A_\phi$ at the inner boundary unchanged have zero canonical charge. Most such gauge transformations vanish at the inner boundary, but they don't need to -- they can lie in the stabilizer.

\subsection{Alekseev-Shatashvili Action for the General Normal Orbit}\label{ASaction}

In this section, we will plug in the general form of the bulk gauge field \eqref{spatialgauge-first} into our two-boundary Chern-Simons action. We can be agnostic about the (periodic) choices regarding $g$ and $G$ at the inner boundary, as long as we allow their product $G_0(t,r,\phi)$ to have a non-trivial profile at the $r \rightarrow \infty$ asymptotic region. The goal is to derive the asymptotic contribution $\langle P; C| \Psi_{\rm out}\rangle$ directly by doing the path integral in temporal quantization (to show that it is the Virasoro character). We present the calculation as a demand that the boundary reduction lies in the Drinfel'd-Sokolov gauge: this is a slightly more general implementation of the calculation in \cite{CJ1, Chua}. The one-loop exact path integral that follows from the resulting AS action is known to be the correct Virasoro character \cite{CJ1}, so we will not repeat that step.

%We take $G_0 \equiv g(t,\phi)G(t,r,\phi)$  with $G(t, r_{\rm in},\phi)=1$ to contain a non-trivial $g(t, \phi)$. In other words, we allow boundary gravitons. 

In this parametrization, direct calculation shows that the total action becomes a sum of chiral WZW actions: 
\beq
S_{1} = S_-[G_0] + S_+[\overline G_0]
\eeq
with 
\beq \label{WZW}
S_\pm[G_0] = \frac{k \ell}{4 \pi} \int_{\partial \mathcal{M}} d^2 x \ {\rm Tr}\left(\partial_\phi G_0^{-1} \partial_\pm G_0 + \frac{4P}{\sqrt{k}} L_0 G_0^{-1}\partial_\pm G_0 - \frac{4P^2}{k} L^2_0\right) \mp \frac{k \ell}{12 \pi} \int_{\mathcal{M}} {\rm Tr}\left(G_0^{-1}dG_0\right)^3
\eeq
where $\partial_\pm = \partial_\phi \pm \partial_t$. 
We now use Gauss decomposition for the group element:
\beq \label{gausspar}
\begin{aligned}
G^{-1}_0(t,r,\phi) &= e^{X L_+} e^{Y L_0} e^{Z L_-} = \begin{pmatrix}
1 & 0\\
-X & 1
\end{pmatrix}.
\begin{pmatrix}
e^{Y/2} & 0\\
0 & e^{-Y/2}
\end{pmatrix}.
\begin{pmatrix}
1 & Z\\
0 & 1
\end{pmatrix}
\\[6pt]
\implies G_0(t,r,\phi) &= e^{-Z L_-} e^{-Y L_0} e^{-X L_+} = \begin{pmatrix}
1 & -Z\\
0 & 1
\end{pmatrix}.
\begin{pmatrix}
e^{-Y/2} & 0\\
0 & e^{Y/2}
\end{pmatrix}.
\begin{pmatrix}
1 & 0\\
X & 1
\end{pmatrix},
\end{aligned}
\eeq
where $X,Y$ and $Z$ are functions of $(t,r,\phi)$. Plugging this in \eqref{spatialgauge-first} we find:
\bea 
&A_r = \begin{pmatrix}
\frac12 \partial_r Y + e^Y Z \partial_r X & \partial_r Z + Z \left(\partial_r Y + e^Y Z \partial_r X\right)\\
-e^Y \partial_r X & -\frac12 \partial_r Y - e^Y Z \partial_r X 
\end{pmatrix}, \nonumber
\\[6pt]
&A_\phi = 
\begin{pmatrix}
\frac12 \partial_\phi Y + e^Y Z \partial_\phi X & \partial_\phi Z + Z \left(\partial_\phi Y + e^Y Z \partial_\phi X\right)\\
-e^Y \partial_\phi X & -\frac12 \partial_\phi Y - e^Y Z \partial_\phi X 
\end{pmatrix} + 
\frac{P}{\sqrt{k}} \begin{pmatrix}
1 - 2 e^Y X Z & 2Z \left(1 - e^Y X Z\right) \\
2e^Y X & - 1 + 2 e^Y X Z
\end{pmatrix}. \nonumber
\eea
With this decomposition, the individual terms in the action \eqref{WZW} evaluate to, 
\bea
&\int_{\partial \mathcal{M}} d^2 x \ {\rm Tr}\left(\partial_\phi G^{-1}_0 \partial_\pm G_0 \right) = \int_{\partial \mathcal{M}} d^2 x \ \left(-\frac12\partial_\phi Y \partial_\pm Y + e^Y \left(2 \partial_\phi X \partial_\phi Z \pm \dot Z \partial_\phi X \pm \partial_\phi Z \dot X\right)\right), \nonumber \\[6pt]
&\int_{\partial \mathcal{M}} d^2 x \ {\rm Tr}\left(\frac{4P}{\sqrt{k}} L_0 G^{-1}_0\partial_\pm G_0 - \frac{4P^2}{k} L^2_0\right) = -\int_{\partial \mathcal M} d^2 x \ \left(\frac{2P}{\sqrt{k}} \partial_\pm Y +  \frac{4P}{\sqrt{k}} e^Y X \partial_\pm Z + \frac{2P^2}{k}\right), \nonumber \\[6pt]
&\frac13 \int_{\mathcal{M}} {\rm Tr}\left(G^{-1}_0dG_0\right)^3 = \int_{\partial \mathcal M} d^2 x \ e^Y \left(\dot X \partial_\phi Z - \dot Z \partial_\phi X\right), \nonumber 
\eea
which leads to the boundary reduced action:
\beq
S_\pm[G_0,P] = -\frac{k \ell}{4 \pi} \int_{\partial \mathcal{M}} d^2 x \ \left(\frac12 \partial_\pm Y \left(\partial_\phi Y + \frac{4P}{\sqrt{k}} \right) + \frac{2P^2}{k} - 2e^Y \partial_\pm Z \left(\partial_\phi X -\frac{2P}{\sqrt{k}} X\right)\right).
\eeq
Next, we demand that the gauge field components we obtained, fall into the DS gauge. Note that choice of gauge is more than an aesthetic choice in the asymptotic region, and this can be viewed as our asymptotically AdS$_3$ demand:
\beq 
A_r = L_0, \qquad A_\phi = e^{r} L_+ - \frac{2\pi \mathcal L(t,\phi)}{k} e^{-r} L_-.
\eeq
This gives the constraints 
\bea \label{constraint}
&\partial_r X = 0, \qquad \partial_r Y = 1, \qquad \partial_r Z = -Z,
\\[6pt]
&e^Y \left(\partial_\phi X - \frac{2P}{\sqrt{k}} X\right)= e^{r}, \qquad \partial_\phi Y + \frac{2P}{\sqrt{k}} = -2 Z  e^r , \qquad \partial_\phi Z - Z^2 e^{r} = - \frac{2\pi \mathcal L(t,\phi)}{k} e^{-r}.\nonumber \\ \label{constraint2}
\eea
We introduce the following parametrization\footnote{The DS-matching constraints involve $X$ only through the ``covariant-derivative'' combination
$\,\partial_\phi X-\frac{2P}{\sqrt{k}}\,X$.
It is therefore natural to introduce a new field $f(t,\phi)$ by \emph{defining} it so that this combination
becomes proportional to $f' X$:
\bea
\partial_\phi X-\frac{2P}{\sqrt{k}}\,X \;\equiv\; -\frac{2P}{\sqrt{k}}\,(\partial_\phi f)\,X .
\label{Dfdef}
\eea
This motivates the re-write $X(t,\phi)=\exp\!\Big(-\frac{2P}{\sqrt{k}}\,(f(t,\phi)-\phi)\Big)$. This also has the nice property that $X$ being single-valued on the circle $X(t, \phi+2 \pi)=X(t, \phi)$ translates to the statement that $f-\phi$ is periodic: $ f(t, \phi+2 \pi)- (\phi+2 \pi)= f(t,\phi)-\phi$, which translates to the winding condition \cite{CJ1}: $f(t,\phi+2 \pi)=f(t, \phi)+2 \pi$.
}
\beq \label{paramterization}
X = \exp \left(-\frac{2P}{\sqrt{k}} \left(f-\phi\right)\right)
\eeq
where $f \equiv f(t,\phi)$ is an element of Diff($S^1$) with periodicity $f(t,\phi+2\pi)=f(t,\phi)+2\pi$. Plugging this in \eqref{constraint2}, solves $Y$ and $Z$ as a function of $f$
\bea \label{solutions}
&Y = \log\left(-\frac{\sqrt{k} e^{r+\frac{2P}{\sqrt{k}}(f -\phi)}}{2P \partial_\phi f}\right), \qquad Z = e^{-r}\left(
\frac{\partial^{2}_\phi f}{2\partial_\phi f} -\frac{P}{\sqrt{k}} \partial_\phi f \right), \\  
&e^{-r} \left(\{f,\phi\} - \frac{2P^2}{k} (\partial_\phi f)^2 + \frac{4 \pi \mathcal L(t,\phi)}{k}\right) = 0 \label{Schwarzian}
\eea
where $\{f,\phi\}$ is the Schwarzian derivative of $f$ with respect to $\phi$ in the last relation, that one gets after plugging in the solution of $Z$ in the last constraint in \eqref{constraint2}. 

With \eqref{paramterization} and \eqref{solutions}, we obtain the chiral Alekseev-Shatashvili action on the outer boundary: 
\beq
S_\pm[G_0,P] = -\frac{C}{48 \pi} \int_{\partial \mathcal M} \ d^{2} x \left(\frac{\partial_\pm (\partial_\phi f) \ \partial^2_\phi f}{(\partial_\phi f)^2} + \frac{24P^2\ell}{C} \partial_\pm f \ \partial_\phi f\right), \quad {\rm with} \ \ C = 6 k \ell = \frac{3\ell}{2G_N}.
\eeq
Equation \eqref{Schwarzian} simply shows that $\mathcal L$ has to lie in the coadjoint orbit of the constant representative, $P^2 \sim b_0$.

The path integral for the AS action can be exactly evaluated and reproduces the Virasoro character of the corresponding normal orbit. This was argued for orbits  below  the BTZ threshold in \cite{CJ1}. The calculation only relies on the normal orbit nature, and the same calculation applies above the threshold as well. In fact, the calculation works {\em more} straightforwardly above the threshold because $P$ is real: the variable change defined in equation (5.5) of \cite{CJ1} makes the field complex when the orbit label is below threshold. So the result for the conical defect is best viewed as an analytic continuation in the orbit label. This analytic continuation has similarities in spirit to how we define exceptional orbit results in Section \ref{sec:exc-orb}. Equivalently, we can view the one loop exact character as being defined for arbitrary $P$, and for the exceptional orbit we simply have to mod out by the extra stabilizer modes. This approach gives the correct characters for all orbits, with $c=C+1$. This is again the ``regulate-and-then-quotient" prescription in effect. We will view the above action as sort of the ``universal" Alekseev-Shatashvili orbit action. 

Together with the discussions in the previous subsections, this completes our discussion of the ingredients in the construction of the annulus state.

\subsection{Fourier Kernel} 
\label{PolCS0}

Changing polarization at the inner boundary will be a major theme for us in the next section. So in this subsection we explain carefully why cutting open a path integral naturally prepares a state in a particular polarization, determined by
the boundary terms in the action. We first review the familiar second-order case,
then contrast it with the first-order structure relevant for radial Chern-Simons.

\subsubsection{Polarization Choice at the Cut: Second vs First Order Actions}\label{pol}

Consider the familiar scalar action (see \cite{Robin} for related discussions):
\begin{equation}
  S[\phi] = \int_{t_i}^{t_f} dt\,
  \Big(\tfrac{1}{2}\dot\phi^2 - V(\phi)\Big).
  \label{eq:second-order-action}
\end{equation}
The variation gives
\begin{equation}
  \delta S = \text{(bulk EOM)} + \Big[ \pi\,\delta\phi \Big]_{t_i}^{t_f},
  \qquad \pi = \dot\phi,
  \label{eq:second-order-boundary-variation}
\end{equation}
so a well-posed variational principle is obtained by fixing $\phi$ at the endpoints
($\delta\phi(t_i)=\delta\phi(t_f)=0$). The canonical conjugate is $\pi$, which is
{\it not} fixed.

Now consider the time-evolution kernel
$\langle \phi_f|e^{-iH(t_f-t_i)}|\phi_i\rangle$,
represented as a path integral with Dirichlet boundary conditions
$\phi(t_i)=\phi_i,\ \phi(t_f)=\phi_f$. If we cut the time interval at an intermediate
time $t_*$ and insert a resolution of the identity,
\begin{equation}
  \mathbf{1} = \int d\phi_*\,|\phi_*\rangle\!\langle\phi_*|,
\end{equation}
we obtain
\begin{equation}
  \langle \phi_f|e^{-iH(t_f-t_i)}|\phi_i\rangle
  = \int d\phi_*\,\Big(\underbrace{\langle \phi_f|e^{-iH(t_f-t_*)}|\phi_*\rangle}
 _{\Psi_{\rm right}[\phi_*]}\Big)
  \Big(\underbrace{\langle \phi_*|e^{-iH(t_*-t_i)}|\phi_i\rangle}
  _{\Psi_{\rm left}[\phi_*]}\Big).
  \label{eq:kernel-factorization}
\end{equation}
Note that both ``pieces" are precisely of the same structure as the original propagator kernel with fixed $\phi$ at the two ends, and therefore should naturally be viewed as Dirichlet (sub-)path integrals. The ``left'' piece
$\Psi_{\rm left}[\phi_*]$ (for example)
is exactly the Schr\"odinger wavefunctional in the $\phi$-basis:
$\Psi_{\rm left}[\phi_*] = \langle \phi_*|\Psi_{\rm left}\rangle$.
In other words:

\medskip
\emph{With the standard second-order action and Dirichlet boundary conditions,
cutting the path integral produces a state naturally in the field (configuration)
polarization.}
\medskip

To obtain the wavefunctional in the conjugate basis, $\tilde\Psi_{\rm left}[\pi_*]
= \langle \pi_*|\Psi_{\rm left}\rangle$, one must perform a Fourier transform
\begin{equation}
  \tilde\Psi_{\rm left}[\pi_*]
  = \int d\phi_*\,
  \underbrace{\langle \pi_*|\phi_*\rangle}_{e^{-i\pi_*\phi_*}}\,
  \Psi_{\rm left}[\phi_*],
  \label{eq:fourier-second-order}
\end{equation}
where the kernel $\langle \pi|\phi\rangle = e^{-i\pi\phi}$ is the usual plane wave. That the Fourier transform is the natural thing to do here is clear from analogy with position and momentum operators in non-relativistic quantum mechanics. But it can also be understood as an insertion (at the cut) of a boundary term that changes the Dirichlet problem to a Neumann problem, by viewing the exponent of the Fourier kernel as an addition to the bulk action. See the discussion\footnote{The case of interest here is what was called the {\em general} Neumann problem in \cite{Robin}. See also \cite{Neumann} for the general Neumann problem in gravity.} in \cite{Robin} for generalities of such boundary terms. The key lesson here is that boundary terms are a crucial ingredient in deciding what is the natural polarization of the wave function(al), when we cut open a path integral.   

The above discussion was in the context of a second order action. We will be interested in Chern-Simons theory, which has a first order action. 
For comparison, consider the first-order action
\begin{equation}
  S[q,p] = \int_{t_i}^{t_f} dt\,
  \Big( p\,\dot q - H(q,p)\Big).
  \label{eq:first-order-action}
\end{equation}
Its variation is
\begin{equation}
  \delta S = \text{(bulk EOM)} + \Big[ p\,\delta q \Big]_{t_i}^{t_f}.
  \label{eq:first-order-boundary-variation}
\end{equation}
A well-posed variational principle is obtained by fixing $q$ at the endpoints
($\delta q(t_i)=\delta q(t_f)=0$); $p$ is then the conjugate momentum. Again,
cutting open the path integral in time naturally gives a wave-functional in the $q$-basis:
$\Psi[q_*] = \langle q_*|\Psi\rangle$. To obtain the $p$-basis wavefunctional
one must Fourier transform with kernel $e^{-ipq}$, exactly as before. Again this Fourier transform can be understood as the insertion of a boundary term \cite{Robin}.

The crucial lesson for us is that in a first-order theory one fixes {\it only one}
member of the canonical pair at the boundary; the other is conjugate and should
not be fixed. The choice of which variable is fixed is encoded in the boundary terms.

\subsubsection{Radially Quantized Chern-Simons}\label{polCS}

Now consider Chern-Simons on a 3-manifold whose slices at fixed
radial coordinate $r$ are 2-tori parameterized by $(t,\phi)$. Choosing $r$ as
Euclidean ``time'', the covariant CS action can be written in radial quantization as
\begin{equation}
  S_{\rm CS} = \frac{k}{4\pi}\int dr\int_{T^2}
  \mathrm{Tr}\Big( A_\phi\,\partial_r A_t - A_t\,\partial_r A_\phi
  + 2 A_r F_{t\phi}\Big),
  \label{eq:CS-radial}
\end{equation}
where $A_r$ plays the role of a Lagrange multiplier enforcing $F_{t\phi}=0$. It is important to note that this bulk form is equivalent to our previous bulk forms, up to boundary terms. 

The canonical $1$-form on a slice is
\begin{equation}
  \Theta = \frac{k}{4\pi}\int_{T^2}
  \mathrm{Tr}\Big( A_\phi\,\delta A_t - A_t\,\delta A_\phi\Big).
  \label{eq:CS-canonical-one-form}
\end{equation}
This quantity depends on the boundary terms in unimportant (for us) ways, but the symplectic form does not:
\begin{equation}
  \Omega = \delta\Theta
  = \frac{k}{2\pi}\int_{T^2}
  \mathrm{Tr}\big( \delta A_\phi \wedge \delta A_t \big).
  \label{eq:CS-symplectic-form}
\end{equation}
The symplectic form is of the familiar particle mechanics form: $(A_\phi,A_t)$ form a canonical pair. This parallel means that we can directly export the Fourier kernel of the $p$-$q$ system to our case. It is a correspondence, and not merely an analogy.

We will find it convenient in the next section, to choose a basis of cycles $\{\theta,y\}$ on the torus. This will be useful when we need our calculation to be  general enough to incorporate the general rotating BTZ black hole. In this case $\phi$ and $t$ will be replaced by $\theta$ and $y$, the coordinates along the canonical cycles. We will define them more explicitly in the next section.

The variation of our canonical bulk action ($S_1$ we discussed earlier) without any extra boundary term at the inner boundary, induces a variation of the following form in terms of the canonical variables:
\begin{equation}
  \delta S\big|_{-}
  = -\,\frac{k}{2\pi}\int_{\partial M_{\rm inner}}
  \mathrm{Tr}\big( A_y\,\delta A_\theta\big).
  \label{eq:CS-inner-variation}
\end{equation}
This is a generalization of some of our earlier expressions in the $\phi$-$t$ setting. So a well-posed variational principle at the inner torus is obtained by imposing
Dirichlet boundary conditions for $A_\theta$:
\begin{equation}
  \delta A_\theta\big|_{-} = 0.
  \label{eq:CS-inner-Dirichlet}
\end{equation}
In other words, $A_\theta$ is the ``configuration'' variable at the inner boundary and $A_y$ is its conjugate momentum. The symplectic form takes the form
\begin{equation}
  \Omega = \delta\Theta
  = \frac{k}{2\pi}\int_{T^2}
  \mathrm{Tr}\big( \delta A_\theta \wedge \delta A_y \big).
  \label{eq:CS-symplectic-form-can}
\end{equation}
Now consider performing the radial path integral from some outer boundary down to
the inner torus, while leaving the inner $A_\theta$ as a fixed but arbitrary configuration. The result is
a wavefunctional
\begin{equation}
  \Psi_{\rm ann}[A_\theta] = \langle A_\theta|\Psi_{\rm ann}\rangle.
  \label{eq:annulus-wavefunctional-Atheta}
\end{equation}
Thus:\emph{With the canonical structure (\ref{eq:CS-symplectic-form-can}) and
the Dirichlet condition (\ref{eq:CS-inner-Dirichlet}), cutting open the radial
Chern-Simons path integral at the inner torus naturally prepares a state in the
$A_\theta$-polarization.}
\medskip

To express the same state in the conjugate polarization, one must perform the
functional analogue of a Fourier transform. The overlap between the two
polarizations,
$\langle A_y|A_\theta\rangle$, is given by the exponential of the generating
functional associated to $\Theta$,
\begin{equation}
  \langle A_y|A_\theta\rangle \;\propto\;
  \exp\!\left\{
    -\frac{i k}{2\pi}\int_{T^2}\mathrm{Tr}\big(A_\theta A_y\big)
  \right\},
  \label{eq:CS-polarization-kernel}
\end{equation}
which is simply the field-theoretic analog of $\langle p|q\rangle = e^{-ipq}$. The
wavefunctional in the $A_y$-polarization is then
\begin{equation}
  \langle A_y|\Psi_{\rm ann}\rangle
  = \int \mathcal{D}A_\theta\,
  \langle A_y|A_\theta\rangle\,
  \Psi_{\rm ann}[A_\theta].
  \label{eq:CS-Fourier-transform}
\end{equation}

In particular, eigenstates of the holonomy of $A_\theta$ along the spatial cycle
(the labels $P$ in the main text) are natural in the $A_\theta$-polarization:
in that basis the inner projector produced by a Wilson line looks like a
delta functional $\delta(P-P_0)$. The conjugate basis, which diagonalizes the
holonomy of $A_y$ (labels $\tilde P$), is related by the kernel
(\ref{eq:CS-polarization-kernel}) and corresponds to a functional Fourier (actually cosine, as we will see)
transform of the $A_\theta$ wavefunctional.

The form of the kernel is natural from the Fourier transform perspective, but the reader may be a bit puzzled: how are we to interpret this term as a boundary term in the action, by absorbing it as an addition to the bulk action? The fact that we are working with a Euclidean path integral, but the Fourier term seems to have  $-i$ in the exponent may seem disturbing. But it turns out that it is precisely this combination that one needs for a correct boundary term -- in fact, the extra term when viewed as a term in the exponent of the path integral is insensitive to whether we are working with Euclidean or Lorentzian signature. Since this is a point that can be confusing, we show this explicitly in Appendix \ref{app:pq-boundary} in the less baroque setting of the Hamiltonian $p$-$q$ path integral -- the Chern-Simons case is precisely isomorphic. 

\subsubsection{Conjugate Holonomies}

As mentioned already, our degrees of freedom at the inner boundary are the two holonomies (and not $A_\phi$ or $A_t$), which form a canonical pair because Chern-Simons is a first order theory. On a torus slice, the CS
symplectic form makes the two pieces of holonomy data canonically conjugate. In a primitive cycle basis $(\theta,y)$ one finds a canonical one-form of the
form $\Theta=\frac{k}{2\pi}\int_{T^2}\Tr(A_\theta\,\delta A_y)$ and hence
$\Omega=\delta\Theta=\frac{k}{2\pi}\int_{T^2}\Tr(\delta A_\theta\wedge \delta A_y)$.
In the constant Cartan sector (where both commuting holonomies may be simultaneously
conjugated into a Cartan), this reduces schematically to
$\Omega\propto \delta P\wedge \delta\widetilde P$, where $P$ labels the holonomy along $C$ and
$\widetilde P$ labels the holonomy along $\widetilde C$. So the two holonomies are the two members
of a canonical pair: fixing one at the boundary selects a polarization and leaves the other
unfixed (as its conjugate momentum). We will determine the details by working in the language of the Chern-Simons gauge field, but by restricting to the constant Cartan sector in the upcoming section.

\section{Partition Functions from the Janus Path Integral}
\label{sec:partition}

In this section, we use our previous discussions to compute various partition functions (characters) from our two-boundary Chern-Simons path integral. 

\subsection{Normal and Exceptional Orbits of the Spatial Cycle}

We first focus on \eqref{inner-delta} which corresponds to projection onto a spatial cycle holonomy state at the inner boundary. This means from \eqref{annulus-state} that we are computing $\langle P;C \big|\Psi_{\rm ann}\big\rangle$ or $\langle \mu;C \big|\Psi_{\rm ann}\big\rangle$ depending on whether the holonomy label at the inner boundary is hyperbolic or not. Both of these are normal orbits. We can of course also consider the case when the projection is onto the vacuum (exceptional orbit). 

As we discussed earlier, for normal orbits the outer torus character is
\begin{equation}
\chi_P(\tau)=\frac{q^{P^2}}{\eta(\tau)},\qquad q=e^{2\pi i \tau},\qquad
h-\frac{c-1}{24}=P^2,
\end{equation}
with the coadjoint parameter $b_0$ related by
\begin{equation}
P^2=2\pi b_0.
\end{equation}
The Virasoro coadjoint orbit through a constant $b_0$ is the classical phase space of chiral boundary gravitons.
Upon quantization, each orbit yields a Virasoro representation together with a one-loop shift of the central charge and the conformal weights. The two-boundary Chern-Simons path integral on $T^2\times I$ makes this structure manifest: the outer DS boundary produces the universal boundary graviton factor, while the inner boundary projects onto a chosen orbit (normal or exceptional) by fixing the spatial holonomy. In the normal orbit case, this trivially reproduces the correct Virasoro character directly from our inner boundary projector. Only the exceptional orbit requires a bit of discussion. 

Our prescription that we should treat all orbits as normal, regulate our calculations accordingly and \emph{then} quotient by the stabilizer leads to one distinction from the results in \cite{CJ1}. Our quantum correction in the central charge is $+1$ instead of $+13$. This means that the one-loop path integral over this orbit produces the vacuum character
\begin{equation}
  \chi_{\rm vac}(\tau) 
  \;=\; q^{-(c-1)/24}\,\frac{1-q}{\eta(\tau)}\,
  \;=\; q^{-c/24}\,\prod_{n=2}^{\infty}\frac{1}{1-q^n},
  \label{eq:vacuum-character}
\end{equation}
where $q=e^{2\pi i \tau}$, $\eta(\tau)$ is the Dedekind eta function, and $c$ is the \emph{quantum} chiral central charge equal  to 
\bea
c=C+1.
\eea
The factor $(1-q)$ reflects the removal of the $n=1$ mode, while $1/\eta(\tau)$ encodes the tower of boundary graviton descendants. Note that it is crucial for the interpretation that both the vacuum and excited primaries all belong to the same holographic CFT, that the central charge has the above shift, and this is precisely the virtue of our {\em regulate and then quotient} prescription.

We will see later in Section \ref{conjugBTZ} that there is an alternate way to obtain the exceptional orbit result via a natural analytic continuation argument in the orbit label. This approach seems more general and we will be able to use it also for the conjugate exceptional orbit. 

\subsection{Normal Orbits of the Conjugate Cycle}
\label{subsec:normal-conjugate-orbit}

In the annulus path integral on $T^2\times I$ the inner torus supports a Hilbert space
$\mathcal H_{T^2}$ obtained by quantizing the moduli space of flat $SL(2,\mathbb R)$
connections on the boundary torus. For a choice of primitive cycle $C$ (which we take to
be the spatial circle) and its dual cycle ${\tilde C}$, flat connections are characterized by a
pair of commuting holonomies $(H_C,H_{{\tilde C}})$ modulo conjugation. In the constant-orbit
sector these are conveniently parametrized by real labels $(P,\tilde P)$, with $P$ the
normal-orbit label associated to the spatial holonomy and $\tilde P$ the label for the dual
cycle.

With the canonical bulk action and Dirichlet boundary condition for $A_\theta$ at the inner
boundary, the annulus path integral naturally prepares a state in the $P$-polarization, as we saw earlier. This is what we called the annulus state. The goal of this subsection is purely kinematical: we evaluate the
Fourier kernel that changes polarization from the spatial holonomy $P$ to its conjugate
$\tilde P$, and show that after imposing Weyl invariance and restricting to $P,\tilde P\ge0$
this kernel reduces to the cosine modular $S$-kernel of the Virasoro normal orbits.

\subsubsection{Flat Connections and the Orbit Label $P$}

For normal orbits with a constant coadjoint representative $L(\phi)=b_0$ on the cylinder,
the Drinfel'd-Sokolov angular component of the connection is
\begin{equation}
a_\phi
=
L_1
-
\frac{2\pi}{k} b_0\,L_{-1},
\end{equation}
with $L_{\pm1}$ as in earlier sections. In the $2\times2$ representation of \cite{Ammon} this is
\begin{equation}
a_\phi
=
\begin{pmatrix}
0 & -q_0 \\
-1 & 0
\end{pmatrix},
\qquad
q_0 := \frac{2\pi}{k} b_0.
\end{equation}
The eigenvalues $\lambda_\pm$ of $a_\phi$ obey
\begin{equation}
\lambda_\pm^2 = s
\qquad\Rightarrow\qquad
\lambda_\pm = \pm\sqrt{\frac{2\pi b_0}{k}}.
\end{equation}
With the normal-orbit label $P$ defined by
$P^2 = 2\pi b_0$, we get
\begin{equation}
\lambda_\pm = \pm \frac{P}{\sqrt{k}}.
\end{equation}
Any constant $a_\phi$ is conjugate to a Cartan element. Writing
$H=L_0$ with $\mathrm{Tr}(H^2)=\tfrac12$, we can go to a gauge where the spatial component
is diagonal,
\begin{equation}
A_\phi = a\,H,
\qquad
a\in\mathbb R,
\end{equation}
and the holonomy along the spatial cycle $C$ is
\begin{equation}
M_C
=
\mathcal P\exp\!\Bigl(-\oint_C A_\phi \Bigr)
=
\exp(-2\pi a H),
\end{equation}
with eigenvalues $\exp(\mp\pi a)$.
Matching these with the DS-gauge eigenvalues $\exp(\mp 2\pi P/\sqrt{k})$ of
$\exp(-2\pi a_\phi)$ fixes
\begin{equation}
a = \frac{2P}{\sqrt{k}},
\qquad\Rightarrow\qquad
A_\phi = \frac{2P}{\sqrt{k}}\,H.
\label{eq:Aphi-Cartan}
\end{equation}
Thus the orbit label $P$ is precisely the Cartan holonomy parameter along the spatial
cycle. We will take $P\ge0$ so that each conjugacy class is represented once.

The choice that $P$ is real means that we are working with hyperbolic holonomy, and these are states above the BTZ threshold. Any $SL(2,\IR)$ Chern-Simons quantization is expected to contain the hyperbolic continuum \cite{Teschner, Killingback}. 

To incorporate the general boundary torus with complex modulus $\tau$ (i.e., the addition of chemical potential in the ensemble), we will work with holonomies around the natural cycles of the torus. The corresponding coordinates are $\theta$ and $y$. They are easier to define via the complex coordinate $z=\phi-i t_E$ used in \cite{Ammon} (but see some important comments regarding our conventions in Appendix \ref{AppConventions}\footnote{Also: we define $\tau \equiv \tau_1-i \tau_2$. The chiral boundary condition becomes $a_\theta = \tau a_y$ in these coordinates. But note that in identifying the Fourier kernel (which is a kinematic object) we do {\em not} impose chirality or any other constraint.}.):
\bea
z = \theta + \tau y, \ \implies \phi = \theta+\tau_1 y, \ \ t_E = \tau_2 y.
\eea
Flatness on $T^2 \times I$ implies that the component of $A$ along the second primitive cycle $\tilde C$ commutes with $A_\theta$ and can be brought to the same Cartan\footnote{This is the analogue of the classical statement that $q$ and $p$ commute in particle mechanics. But upon quantization they have a non-trivial Fourier kernel.}. Working in coordinates $(\theta,y)$ adapted to the cycles $(C,{\tilde C})$, with
$C:\theta \sim \theta + 2\pi$ and ${\tilde C}:y \sim y + 2\pi$, we can therefore write
\begin{equation}
  A_\theta = \frac{2P}{\sqrt{k}}\,H,
  \qquad
  A_y = \frac{2\tilde P}{\sqrt{k}}\,H.
  \label{double}
\end{equation}
The pair $(P,\tilde P)$ provides local Darboux coordinates on the moduli space of flat connections on the torus in this constant-connection sector\footnote{The normalization of $\tilde P$ in $A_y$ is chosen to match that of $P$ in $A_{\theta}$. This also has the virtue that it makes various formulas nicer, and also ends up matching with the standard 2D CFT/Liouville conventions.}.
In the next subsection we discuss some useful facts about these coordinates on the general torus with modulus $\tau$, by expanding
it in the dual one-form basis $\omega_C,\omega_{\tilde C}$. 

\subsubsection{Cycles and Canonical One-forms on the Torus}

We now discuss some useful facts for describing gauge fields on the
 torus with a general complex modulus
$\tau = \tau_1 - i \tau_2$ with $\tau_2 > 0$. The real part $\tau_1$ encodes the angular
potential (chemical potential for angular momentum) and the imaginary part $\tau_2$ the
inverse temperature (up to an overall scale). 

Let $(\phi,t_E)$ be real coordinates on the boundary torus. We choose the fundamental
identifications to be
\begin{equation}
(\phi,t_E) \sim (\phi+2\pi,t_E),
\qquad
(\phi,t_E) \sim \bigl(\phi+2\pi\tau_1,\;t_E+2\pi\tau_2\bigr),
\label{eq:torus-identifications}
\end{equation}
so that the complex coordinate $w = \phi - i t_E$ obeys
$w \sim w + 2\pi \sim w + 2\pi\tau$, with $\tau=\tau_1 - i\tau_2$. The two primitive cycles we will use as our canonical
basis are then
\begin{itemize}
\item $C$: the image of the segment $(\phi,t_E) = (s,0)$, $s\in[0,2\pi]$ (spatial cycle);
\item ${\tilde C}$: the image of the segment
$(\phi,t_E) = (2\pi\tau_1\,u,\;2\pi\tau_2\,u)$, $u\in[0,1]$ (dual cycle).
\end{itemize}
It is convenient to introduce one-forms $\omega_C,\omega_{{\tilde C}}$ on the torus that are dual
to these cycles in the sense
$\oint_C \omega_C = 2\pi$, $\oint_{{\tilde C}} \omega_C = 0$ and
$\oint_C \omega_{{\tilde C}} = 0$, $\oint_{{\tilde C}} \omega_{{\tilde C}} = 2\pi$.
Solving these conditions with the identifications \eqref{eq:torus-identifications} gives
\(
\omega_C = d\phi - \frac{\tau_1}{\tau_2}\,dt_E,
\qquad
\omega_{{\tilde C}} = \frac{1}{\tau_2}\,dt_E.
\)
Indeed, along $C$ we have $d\phi=2\pi$, $dt_E=0$, so $\oint_C \omega_C = 2\pi$ and
$\oint_C \omega_{{\tilde C}}=0$. Along ${\tilde C}$ we have $d\phi=2\pi\tau_1$, $dt_E=2\pi\tau_2$,
so $\oint_{{\tilde C}}\omega_C = 2\pi\tau_1 - (\tau_1/\tau_2)2\pi\tau_2 = 0$ and
$\oint_{{\tilde C}}\omega_{{\tilde C}} = (1/\tau_2)2\pi\tau_2 = 2\pi$.

The wedge product of these dual one-forms is
\(
\omega_C\wedge\omega_{{\tilde C}}
=
\left(d\phi - \frac{\tau_1}{\tau_2}dt_E\right)\wedge\frac{1}{\tau_2}dt_E
=
\frac{1}{\tau_2}\,d\phi\wedge dt_E.
\)
The area of the fundamental parallelogram spanned by the two lattice vectors
$(2\pi,0)$ and $(2\pi\tau_1,2\pi\tau_2)$ in the $(\phi,t_E)$-plane is
\(
\mathrm{Area}(T^2)
=
\det\begin{pmatrix}
2\pi & 2\pi\tau_1\\[2pt]
0    & 2\pi\tau_2
\end{pmatrix}
=
(2\pi)^2\tau_2. 
\)
Therefore
\(
\int_{T^2}\omega_C\wedge\omega_{{\tilde C}}
=
\frac{1}{\tau_2}
\int_{T^2} d\phi\wedge dt_E
=
\frac{1}{\tau_2}\,(2\pi)^2\tau_2
=
(2\pi)^2.
\)
This shows explicitly that the basic volume factor appearing in the Fourier kernel is a
purely topological quantity: it computes the oriented intersection number of $C$ and
${\tilde C}$ times $(2\pi)^2$ and is independent of the complex structure modulus $\tau$.
In particular, turning on a chemical potential (nonzero $\tau_1$) does not affect the value of
$\int_{T^2}\omega_C\wedge\omega_{{\tilde C}}$.

Restricting to the Cartan zero-modes, it is natural to expand the
constant connection on the torus in this dual basis as
\bea
A\bigl|_{T^2}
=
\frac{2P}{\sqrt{k}}\,H\,\omega_C
+
\frac{2\tilde P}{\sqrt{k}}\,H\,\omega_{{\tilde C}}. \label{double-cycle-form}
\eea
By construction, this choice makes the holonomy along $C$ depend only on $P$ and the
holonomy along ${\tilde C}$ only on $\tilde P$, for any $\tau$. In components with respect to
$(\phi,t_E)$ one finds
\(
A_\phi = \frac{2P}{\sqrt{k}}\,H,
\qquad
A_{t_E} = \frac{2}{\sqrt{k}\,\tau_2}\bigl(\tilde P - \tau_1 P\bigr)H,
\)
but in the canonical one-form and Fourier kernel it is the coefficients of $A$ along
$\omega_C$ and $\omega_{{\tilde C}}$ that enter. The only $\tau$-dependence there is through
$\int_{T^2}\omega_C\wedge\omega_{{\tilde C}} = (2\pi)^2$, which we have just shown to be
independent of $\tau$.

\subsubsection{Evaluating the Fourier Kernel}

We discussed in previous sections that the radial quantization of Chern-Simons theory on $T^2\times I$
induces the overlap between the two polarizations is given by the functional Fourier
kernel
\begin{equation}
\label{eq:Fourier-functional}
\langle A_y|A_\theta\rangle
\propto
\exp\!\left(
-\frac{i k}{2\pi}
\int_{T^2}
\mathrm{Tr}\!\left(A_\theta A_y\right)
\right),
\end{equation}
see equation \eqref{eq:CS-polarization-kernel}. In our concrete setting, $(\theta,y)=\big(\phi - \frac{\tau_1}{\tau_2}\,t_E, \frac{t_E}{\tau_2}\big)$ on the 
torus. Substituting the Cartan zero-modes \eqref{double} 
into $A=A_\theta d\theta + A_y dy$ reproduces \eqref{double-cycle-form} and
\begin{equation}
\int_{T^2}
\mathrm{Tr}\!\left(A_\theta A_{y}\right)
=
\mathrm{Tr}\!\left( H^2 \right)\,
\frac{4 P \tilde P}{k}
\int_{T^2}
\omega_C\wedge\omega_{{\tilde C}}
=
\frac12\cdot \frac{4 P \tilde P}{k}\cdot (2\pi)^2
=
\frac{8\pi^2}{k}\,P\tilde P.
\end{equation}
Using this in \eqref{eq:Fourier-functional} we obtain the plane-wave kernel on the $P$-line,
\begin{equation}
K(P,\tilde P)
:=
\langle \tilde P|P\rangle
\propto
\exp\!\left(
-\frac{i k}{2\pi}\cdot \frac{8\pi^2}{k}\,P\tilde P
\right)
=
\exp\!\left(
- i 4\pi P\tilde P
\right).
\label{eq:plane-wave-kernel}
\end{equation}
Different choices of orientation of the cycles or of the radial coordinate would flip the sign
of the exponent, but this overall phase is immaterial once we restrict to Weyl-invariant
wavefunctions and project to the half-line.

\subsubsection{Weyl Reflection and the Cosine Kernel}

The Cartan generator $H=L_0$ has Weyl group $W\cong\mathbb Z_2$ acting as $H\to -H$.
Our Cartan representative $A_\theta = (2P/\sqrt{k})H$ therefore has a Weyl-reflected partner
with label $-P$. The conjugacy class of the holonomy, and hence the Virasoro data $(c,h)$,
depend only on $P^2$, so physical quantities such as the character $\chi_P(\tau)$ are Weyl
invariant,
\begin{equation}
\chi_P(\tau) = \chi_{-P}(\tau).
\end{equation}
It is natural to restrict attention to Weyl-invariant wavefunctions on the full line
$P\in\mathbb R$, i.e.\ to even functions
\begin{equation}
\Psi_{\text{phys}}(P) = \Psi_{\text{phys}}(-P),
\end{equation}
and to regard $P\ge0$ as a fundamental Weyl chamber\footnote{An identical requirement is made when quantizing $SU(2)$ flat connections on the torus as well, and results in the range being restricted to a half-circle instead of a circle.  What we are finding here can be viewed as the non-compact analogue.}. For Weyl-invariant states only the even part is relevant, and the physical transform on the
half-line $P,\tilde P\ge0$ is a cosine transform:
$\tilde\Psi(\tilde P) \sim
\int_{-\infty}^{\infty}\mathrm dP\;
\mathrm e^{-i4\pi P\tilde P}\,\Psi(P)
\sim 
\int_0^\infty \mathrm dP\,
\cos(4\pi P\tilde P)\,\Psi_{\text{even}}(P)$.

We can also fix the normalization of the kernel $S(P,\tilde P) \sim \cos(4\pi P\tilde P)$, by demanding that the transform is unitary on $L^2(\mathbb R_+,\mathrm dP)$:
\begin{equation}
\int_0^\infty \mathrm dP\;
S(P,\tilde P)\,S(P,\tilde P')
=
\delta(\tilde P-\tilde P').
\end{equation}
Using a standard formula \cite{NISTDLMF} to write the orthogonality relation for cosines on the half-line in the form
\begin{equation}
\int_0^\infty \mathrm dP\;\cos(4\pi P\tilde P)\,\cos(4\pi P\tilde P')
=
\frac18\,\delta(\tilde P-\tilde P'),
\qquad
\tilde P,\tilde P'>0,
\end{equation}
we see that unitarity is achieved by
\begin{equation}
S(P,\tilde P)
=
2\sqrt2\,\cos(4\pi P\tilde P),
\qquad
P,\tilde P\ge0.
\label{eq:cosine-kernel}
\end{equation}
This is precisely the modular $S$-kernel that appears in the Virasoro normal-orbit
characters:
\begin{equation}
\chi_{\tilde P}\!\left(-\frac{1}{\tau}\right)
=
\int_0^\infty \mathrm dP\;
S(P,\tilde P)\,\chi_P(\tau),
\end{equation}
with $\chi_P(\tau)=q^{P^2}/\eta(\tau)$ and $q=\mathrm e^{2\pi i\tau}$. Note that in our calculation, we obtain the right hand side {\em directly} when we project onto a conjugate label. The left hand side follows simply as a mathematical identity. This means that we could have obtained the left hand side by using a modular S-transform on the spatial cycle character with the $\tilde P$-label. This was the philosophy adopted in \cite{CJ1} for the special case when the orbit label was exceptional, to obtain the BTZ character. We will discuss the BTZ case in a later subsection.

The two-boundary Chern-Simons path integral on the thickened torus thus reproduces the Virasoro modular
$S$-transform purely from the Fourier kernel. We have done this simple calculation in some detail, to drive home the point that the kernel (including the normalization) follows from general expectations from the bulk quantization.

\subsection{Quantization of the Moduli Space of Flat $SL(2,\IR)$ Connections}
\label{subsec:holonomy-phase-space}

The previous subsection showed that the change of polarization between the spatial holonomy
label $P$ and its conjugate $\tilde P$ is governed by the cosine kernel
\eqref{eq:cosine-kernel}, which matches the Virasoro modular $S$-kernel. In this subsection
we collect some conceptual lessons behind that calculation. We emphasize four points:
\begin{itemize}
\item In the Janus cobordism path integral on $T^2\times I$ the holonomy is a fluctuating
degree of freedom, not a fixed parameter;
\item The moduli space of flat connections on a torus and its quantization provide the
natural holonomy phase space $(P,\tilde P)$;
\item Brown-Henneaux boundary conditions restrict classical configurations to a Lagrangian
submanifold of this phase space, but this still defines a perfectly good state in
$\mathcal H_{T^2}$ that can be read in either polarization;
\item This picture dovetails with our previous discussions, where we contrasted the
continuous holonomy phase space with the discrete spectrum of primaries in a holographic CFT.
\end{itemize}

\subsubsection{Holonomy Fluctuations in the Janus Cobordism}

The Janus (thickened-torus) path integral with an outer Drinfel'd-Sokolov boundary and
an inner Dirichlet cut prepares an annulus state
as a sum over holonomies. This is a key conceptual statement:
\begin{itemize}
\item The path integral is an integral over holonomy labels $P$, with weights
$\chi_P(\tau)$ supplied by the outer Alekseev-Shatashvili path integral.
\item The outer boundary conditions fix the form of the connection (DS gauge, chirality, complex structure $\tau$), but they do \emph{not} fix the value of the
holonomy. The orbit label $P$ is a dynamical variable even asymptotically. This is a somewhat unfamiliar situation in AdS/CFT where we are usually treating the background as classical, and therefore the asymptotic charges are fixed. 
\end{itemize}
Only if we subsequently insert a projector at the inner boundary--for instance by
re-inserting a Wilson line with a definite spatial holonomy class--do we pick out a
particular primary. The bare Janus cobordism by itself produces a superposition.

In this sense it is misleading to think of $a_\phi$ (or the associated holonomy) as a
classical parameter already fixed at infinity. For each classical configuration in the
path integral, $a_\phi$ certainly has a definite value, but the quantum state is a functional
of that value and the path integral sums over all allowed $a_\phi$ consistent with the
boundary conditions.

\subsubsection{Holonomy Phase Space}

On the thickened torus $T^2\times I$ with a worldline excised, the Chern-Simons equations
of motion imply that the connection is flat away from the worldline. As we discussed, a flat connection on a torus is completely specified (up to gauge) by a pair of commuting
holonomies along two primitive cycles. For our choice of cycles $(C,{\tilde C})$ the moduli
space of flat $SL(2,\mathbb R)$ connections is a two-dimensional phase space,
\begin{equation}
\mathcal M_{\text{flat}}(T^2)
\sim
\{(H_C,H_{{\tilde C}})\ \text{commuting}\}/\text{conjugation},
\end{equation}
which in the constant-sector we parametrize by the real labels $(P,\tilde P)$ introduced
above. Because the connection is flat, these holonomy labels are radially conserved: the
holonomies measured on the inner and outer tori are the same for a given state.

Radial constancy, however, does \emph{not} mean that $P$ and $\tilde P$ are fixed numbers
in the quantum theory. It means that the classical equations restrict the configuration
space to the moduli space $\mathcal M_{\text{flat}}(T^2)$, and that any choice of radial
slice provides a good set of coordinates on that space. Quantization then promotes
$(P,\tilde P)$ to operators acting on the torus Hilbert space $\mathcal H_{T^2}$, with a Fourier kernel between them because of their symplectic structure.

\subsubsection{Brown-Henneaux Boundary Conditions and Lagrangian Submanifolds}

Drinfel'd-Sokolov gauge and Brown-Henneaux boundary conditions at the outer torus impose
a specific relation between the angular and time components of the boundary connection
($a_t=a_\phi$ for $\mu=1$). The definition of the asymptotic boundary in Euclidean signature also involves fixing the complex structure $\tau$ of the boundary torus. {\em Classically}, this means that for each orbit label $P$ the two holonomies $H_C$ and $H_{{\tilde C}}$ are built from the same constant DS matrix $a$, and once the complex structure $\tau$ is specified, their conjugacy classes are not independent. 

From the point of view of the holonomy phase space $\mathcal M_{\text{flat}}(T^2)$, these
boundary conditions cut out a one-dimensional Lagrangian submanifold
$\mathcal L_{\text{BH}}\subset\mathcal M_{\text{flat}}(T^2)$, parametrized by $P$.
Schematically,
\begin{equation}
\mathcal L_{\text{BH}}
=
\{(P,\tilde P)\mid \tilde P = f_\tau(P)\},
\end{equation}
for some function $f_\tau$ determined by the chirality condition and the choice of modulus
$\tau$. Because $\Omega\propto \mathrm dP\wedge\mathrm d\tilde P$, any single-valued relation
between $P$ and $\tilde P$ defines a middle-dimensional Lagrangian submanifold. Quantizing this configuration yields a particular state $|\Psi_{\text{ann}}(\tau)\rangle$
in the full Hilbert space $\mathcal H_{T^2}$ which is the quantization of the Lagrangian
brane $\mathcal L_{\text{BH}}$. In the $P$-basis this state is precisely the annulus state
with coefficients $\chi_P(\tau)$. Semiclassically, its support is localized on
$\mathcal L_{\text{BH}}$.

Crucially, this does \emph{not} reduce the Hilbert space or eliminate the conjugate variable
$\tilde P$\footnote{Ultimately, this is again a result of the fluctuating holonomy. If we were to fix it (as we would classically), the conjugate holonomy would also be fixed by the asymptotic boundary conditions and the fixed torus modulus.}. It simply specifies which vector in $\mathcal H_{T^2}$ is prepared by the
Janus path integral. That vector can be expanded in any complete basis, including the
$\tilde P$-basis, using the Fourier kernel. In other words, localization of classical configurations on a Lagrangian submanifold
$\mathcal L_{\text{BH}}$ is perfectly compatible with reading the corresponding quantum
state in either polarization; the change of polarization is controlled entirely by the
symplectic structure and the topology of the torus, kinematically.

\subsubsection{CFT Spectrum and Modular Bootstrap}\label{sec:mod-bootstrap}

From the Chern-Simons point of view it is natural to treat $(P,\tilde P)$ as continuous coordinates
on the holonomy phase space $\mathcal M_{\text{flat}}(T^2)$, and the cosine kernel
\eqref{eq:cosine-kernel} as a Fourier transform between two polarizations. This is
the picture we have used in deriving the modular $S$-kernel and in interpreting the BTZ
character as a sum over primaries in later sections.

A microscopic holographic CFT, however, has a \emph{discrete} set of primaries $\{h_n\}$,
with $P_n^2 = h_n - (c-1)/24$. This continuum-to-discreteness tension can be understood as the usual tension that makes modular bootstrap possible. To see this, let us write the CFT partition function as a sum over characters 
$Z(\tau,\bar \tau) = \sum_{P, \bar P}N_{P, \bar P} \, \chi_P(\tau) \chi_{\bar P}(\bar \tau)$, 
where to emphasize that the spectrum is discrete, we have now written it as a sum rather than an integral (unlike before). In terms of a spectral density $\rho(P,\bar P)$ such as the one in \eqref{distrib-Z}, this means we have something like
$ \rho(P,\bar P)
= \sum_{P_*,\bar P_*} N_{P_*\bar P_*}\,
\delta(P-P_*)\,\delta(\bar P-\bar P_*)$.
Modular invariance $Z(-1/\tau,-1/\bar\tau) = Z(\tau,\bar\tau)$, and writing the $S$-transformed characters in terms of modular kernels $S(P,P')$ results in the equality, 
\bea
\rho_S(P', \bar P') = \rho(P, \bar P),
\eea
where we have defined the S-transformed density:
\[
\rho_S(P',\bar P')
= \int dP\,d\bar P\;\rho(P,\bar P)\,S(P,P')\,S(\bar P,\bar P')\,.
\]
The fact that this is a stringent constraint on the spectrum is clearer when we write it as
\bea
\sum_{P_*,\bar P_*} N_{P_*\bar P_*}\,
\delta(P'-P_*)\,\delta(\bar P'-\bar P_*)
\;=\;
\sum_{P_*,\bar P_*} N_{P_*\bar P_*}\,
S(P_*,P')\,S(\bar P_*,\bar P')\,, \label{PoissonResum}
\eea
which is to be understood as an equality of distributions in the variables
$(P',\bar P')$. 

In the next subsection we will show that the modular kernel for the vacuum row shows up naturally in our discussion of the BTZ black hole. Tauberian modular bootstrap of \cite{Zhiboedov, Pal} can be used to argue that the asymptotic behavior of the vacuum modular $S$-kernel is essentially the same data as the density of the high lying spectrum, within averaging windows\footnote{It is important to emphasize that while the $S$-kernel captures ``averaged" information about the spectrum in this sense, it is more than just {\em some} naive coarse-graining of the spectrum. It shows up as an {\em exact} object when we demand modular invariance of $Z(\tau, \bar \tau)$, by showing up in the modular transforms of the characters.}. But it seems possible that this relation \eqref{PoissonResum} may be hiding more clues about the structure of holography in AdS$_3$. It will be very interesting if one could more directly understand the physics underlying the emergence of individual discrete spectra from the universality of the continuum modular $S$-kernel. 

We note that a parallel structure (with some crucial differences) exists even in the case of compact Chern-Simons theories. This is discussed in some detail, in Appendix \ref{sec:compact_vs_noncompact}. $SU(2)_k$ Chern-Simons theory defines chiral characters and $S$-kernels, but to define a full non-chiral CFT, one again needs discrete spectrum information.

\subsection{Conjugate Exceptional Orbit: The BTZ Black Hole}\label{conjugBTZ}

Our results so far show that the bulk theory leads to the cosine kernel, 
\begin{equation}
  \langle \tilde P; \tilde C | P; C\rangle \equiv S(\tilde P,P) = 2\sqrt{2}\,\cos(4\pi P\tilde P),
  \label{eq:co$S$-kernel}
\end{equation}
when we project the annulus state onto a holonomy on the conjugate cycle:
\begin{equation}
 \langle \tilde P; \tilde C|\Psi_{\text{ann}}\rangle = 
  \int_0^\infty dP\;S(\tilde P,P)\,\chi_P(\tau) =\chi_{\tilde P}\!\left(-\frac{1}{\tau}\right).
  \label{eq:S-on-generic}
\end{equation}
Note that the last equality is a mathematical identity, that follows from the cosine kernel integral and the expressions for the characters. In other words, we are {\em deriving} that the result has a natural interpretation as a character on the $S$-transformed solid torus. The result is what one would obtain when one places the dual cycle on a hyperbolic holonomy state.

These observations suggest that we can ask the following question: what happens if we place the inner boundary in a state corresponding to the {\em exceptional orbit}? The formula above is suggestive that we should end up getting the $S$-transformed vacuum character, also called  the BTZ character. This result would be particularly striking: in our bulk language, it will be naturally interpreted as expressing the character associated to the bulk geometry with a smoothly contractible thermal cycle as a \emph{sum} over super-threshold states, all of which have a defect at the origin of the spatial cycle.

The technical difficulty is that the state in which we wish to place the dual cycle state requires analytic continuation to define. This is related to the fact that the exceptional orbit corresponds to the vacuum character, which is a degenerate representation with null states at level 1. But there is a natural guess: we can adopt the standard analytic continuation familiar from 2D CFT to define the modular $S$-kernel row relevant for the vacuum\footnote{Note that by reproducing the modular kernel from the bulk, we have established already that its analytic continuation properties are isomorphic to those of the modular kernel in 2D CFT. We also note that the label space in the dual cycle spans the same Hilbert space. This is familiar from $SU(2)_k$ Chern-Simons theory on $T^2$ where the modular $S$-matrix is usually denoted by $S_{ij}$ and both its legs span the same set of spin labels. For classic discussions of the torus Hilbert space and modular transformations in compact Chern-Simons theory, see e.g. \cite{WittenJones,Seiberg,Axelrod,Jeffrey:1992tk}.}. This suggests that the modular kernel one obtains when we project the annulus state onto the conjugate exceptional orbit at the inner boundary (which we will denote by $| 0; \tilde C\rangle$) should be the standard one in 2D CFT that results in $\chi_{\text{vac}}\!\left(-\frac{1}{\tau}\right)$:
\begin{equation}
  \langle 0; \tilde C|\Psi_{\text{ann}}\rangle
  \;=\;
  \int_0^\infty dP\;S_{0P}\,\chi_P(\tau),
  \label{eq:vacuum-row-goal}
\end{equation}
with the vacuum row of the kernel given by the standard sinh-sinh form
\begin{equation}
  S_{0P}
  \;=\;
  4\sqrt{2}\,\sinh(2\pi bP)\,\sinh(2\pi P/b),
\ \  P>0, \ \ c\equiv1+6\Big(b+\frac{1}{b}\Big)^2,
  \label{eq:S0P-sinh-sinh}
\end{equation}
in our $P>0$ conventions. It is a mathematical identity, that this integral is simply the $S$-transform of the Virasoro vacuum character, $\chi_{\text{vac}}\!\left(-\frac{1}{\tau}\right).$

\subsubsection{Exceptional Orbit as a Degenerate Virasoro Representation}\label{sec:exc-orb}

The manipulations required to get \eqref{eq:vacuum-row-goal} are essentially isomorphic to those in 2D CFT, but we quickly go through them below because the logic is slightly more bulk-based.

We start by reproducing the exceptional orbit case of the {\em spatial} cycle\footnote{We previously derived it using our regulate-like-a-normal-orbit, but quotient-like-an-exceptional-orbit prescription. Here we obtain it as a projection on to a suitable bra.} using this ``projecting onto the correct analytically continued bra" perspective. Once we reproduce this, the same bra, but for the {\em dual} cycle, should reproduce \eqref{eq:vacuum-row-goal}. 

We will follow standard conventions for degenerate representations in the ``Liouville parameterization". Defining $Q\equiv(b+1/b)$ with $c=1+6 Q^2$, degenerate representations are labeled by integer pairs $(m,n)$ with
\begin{equation}
  P_{m,n} \equiv \frac{i}{2}\Bigl(mb + \frac{n}{b}\Bigr),
\end{equation}
so that the scaling dimensions take the form
\begin{equation}
  h_{m,n} = \frac{Q^2}{4} + P_{m,n}^2.
\end{equation}
The vacuum representation is the $(m,n)=(1,1)$ degenerate case, which has $h_{1,1}=0$ and\footnote{It is worth noting here as an aside, that under the standard Liouville parametrization $c=1+6Q^2$ for the central charge, this leads to $P^2= -\frac{c-1}{24}=-\frac{C}{24}$, which is precisely the definition of the exceptional orbit.}
\begin{equation}
  P_{1,1} = \frac{iQ}{2}.
\end{equation}
Because the vacuum module has a null at level 1, its character is not simply $\chi_{iQ/2}$ but rather a  difference of two analytically continued non-degenerate characters. It is trivial to check that
\begin{equation}
  \chi_{\text{vac}}(\tau)
  \;=\;
  \chi_{iQ/2}(\tau) - \chi_{i(b-1/b)/2}(\tau),
  \label{eq:vac-as-difference}
\end{equation}
where $\chi_{iQ/2}$ and $\chi_{i(b-1/b)/2}$ are the analytic continuations of $\chi_P$ to purely imaginary $P$. This reduces to
\begin{equation}
  \chi_{\text{vac}}(\tau)
  = \frac{1-q}{\eta(\tau)}.
\end{equation}
The key lesson for our purposes from these standard facts is that it suggests a natural definition of an inner-boundary \emph{exceptional} bra in the spatial holonomy basis:
\begin{equation}
  \langle  0; C|
  \;\equiv\; \left\langle {\frac{iQ}{2}};C\right| - \left\langle {\frac{i\left(b-\frac{1}{b}\right)}{2}};C\right|,
  \label{eq:exceptional-bra}
\end{equation}
If we project on to this state, we get the vacuum character that we produced earlier without explicitly invoking this state, but by using the normal orbit philosophy followed by quotienting of the zero modes. 

We will be interested in the same form for the state, but in the {\em conjugate cycle} basis, when projecting onto the conjugate exceptional orbit.

\subsubsection{Analytic Continuation and Projection onto Conjugate Exceptional Orbit}

The key dynamical input from the bulk is that the two-boundary CS kernel $S(\tilde P,P)$ in \eqref{eq:co$S$-kernel} is an analytic function of the inner orbit label $\tilde P$. This allows us to evaluate the $S$-transform \eqref{eq:S-on-generic} not only for real (hyperbolic) $\tilde P$, but also for imaginary $\tilde P$ corresponding to degenerate orbits.

Using the annulus state and the definition of the exceptional bra \eqref{eq:exceptional-bra}, we can write the overlap of the annulus state with the conjugate exceptional orbit as
\begin{equation}
  \langle 0; \tilde C|\Psi_{\text{ann}}(\tau)\rangle
  =
  \Bigl(\langle {iQ/2}; \tilde C| - \langle {i(b-1/b)/2}; \tilde C|\Bigr)
  |\Psi_{\text{ann}}(\tau)\rangle.
\end{equation}
By construction, $\langle \tilde P; \tilde C|\Psi_{\text{ann}}(\tau)\rangle = \chi_{\tilde P}(-1/\tau)$ for any (real or analytically continued) $\tilde P$, so
\begin{equation}
  \langle 0; \tilde C|\Psi_{\text{ann}}(\tau)\rangle
  =
  \chi_{iQ/2}\!\left(-\frac{1}{\tau}\right)
  - \chi_{i(b-1/b)/2}\!\left(-\frac{1}{\tau}\right)
  =
  \chi_{\text{vac}}\!\left(-\frac{1}{\tau}\right),
\end{equation}
where in the last equality we used \eqref{eq:vac-as-difference}. We can explicitly evaluate the form of the vacuum kernel by evaluating the intermediate step in the above sequence explicitly. For imaginary $\tilde P = i y$ one has
\begin{equation}
  \cos(4\pi P\tilde P) = \cosh(4\pi Py),
\end{equation}
and plugging this in the intermediate expressions and combining the hyperbolic cosines, we get precisely
\begin{equation}
  \chi_{\text{vac}}\!\left(-\frac{1}{\tau}\right)
  =
  \int_0^\infty dP\;4\sqrt{2}\,
  \sinh(2\pi bP)\,\sinh(2\pi P/b)\,\chi_P(\tau),
\end{equation}
which is the form that we set out to demonstrate.

Note that the two crucial facts we have used can be viewed as statements about the two-boundary CS path integral and its changes of polarization. We used (i) the analyticity of the CS kernel in the inner orbit label $\tilde P$ and (ii) and a natural definition of the exceptional orbit state (which, notably, applies equally well to the {\em spatial} cycle states). This was sufficient to derive the sinh-sinh vacuum row $S_{0P}$. 

\subsubsection{Bulk Fourier Transform as Boundary Modular Transform}

It is useful to emphasize what is being used here from the CFT side. Independently of the bulk, Virasoro characters obey a modular $S$-transformation of the form
\begin{equation*}
  \chi_{\tilde P}\!\left(-\frac{1}{\tau}\right)
  = \int_0^\infty dP\, S(P,\tilde P)\,\chi_P(\tau)\,,
  \qquad
  S(P,\tilde P) = 2\sqrt{2}\,\cos(4\pi P\tilde P)\,.
\end{equation*}
In other words, the family of characters $\{\chi_P(\tau)\}$ is closed under the modular $S$-kernel $S(P,\tilde P)$, and the parameter $\tilde P$ again labels the same continuum of Virasoro representations, now viewed in the $S$-transformed channel.

From the bulk Chern-Simons point of view, the two-boundary Chern-Simons path integral defines an annulus state $|\Psi_{\rm ann}\rangle \in \mathcal{H}_{T^2}$ which can be expanded either in the $P$-basis or in the conjugate $\tilde P$-basis. The overlap
\begin{equation*}
  \langle \tilde P; \tilde C | \Psi_{\rm ann} \rangle
  = \int_0^\infty dP\, K(P,\tilde P)\,\chi_P(\tau)
\end{equation*}
shows that the bulk change-of-basis kernel $K(P,\tilde P)$ between holonomy eigenstates coincides precisely with the CFT modular kernel $S(P,\tilde P)$ above. In this sense, the bulk Fourier transform on the torus moduli space {\em is} the modular $S$-kernel acting on the family of Virasoro characters. Once this identification is made, our analytic continuation arguments on the bulk and CFT sides are tautologically the same.

\subsection{T-shift $\tau \to \tau + 1$: Holonomy Energy vs.\ Graviton Energy}
\label{sec:T-shift}

So far we have focused on the modular $S$-transform, whose bulk avatar was
the Fourier transform between holonomy polarizations on the thickened torus. In this
subsection we briefly discuss the other generator of the modular group, the Dehn twist
$T:\tau\to\tau+1$, and its interpretation in the Chern-Simons holonomy picture.

From the CFT side, the $T$-move acts on a highest-weight character as
\begin{equation}
  \chi_h(\tau+1)
  \;=\;
  e^{2\pi i(h - c/24)}\,\chi_h(\tau)\,.
  \label{eq:T-on-chi-h}
\end{equation}
Using the by now familiar definition of the normal orbit in terms of $P$ we can rewrite the exponent as:
\begin{equation}
  h - \frac{c}{24}
  \;=\;
  P^2 - \frac{1}{24}\,.
  \label{eq:hminuscover24-P}
\end{equation}
Therefore the $T$-phase in the $P$-basis is
\begin{equation}
  \chi_P(\tau+1)
  \;=\;
  e^{2\pi i(P^2 - 1/24)}\,\chi_P(\tau)\,,
  \label{eq:T-on-chi-P}
\end{equation}
and the corresponding kernel is diagonal:
\begin{equation}
  K_T(P',P)
  \;=\;
  e^{2\pi i(P^2 - 1/24)}\,\delta(P'-P)\,.
  \label{eq:T-kernel-P}
\end{equation}

This result is natural from the bulk perspective as well: the time translation by a unit is controlled by the energy operator, which is just $L_0-c/24$ in the CFT and translates to $P^2-1/24$ via its relation to quantized normal orbits. It is also worth noting that the form naturally shows a split between the holonomy energy contribution ($P^2$) and the universal descendant energy contribution (-1/24). This is more than an analogy: the $\tau \rightarrow \tau+1$ leads directly to $\frac{1}{\eta(\tau)}  \rightarrow e^{-2\pi i/24}\frac{1}{\eta(\tau)}$. This split between primaries and descendants is a very general feature of our perspective: it leads not just to the split in energies, but also to how the states themselves are ``localized" in the bulk. We turn to this in the next section.

Together with the $S$-kernel $S(P,P')$ this diagonal $T$-kernel generates the full
$PSL(2,\mathbb{Z})$ action on the continuum of normal-orbit characters. We will review 
the combined modular action and its role in the spectrum of the Maloney--Witten ``pure gravity path integral" in
Appendix  \ref{App:Maloney-Witten}.

\section{Cardy Split: Horizon-vs-Boundary as Primary-vs-Descendant}
\label{sec:CardySplit}

In this section we sharpen the interpretation of the two-boundary Chern-Simons path integral as furnishing a split of the total central charge
\begin{equation}
c = (c-1)_{\text{horizon}} + 1_{\text{boundary}}
\label{eq:CardySplitStatement}
\end{equation}
between holonomy edge modes localized on the stretched horizon and boundary gravitons at the asymptotic boundary.  There are two related ingredients:
\begin{itemize}
  \item An \emph{exact} factorization of the vacuum modular $S$-transform into a ``horizon'' piece and a universal ``boundary'' piece, valid for all $\tau$ in the upper half-plane.
  \item The \emph{Cardy limit} $t \to 0^+$ with $\tau = i t$, in which the leading $\sim 1/t$ growth of $\log \chi_{\mathrm{vac}}(-1/\tau)$ can be cleanly attributed to $(c-1)$ and $1$, respectively.
\end{itemize}

\subsection{Horizon-Boundary ``Factorization"}

We can write the BTZ character as 
\begin{equation}
\chi_{\mathrm{vac}}\!\left(-\frac{1}{\tau}\right)
= \left[ \int_0^\infty dP\; S_{0P}(c)\,q^{P^2} \right] \times \frac{1}{\eta(\tau)}\,.
\label{eq:exact-factorization}
\end{equation}
It is convenient to define
\begin{equation}
Z_{\text{hor}}(\tau) \equiv \int_0^\infty dP\; S_{0P}(c)\,q^{P^2}\,,
\qquad
Z_{\text{bdy}}(\tau) \equiv \frac{1}{\eta(\tau)}\,,
\label{eq:Zhor-Zbdy-def}
\end{equation}
so that
\begin{equation}
\chi_{\mathrm{vac}}\!\left(-\frac{1}{\tau}\right)
= Z_{\text{hor}}(\tau)\,Z_{\text{bdy}}(\tau)\,.
\label{eq:vacuum-split-ZhorZbdy}
\end{equation}

The factor $Z_{\text{bdy}}(\tau) = 1/\eta(\tau)$ is the partition function of a \emph{single} chiral boson, i.e.\ of a $c_{\text{bdy}} = 1$ CFT, and we have already identified it with the contribution of boundary gravitons living at the asymptotic DS boundary. The horizon factor $Z_{hor}(\tau)$ arises from the kernel, and therefore is naturally attributed to the inner boundary. 

It is also true simultaneously, that there are strong CFT arguments (see e.g., \cite{Shouvik}), to think that the total central charge $c$ gets a contribution $c-1$ from primaries and the remaining $+1$ from the descendants. Since we have already accounted for the descendants, it is natural to attribute the primaries to the factor $Z_{\text{hor}}(\tau)$ and the inner Wilson-line boundary. The split also suggests that we should assign
\begin{equation}
c_{\text{hor}} = c - c_{\text{bdy}} = c-1\,
\label{eq:c-horizon}
\end{equation}
to the inner boundary. We will see this as a precise Cardy split, below. 

Before we proceed, let us emphasize that we are {\em not} suggesting that the inner boundary $Z_{\rm hor}$ is truly a decoupled theory. Clearly a fully consistent holographic CFT requires both pieces. What the factorization of the partition function provides, is intuition about how the degrees of freedom are localized in the bulk. Because the holonomies (charges) are visible at any radial cut, they are usually viewed as living at the boundary -- what we are emphasizing is that {\em precisely} because they are visible at any cut, there are other interpretations that are available and maybe more useful in some contexts. Note in particular that the descendants do {\em not} allow this interpretation: they necessarily live at infinity.

\subsection{Cardy Limit and the $P$-saddle}

We follow the notations of \cite{Pal} in this discussion\footnote{We have used the convention $\tau=\tau_1-i\tau_2$ in most of this paper. This arises from our definition of the complex boundary coordinate $z$ (see Appendix \ref{AppConventions}). Flipping the sign would require a change of the chirality condition of the asymptotic CS gauge field from $a_\phi=a_t$, which we preferred not to do.  Note that with either sign choice, $\tau_2 \ge 0$. In this Appendix however, we use $\tau=\tau_1+i \tau_2$ to match with various standard CFT conventions, because the chirality of the bulk gauge field never directly enters. The discussions here are self-contained.}. The torus partition function for a unitary, modular invariant $2$D CFT is: 
\beq
Z(\tau,\overline \tau) = {\rm Tr}_{\mathcal H_{\rm CFT}} \left(q^{L_0 - \frac{c}{24}} \ {\overline q}^{\overline L_0 - \frac{c}{24}} \right), \quad q = e^{2 \pi i \tau} 
\eeq
where $L_0$ and $\overline L_0$ are the standard Virasoro algebra generators and $\mathcal H_{\rm CFT}$ is the CFT Hilbert space. In this subsection we will treat the CFT partition function $Z(\tau,\bar \tau)$ as a function of independent complex variables $\tau$ and $\bar \tau$ (as in modular bootstrap).  But for the thermodynamic Cardy limit we specialize to the slice of $\tau$-$\bar \tau$ space where the modular parameter is related to the real left and right moving (independent) inverse temperatures \cite{Pal} via
\beq
\begin{aligned}
& \tau = \frac{i \beta_L}{2 \pi}, \quad \overline \tau = -\frac{i \beta_R}{2 \pi} 
\\[6pt]
{\rm with}, \quad &\beta_L = \beta \ (1+\Omega), \quad \beta_R = \beta \ (1-\Omega).
\end{aligned}
\eeq
Here $\beta$ and $\Omega$ are the inverse temperature and the Lorentzian angular velocity of the rotating BTZ metric. Note that on this slice, $\beta_L$ and $\beta_R$ are manifestly real.

The high-temperature/Cardy limit corresponds to $\beta_{L,R} \to0$. For our purposes, it is convenient to work in the modular-transformed variable 
\begin{equation}
  \widetilde {\tau} \equiv -\frac{1}{\tau} = \frac{2\pi i}{\beta_L}
  \label{Cardyydef}
\end{equation}
In this regime, the vacuum character of CFT obeys the usual Cardy growth: 
\beq
  \log \chi_{\mathrm{vac}}\!\left(-\frac{1}{\tau}\right) \approx \frac{\pi^2 c }{6 \beta_L} + \mathcal O\left(e^{-\frac{4 \pi^2}{\beta_L}}\right)
  \label{CardyvacCFT}
\eeq

Our goal is to show that the total Cardy coefficient, naturally decomposes into a $(c-1)$ and ``$1$" piece in accordance with the horizon-boundary split we mentioned at the beginning of this section, from the $Z_{\mathrm{hor}}$ and $Z_{\mathrm{bdy}}$ respectively. 

\paragraph{Boundary Contribution.}

The boundary partition function in \eqref{eq:exact-factorization}, has the asymptotics: 
\beq \label{Zbdryy}
\log Z_{\rm bdy}(\tau) = -\log \eta(\tau) \approx \frac{\pi^2}{6 \beta_L} + \frac12 \log \left(\frac{\beta_L}{2\pi}\right) + \mathcal O\left(e^{-\frac{4 \pi^2}{\beta_L}}\right).
\eeq 
This uses the identity
\begin{equation}
  \eta\left(-\frac{1}{\tau}\right) = \sqrt{-i\tau}\,\eta(\tau)
  \label{eta}
\end{equation}
and the small $\beta_L$ expansion of $\eta\left(-1/\tau\right)$ in the second equality.  

Hence, the leading Cardy contribution from boundary gravitons may be interpreted as the contribution of a single chiral boson with $c_{\mathrm{bdy}} = 1$. Unsurprising, because $1/\eta(\tau)$ is precisely the partition function of a chiral boson.

\paragraph{Horizon Contribution and the $P$-saddle.}

The horizon partition function in \eqref{eq:exact-factorization}, is
\begin{equation} \label{Zhor}
  Z_{\mathrm{hor}}(\tau) = \int_0^\infty dP\; S_{0P}(c)\,e^{2\pi i\tau P^2} = \int_0^\infty dP \; e^{\Phi(P)}
\end{equation}
where $\Phi(P) \equiv \log S_{0P}(c) - \beta_L P^2 $ in the second equality.  The vacuum row of the modular kernel, which we repeat here for convenience, is
\beq 
S_{0P} = 4\sqrt{2} \sinh\left(2\pi P b\right) \sinh\left(2\pi P/b\right), \quad \, \hspace{0.1cm} {\rm and} \, \hspace{0.1cm} c=1+6 \left(b + \frac1b\right)^2.
\eeq

In the semiclassical regime relevant for the Cardy limit, the integral is dominated by large $P$. This is because for any fixed $\beta_L > 0$, the competition between the Gaussian suppression and the monotonic growth of $S_{0P}$ produces a unique peak at finite $P$. As $\beta_L\to0$, this peak shifts to parametrically large $P$, implying that the integral receives dominant contributions from this region. Moreover, increments in $b$ (which translates to increments in $c$) makes the situation even better, i.e. shifts the peak to even larger $P$.  In this regime, i.e. $P \rightarrow \infty$, 
\begin{equation}
  \log S_{0P}(c) \approx 2\pi Q P, \qquad
  Q^2 = \frac{c-1}{6}.
  \label{eq:S0P_asymptotic}
\end{equation} 
The saddle point is determined by
\begin{equation}
  \frac{d}{dP} \, \Phi(P) = 0 \quad\Longrightarrow\quad
  P_* = \frac{\pi Q}{\beta_L}.
  \label{Psaddle}
\end{equation}
and hence the horizon partition function, up to the first sub-leading correction, is\footnote{Note that the $Z_{\rm hor}(\tau)$ integral \eqref{Zhor} is exactly doable: either directly using Gaussian integrals or by using the fact that it can be written as $\chi_{\rm vac}(-1/\tau)\eta(\tau)=\chi_{\rm vac}(-1/\tau)\eta(-1/\tau)/\sqrt{-i\tau}$, which only contains explicitly known functions. In the small-$\beta_L$ limit, this reproduces the same expressions as our saddle result. We stick with the saddle point evaluation because the location of the saddle is a piece of physics worth knowing. Note also that our main point here is not the split between primaries and descendants, but to {\em attribute} the primaries to the horizon by relating them to the kernel, which we know from our bulk calculations has its origins at the ``stretched horizon".}: 
\begin{equation}
  \log Z_{\rm hor} (\tau) \approx \Phi\left(P_*\right) -\frac12 \log \left(-\Phi''\left(P_*\right)\right) = \frac{\pi^2 (c-1)}{6\beta_L} - \frac12 \log \left(2 \beta_L\right).
  \label{Zhory}
\end{equation}

\paragraph{Combining Horizon and Boundary.}

Clearly, combining the leading boundary and horizon pieces in~\eqref{Zbdryy} and~\eqref{Zhory}, we reproduce the Cardy behavior~\eqref{CardyvacCFT} in the regime where the $S$-transformed modulus $\tilde\tau=-1/\tau$ has large imaginary part. In other words, in the Cardy regime $\mathrm{Im}\!\left(-\frac{1}{\tau}\right)\to\infty$,
the two-boundary Chern-Simons path integral with an excised inner Wilson-line boundary explains how the total central charge $c$ of the BTZ black hole naturally decomposes into $(c-1)$ units carried by the stretched-horizon and $1$ unit carried by the universal boundary gravitons. Our derivation is valid for any BTZ modulus $\tau$ that lies in this high-temperature regime.

\subsection{Generic BTZ Moduli, Cardy Limit and Holography}

It is worth emphasizing that only the Cardy statement  requires the high temperature limit.  The underlying factorization on the other hand, is an \emph{exact identity}, valid for any $\tau$ in the upper half-plane.  The interpretation of $Z_{\text{bdy}}$ as a $c=1$ theory of boundary gravitons and of $Z_{\text{hor}}$ as a $(c-1)$ ``theory" of primaries living on the stretched horizon therefore holds for any BTZ modulus $\tau$. Changing $\tau$ only changes the Boltzmann weights $q^{P^2}$, not the split of degrees of freedom.

What the high temperature limit does is to single out a regime in which the total partition function is dominated by high-energy states and admits a simple Cardy-type form.  In this regime the saddle-point analysis of the $P$-integral becomes reliable, and the decomposition $c = (c-1) + 1$ is visible directly at the level of the Cardy coefficient.

We also add that while it is eminently natural to attribute the primaries to the horizon, there is also a sense in which it is natural to view them as being associated to the entire geometry (or indeed even the boundary). This is because primaries are holonomies, and therefore they are visible at {\em any} radial cut in the geometry. This is the reason why this picture does not run into tension with the usual holographic expectations in AdS/CFT.

\section{A Smooth Horizon without a Smooth Horizon}
% (Left intentionally blank for now.)

The general idea that a black hole may not have an interior in some suitable sense, has a long history. In this paper, we will treat 't Hooft as the originator of the modern\footnote{``Modern", means that this was after the nature of coordinates and geodesics at the horizon, was widely understood. In the initial era of general relativity, there were various ideas about the horizon (including from Einstein), which we will view as pre-history.} version of this thought \cite{tHooft}, even though ``membrane paradigm"-like pictures can be traced to the early 70's. Kip Thorne \cite{KipThorne} credits an observation by Hanni-Ruffini \cite{Hanni} and Cohen-Wald \cite{CohenWald} as the progenitor of the membrane paradigm: an electric charge  dropped into a black hole will appear to an asymptotic observer to be remaining just outside the horizon even at arbitrarily late times, and therefore the flux lines should begin at a surface right outside the horizon. 

There are two significant modern avatars of this philosophy. One is the firewall paradox of Mathur \cite{Mathur} and AMPS \cite{AMPS1, AMPS2, MarolfPolchinski}. Roughly speaking the firewall arguments note that there is tension between assuming that the horizon remains smooth on very long time scales (depending on the specific paradox, these time scales can be ${\mathcal O}(N)$ or ${\mathcal O}(e^N)$, where $N$ is inversely related to Newton's constant) and unitarity of local bulk effective field theory. We will not say much about the firewall paradox in this paper, but see \cite{Burman1, Burman2, VyshnavAlgebra} for closely related comments. We will simply note that our conclusions are broadly aligned with firewall expectations, because our inner boundary for individual microstates is not smooth.

The second is the fuzzball program, originating with Lunin and Mathur \cite{Lunin} and significantly expanded by Bena, Warner and collaborators \cite{FuzzRefs}. Large classes of classical supergravity solutions have been constructed as candidate microstates for black holes in string theory. They have the same charges as the black hole, but the geometry ``caps off" and therefore does not have a horizon. The currently known fuzzball solutions are not generic enough to capture the leading entropy of the 3-charge D1-D5-P black hole. A problem that is sometimes considered deeper, is that the general fuzzball solutions constructed so far have non-trivial angular profile -- but there are reasons to think that the microstates responsible for the entropy of a spherically symmetric BPS black hole are in fact spherically symmetric \cite{Sen0, SenQEF,Dabholkar, SenChow}\footnote{Note that our construction in this paper is compatible with this stronger restriction, while at the same time allowing the possibility of non-smooth (in the {\em radial} direction) microstates. The analogue of spherical symmetry in our setting is the fact all normal orbits have a $U(1)$ stabilizer.}. A possibly closely related feature is that all these are classical supergravity solutions. What we find note-worthy about fuzzballs however, is that they show that string theory and supergravity have mechanisms that allow the violation of no-hair theorems in classical general relativity.  It  seems to us that the fact that such solutions {\em exist}, should not just be brushed aside as an inconvenient fact.

But even if one grants that the fuzzball program will eventually be able to construct generic  microstates in the bulk by allowing more general (stringy, non-smooth, finite-$N$, ...) classes of microstates, there still remains the question of how a smooth horizon can emerge from them. This is a place where the fact that these solutions are classical (at least in their current avatar), is not merely a philosophical nuisance --  it comes with concrete limitations. Our attitude in this paper has been that a smooth horizon can emerge as a {\em sum} over microstates. When working with supergravity solutions (with or without brane sources) it is not clear how one can do such a sum. We need a {\em Hilbert space} of microstates to be able to sum over them\footnote{There have been papers which geometrically quantize the D1-D5(-without-P) phase space \cite{Rychkov} and precisely \cite{Avinash} match the 2-charge entropy. This uses the classical supergravity symplectic structure.}. This may also give us some hints on what kind of configurations and states one should include, in our sum over states. 

\subsection{Discussion}

The main physics punchline of this paper is an explicit illustration in AdS$_3$/CFT$_2$ that (partition functions of) states with smooth Euclidean horizons\footnote{More generally, classes of states with contractible thermal cycles which include the smoothly contractible case as a special case.} can be understood in terms of sums over states which are manifestly {\em not} smooth. This significantly strengthens the case that a smooth horizon is an emergent phenomenon. It also (more importantly, in our view) gives ``mechanical" intuition on how to think about the smooth horizon as a quantum system. We discuss some of the ideas that emerge from these results.

\begin{itemize}

\item Our calculations were robust, and without any real moving parts. This allowed us to view the BTZ character as a very concrete sum over heavy bulk primaries. This robustness is a bulk spin-off of the fact that modular invariance is a basic constraint in 2 dimensional CFTs. Even though it is most powerful in 2D CFTs, there are analogues of modular invariance in higher dimensions as well. In fact, it is natural to demand that partition functions of Euclidean quantum field theories should be invariant under transformations of background fields (usually the fixed metric) under diffeomorphisms. If one allows these diffeomorphisms to also contain those that are not continuously connected to the identity, we have modular invariance. The trouble is that a clean switching of cycles that is a large diffeomorphism for CFT$_2$ on the boundary torus, is no longer available in higher dimensions because the boundary theory is usually living on $S^{d-1}\times S^1$ where $S^1$ is the thermal circle. But if one instead considers the theory on (say) a $T^{d-1} \times S^1$ geometry, where now the spatial geometry is a $(d-1)$-torus, there are natural candidates for modular invariance. It will be interesting if modular-like properties of (supersymmetric) partition functions can be put to use to make useful statements about (say) the 1/16 BPS states of ${\mathcal N}=4$ SYM. 

\item We did not need a {\em full} understanding of the UV complete spectrum. In other words, what precise primaries are part of the spectrum of the {\em specific} holographic CFT is a question that we did not have to answer. We only needed some basic information on their distribution, that follows from modular invariance. The $S_{0P}$ kernel is essentially a density of states of this type: we could get away with the approximation that our microstate distribution lies ``on an average" \cite{Zhiboedov} in that continuum. To the extent that the modular kernel is a coarse-grained approximation for a CFT spectrum on the cylinder \cite{Zhiboedov}, this suggests that in a genuine holographic CFT, the smoothness of the horizon may also be an approximation.

\item In this paper, we worked with modules of conventional Virasoro. Clearly, there are natural generalizations when we have extra symmetries. Extended chiral algebras of various kinds are a possibility. Higher spin algebras may add interesting new subtleties due to bigger gauge invariance and gauge dependence of horizons and singularities \cite{Ammon, Shubho1, Shubho2}. One case that is straightforward and of immediate physical interest is the case of 2D superconformal algebras which are of interest for the D1-D5 system. We expect that a picture similar to what we presented here will play a role, but with  superconformal characters replacing the Virasoro characters. For the gravity multiplet one can write down a Chern-Simons formulation using a $PSU(1,1|2)$ superalgebra \cite{Justin}, and a natural guess is that a boundary reduction like that done in \cite{CJ1} will lead to the characters of the superconformal algebra (and not simply Virasoro). 

\item The modular S-transform is telling us that a sum over primaries supported on a continuum can reproduce the modular image of the vacuum. This is almost certainly\footnote{The Poincare series is an infinite sum, so this claim may not be watertight.} the underlying reason for the continuous spectrum found by Maloney and Witten \cite{MaloneyWitten} in their partition function for 2D pure gravity. To quickly review: In \cite{MaloneyWitten, MaloneyKeller}, the authors tried to {\em define} the quantum gravity partition function by summing over the modular images of the vacuum character. The naive sum can  lead to a negative density of states, but this can be cured (for example, by adding conical defects to the spectrum \cite{Benjamin}). But one problem remains: the spectrum is continuous. It has sometimes been speculated that this continuum should be understood as due to an ensemble average over CFTs \cite{Maloney-Ensemble}. Our results here are consistent with the suggestion that Einstein gravity is an ensemble average in the following sense: modular invariance by itself seems to only require a continuum sum, and it is this aspect that is captured by Einstein gravity. But it should be emphasized that a specific UV complete holographic CFT on the cylinder is expected to require more data -- it involves specifying a {\em discrete} delta function supported density of states in \eqref{CFTstate}. In that sense, we are not committed to pure Einstein gravity (with or without conical defects) in this paper. Our discussions are compatible with a holographic CFT with a discrete spectrum, as long as such a CFT {\em exists}: the constraint here is that such a CFT spectrum has to be compatible with modular bootstrap and other consistency requirements, as discussed in Section \ref{sec:mod-bootstrap} and Appendix \ref{sec:compact_vs_noncompact}.

\item Even though we have not discussed them in the main text, the other modular images can also be written in terms of $S$(and $T$)-kernels. We have already discussed the $S$ and $T$ transforms: the sum over the full set of modular images is reviewed in Appendix \ref{App:Maloney-Witten}, making connections with known previous work \cite{Benjamin}.

\item  It has recently been argued that pure Einstein gravity can be reformulated in terms of a Virasoro TQFT with a continuous spectrum, rather than directly as an $SL(2,R) \times SL(2,R)$ Chern-Simons theory \cite{CollierEberhardtZhangVirasoroTQFT,CollierEberhardtZhangGravityFromVirasoroTQFT,EberhardtCrossingNotes, HartmanViro}. The TQFT language is naturally adapted to our approach. Our ``Janus'' path integral can be viewed as as an instance of a functorial TQFT construction: a 3-manifold $M$ with boundary $\partial M=\Sigma_{\rm in}^*\sqcup \Sigma_{\rm out}$ defines a linear map $Z(M):\mathcal H_{\Sigma_{\rm in}}\to \mathcal H_{\Sigma_{\rm out}}$, and in particular the cylinder $\Sigma\times I$ implements the identity map in a fixed polarization \cite{AtiyahTQFT}. In our setup $M\simeq T^2\times I$ (with two torus boundaries), and the bulk path integral can be viewed either as a state in $\mathcal H_{T^2}\otimes \mathcal H_{T^2}^*$ or as an operator on $\mathcal H_{T^2}$. TQFT provides a framework in which the appearance of the modular kernel is kinematical. In the Virasoro TQFT proposal, the Hilbert spaces $\mathcal H_\Sigma$ are built from the quantization of Teichm\"uller spaces/spaces of Virasoro blocks, and mapping-class-group actions are represented by unitary integral transforms. The modular transformation exchanging the $A$- and $B$-cycles of $T^2$ is realized by the modular $S$-kernel \cite{CollierEberhardtZhangVirasoroTQFT,CollierEberhardtZhangGravityFromVirasoroTQFT,EberhardtCrossingNotes}. Concretely, while $T^2\times I$ gives the identity operator when both boundaries are expressed in the same polarization, expressing the same cobordism between \emph{conjugate} polarizations yields the $S$-kernel as the overlap kernel. 
This precisely mirrors our Chern-Simons Fourier-transform derivation. From this viewpoint, our genus-one construction can be regarded as the torus avatar of a TQFT modular functor.

\item The philosophy of the two approaches on the other hand is different. Virasoro TQFT spectrum contains the hyperbolic continuum, and aims to define bulk path integrals for various topologies by cutting and sewing via the Moore-Seiberg rules for the modular functor \cite{MooreSeibergCQCFT,MooreSeibergPolyEq}. Together with a canonical left-right pairing, this implements pure Einstein gravity as a bulk TQFT. We have instead adopted the perspective that the role of the continuum is the same as that of the $S$-kernel in any $c > 1$ 2D CFT with a discrete (but infinite) set of primaries: despite the $S$-kernel having support in the continuum, the true spectrum of the given CFT on a cylinder can be discrete. We view Chern-Simons/TQFT on our janus cobordism as the natural language for describing Virasoro characters from the bulk, with the left-right paired spectrum being specified as the extra UV input. We discuss these questions a bit more by making comparisons to compact Chern-Simons and rational CFTs in Appendix \ref{sec:compact_vs_noncompact}.

\item The picture that the holonomy labels are bulk sources (i.e., Wilson lines) even above the BTZ threshold, was a crucial ingredient in this work. This mental picture has its origins in \cite{Datar} where it was noticed that one can construct heavy semi-classical Virasoro blocks via a geodesic Witten diagram  prescription, with a source at the Euclidean horizon. Previous calculations of (semi-classical) Virasoro blocks used geodesic Witten diagrams on conical defects, together with an analytic continuation prescription to reach super-threshold values of parameters \cite{River1,River2}. The fact that there exists a natural geodesic Witten diagram prescription in BTZ \cite{Datar} was a clear gesture that worldline sources may be meaningful even above the BTZ threshold.

\item  Our results raise the possibility that new smooth topologies can be {\em apparent}, and may be {\em simulated} by a sum over states\footnote{Let us be precise about what we mean here: the states that we sum over to get a smooth geometry are defect geometries. In other words, by summing over states with singularities due to sources and no contractible thermal cycle, we can mimic the physics of a geometry with a smoothly contractible thermal cycle.}. This would mean that the argument one often hears, that the existence of Hawking-Page transition is evidence that one should include alternate topologies in the gravity path integral, requires qualifiers. It should really only be understood as a useful tool when one {\em starts from} the strict large-$N$ limit. While a sum over topologies of this kind may be a good mimic of the physics in the large-$N$ limit, the truly fundamental operation is a sum over the {\em actual} spectrum of the CFT. This is again related to the fact that the continuum density of states, and the precise spectrum of microstates are fairly distinct aspects of the physics, in our discussion. This should have ramifications for Euclidean wormholes. We find it quite plausible that wormholes are simply the price one has to pay, when trying to reproduce the universal physics of a dense discretuum of heavy states, when one starts from the strict large-$N$ limit (instead of the actual discrete spectrum). Note that one needs non-perturbative in $1/N$ effects to reach finite $N$, if one insists on starting at $N=\infty$.

\item The calculations in \cite{Jensen,MSY} used  ``nearly" AdS$_2$ gravity to study near-extremal black holes and their near-AdS$_2$ throats. The authors argued that a boundary reparameterization mode can be used to capture the dynamics of this  throat. They wrote down a Schwarzian theory for this mode, from which they were able to derive the entropy. In fact, the exact partition function can be computed using localization \cite{StanfordWitten}. In some ways our calculation can be viewed as a higher dimensional generalization of these results, with its close connection to the Alekseev-Shatashvili action (which is a cousin of the Schwarzian). But the AS action by itself did not produce \cite{CJ1} the partition function of the BTZ black hole directly: it required the use of modular S-transform from the vacuum to reach the BTZ character. This made the nature of microstates, obscure. In this paper, we have understood that there is in fact a microstate-based understanding of  the BTZ partition function via the modular $S$-kernel appearing as an explicit density of primary states. It is also worth noting that our calculation looks at the black hole ``from the outside": the \cite{MSY} calculation is phrased ``from the inside" where the entropy arises from an effective theory of re-parameterizations in the asymptotic region of a (cut-off) AdS$_2$ geometry. In this sense, our perspective is closer to the approach adopted by Carlip \cite{Carlip}, who tried to reproduce Schwarzschild entropy by cutting out a small region around the tip of the cigar along some closed curve, and looking for a Schwarzian for the reparameterization mode. Even though very different superficially, we believe the general idea/hope here is the same as ours. One crucial distinction we will point out is that we are actually summing over {\em primaries} (ie., holonomies), and not over reparameterization modes which are more akin to (boundary) gravitons or fluid modes. 

\item This last point we believe is a crucial distinction between many other approaches to ``microstates from edge modes" and our results. One possible interpretation of the latter idea is to view black hole microstates in terms of diffeomorphisms that become global symmetries on a cut-off surface -- presumably, this surface is to be viewed as a putative stretched horizon. It is important to note that in our approach, all of the local inner boundary excitations were gauged. All that was left was the holonomy degree of freedom -- but this is precisely the charge visible from infinity! In our approach, the natural picture is that a dense sliver of such holonomy/primary states is what is directly responsible for the black hole entropy. This is of course very natural, from the perspective of the high-lying spectrum of a holographic CFT. The (continuum-approximated) density of such states is captured by the modular $S$-kernel.

\item It is noteworthy that our picture of black hole microstates correspond to states with high holonomy in AdS$_3$/CFT$_2$. This is reminiscent of old ideas about the string-black hole transition which argues that black holes should be viewed as highly excited wound balls of ``long strings" \cite{Susskind, Polchinski}. It is also worth pointing out that the winding condensation picture (see e.g., \cite{Kutasov, Itzhaki, Sarthak} for relatively recent discussions) will be an emergent idea from our perspective here, because the (smooth) contractibility of the thermal circle is emergent.  

\item Though we did not emphasize it there, the saddle for the holonomy \eqref{Psaddle} that we found in Section \ref{sec:CardySplit} is precisely the one corresponding to the smooth horizon BTZ black hole (in the large-$c$/large-$P$ limit). This is satisfying. Interestingly, we can also compute the fluctuations in the saddle by treating the modular $S_{0P}$ kernel as a literal distribution of states in the integral. We will discuss the details in a different context \cite{CK}, but the result matches the ``quantum fluctuations in the area" result computed recently using semi-classical gravity methods \cite{Jude, Zurek}. 

\item We have sometimes called our cut in the solid Euclidean torus, which is essentially the excised origin of the spatial disc, a ``stretched horizon". But it should be emphasized that we did not need a detailed Planck scale specification at the horizon to get our results in this paper. This is partly because the questions that we addressed here were those which were accessible to the continuum density of states. The individual states that we are summing over are worldline-sourced bulk primaries: the density of states shows up essentially as an extra input from the S-kernel\footnote{Our use of the phrase ``black hole microstate" in this paper should be understood in this context.}. This is consistent with the fact that to see exponential black hole microstate {\em level spacing} from the {\em Lorentzian} bulk, one needed at least a toy model for the UV-complete Lorentzian microstates. See e.g., \cite{Pradipta, Burman1, Burman2, FuzzRandom}. The density of states was directly related to the normal mode spectrum of the stretched horizon in these papers\footnote{It was pointed out in \cite{Burman1, Burman2} that to reproduce Hartle-Hawking correlators before the Page time while also resolving Maldacena's information paradox \cite{eternal}, generic features of the spectrum is sufficient.}. In the present paper on the other hand, the spectral density arose quite indirectly from the modular S-kernel. While the S-kernel is not an extra ``ad-hoc" input, and is indeed controlled by deep consistency conditions, it certainly lacks the immediacy of the normal modes obtained from the stretched horizon. To reproduce finite-$N$ effects in Lorentzian signature, we will need the discrete distribution of primaries in the CFT (or a Planckian toy model for it). 

\item The specification of a bulk state given the CFT scaling dimension, and the specification of the theory via a discrete spectrum of such scaling dimensions, seem to be {\em almost} decoupled problems in our presentation. Of course, this apparent disconnect is superficial. Modular bootstrap (for example) constrains the choice of discrete spectra. Nonetheless, it may be useful to think of 3D quantum gravity as a theory of defects, with 3D Einstein gravity (aka Chern-Simons theory) serving to provide a bulk description for the flat-connections/metric sourced by the defects. A crucial ingredient is that $c > 1$, and that the continuum of hyperbolic holonomies is part of the classical moduli space. Liouville theory serves as the natural model for the $c >1$ continuum, but by itself it lacks the discreteness of the spectrum and a normalizable vacuum \cite{Toumbas}. To summarize: Einstein gravity (and Chern-Simons theory) know about the universal continuum distribution of heavy microstates. We also have relatively good control over each heavy primary from the bulk via Wilson line sources. But assembling a collection of the latter into a single fully consistent discrete spectrum (that approximates the continuum for coarse-grained questions), is a key piece of our ignorance about AdS$_3$ holography.

\item Finally: we have called the excised Wilson line at the core of the geometry, a ``boundary" (because of obvious reasons) or a ``stretched horizon" (because when the re-inserted Wilson line has hyperbolic holonomy, this is a super-threshold microstate). What it really is, is a cut in the {\em bulk radial} path integral. It should be clear from everything we have discussed that it is only after re-inserting a suitable Wilson line that this entity has an interpretation as a wave functional or a partition function.

\end{itemize} 

It is noteworthy that there is a natural split between light states (boundary gravitons) and heavy primaries (horizon states) in our approach. But there is also a natural split between sub-threshold (conical defect) and super-threshold (BTZ-like defect) states. In higher dimensions, we have local bulk fluctuations, which share some features of both. We will discuss the question of higher dimensions as well as the related question of coupling matter to Chern-Simons gravity, through the lens of large-$N$ factorization, in the remainder of this section.

\subsection{Bulk Matter in Large-$N$ Theories}\label{WilsonSpool}

One important question that we have not addressed directly in this paper is that of the coupling of matter corresponding to a bulk local field theory, to AdS$_3$ gravity. There are a few different aspects to this question, so let us approach this in the following way.

We have exploited the fact that primaries are labeled by holonomies, to describe them through Wilson lines at the origin of the spatial disc. Since Chern-Simons theory is topological, this was sufficient to describe the CFT partition function using the projector on to the CFT bra \eqref{CFTstate} at the inner boundary. So in our language, bulk field theory matter with a light scaling dimension is just another primary that is part of the spectrum, and goes into the definition of the CFT bra \eqref{CFTstate}. A correlated fact is that the partition functions of holographic CFTs with bulk field theory matter can nonetheless {\em still} be written as sums of left-right paired Virasoro characters. 

While true, the above discussion does not exploit the local bulk field theory structure to constrain the features of the full CFT partition function. One such crucial input is that a bulk local field theory corresponds to more than one primary. In a genuine holographic CFT, towers of primaries are often {\em not} independent. In large-$N$ theories, multi-trace primaries can be constructed as products of single traces. In the conformal bootstrap language (specialized to the AdS$_3$/CFT$_2$ setting), this implies that there are towers of primaries which correspond to bound states of the light primary: see Figure 2 of \cite{FKW} and the discussion nearby, for some bulk intuition on these bound states\footnote{The idea is that if we have (say) a light primary or a conical defect in the spectrum, then bound states of such states should also be part of the spectrum. Figure 2 in \cite{FKW} is good for semi-classical intuition, but it should not be taken too literally as (say) two conical defects literally spinning around each other. The analogy here is with the electron ``spinning" around the nucleus of an atom: it is a good classical picture, but we are considering an eigenstate with no non-trivial time evolution. The bound state primaries are eigenstates of the CFT.}. These ``multi-trace" bound states are to be viewed as multi-particle states built from the seed primary and in effect generate a Fock space (approximately, up to large-$N$/large-$c$ suppressed effects). This means that a bulk EFT type description is a placeholder for towers of primaries. 

It is tempting to speculate that the dynamics of such towers of primaries, formulated in our bulk language via Wilson line insertions, can equivalently be described via a bulk quantum field theory (EFT) of the seed field. A full discussion of Lorentzian time evolution in such a picture is likely to require new ideas, but there is some evidence for it in Euclidean signature. It is known that one loop torus partition functions of bulk fields \cite{Giombi} can be reproduced using the so-called Wilson spool \cite{Spool1, Spool2, Spool3} in the Chern-Simons language. This is clearly a direction that needs more investigation.

Let us conclude this subsection by noting two distinct ways in which coupling bulk field theory to pure AdS$_3$ gravity is of conceptual interest. Firstly, such an endeavor may offer insight on higher dimensional holography: in both cases, one is introducing local propagating degrees of freedom. We will have more to say about this in the next section. Secondly, it is also of interest in connecting with explicit holographic CFTs, like the (4,4) SCFT of the D1-D5 system. We mentioned in one of the bullet points of the previous subsection that the gravity multiplet in this case allows a Chern-Simons formulation. But such a supergravity by itself is of not much interest, because the D1-D5 system contains matter multiplets as well. But since we expect to be able to write down the SCFT partition function as a sum over characters of the superconformal algebra, it seems possible that even with matter sector included, the full partition functions (and presumably suitable indices) can be written as projections on to suitable states at the inner boundary. 

\subsection{Higher Dimensional Holography with an Inner Boundary}

In any dimension and for any metric, one can go to the gauge
\bea
ds^2=d\rho^2 + g_{ab}(\rho,x)dx^adx^b,
\eea
by fixing $g_{\rho\rho}=1$ and $g_{\rho a}=0$ using the available diffeomorphisms \cite{Budhaditya}. This is  essentially the Fefferman-Graham gauge; we emphasize that this form does not require being on-shell (i.e., it need not be a solution of Einstein's equations). A natural holographic screen is simply the surface $\rho=\rho_0$ for some $\rho_0$. In AdS we take $\rho_0 \rightarrow \infty$ as the natural screen, which is a canonical thing to do, in diffeomorphism invariant theories. In AdS$_3$ this structure becomes particularly nice, because it is straightforward to write down a closed form solution for Einstein's equations in this gauge. So Fefferman-Graham gauge can be used to explicitly integrate Einstein's equations, and we get the Banados solution. The Chern-Simons translation of this was our Drinfel'd-Sokolov gauge field.  For all the Banados solutions except the vacuum (ie., conical defects, super-threshold defects and BTZ), the location $\rho=0$ is special\footnote{For BTZ, note that $\rho=0$ is the horizon, not the singularity.}. This paper has its origins in taking this as a hint about the UV nature of Euclidean microstates.  In this section, we outline some general questions that we believe will have to be answered, for a picture similar to what we have suggested in 2+1 dimensions to be useful for  quantum gravity in higher dimensions. 

It has been known since DeWitt's classic trilogy \cite{Bryce} that the natural observables of quantum gravity are asymptotic. These are believed to be S-matrices in flat space and CFT correlators in AdS. This raises the question: why then is it that we were able to define states inside the bulk in a way that lead to reasonable answers?

%\footnote{It is worth noting here that even to define these observables, we may need a carefully defined notion of ``light" and ``heavy". A heavy operator at the asymptotic region may make sense (``black hole scattering"), but it clearly has to confront  multiple conceptual and technical difficulties. In AdS, stable black hole microstates exist, but since they are heavy and backreact, it will be difficult to make sense of CFT correlators involving multiple black hole states, as asymptotically AdS bulk correlators. (But let us note that for half-BPS states, the LLM geometries suggest a natural way to modify asymptotics for heavy states.)  In flat space, if our usual ideas about Hawking radiation are correct, it is not clear what heavy asymptotic states might even mean, only BPS protected heavy states would be truly stable.}

The reason is simply that primaries are described by holonomies, and they are visible at {\em any} radial location in the geometry. Because the radial direction is in effect a gauge redundancy in holography, we can fix that redundancy in more than one way\footnote{Because of the presence of Virasoro descendants, this statement requires some explanation in 2+1 dimension. We discuss that below. In this sense, 2+1 d is more complicated than higher dimensions.}. We viewed boundary conditions as mostly specifying descendant data at infinity in radial quantization, with the primary being specified at the inner radial cut. This was possible because the descendant structure is essentially universal across primaries, and the primaries are associated to charges conserved across radii. This also means that there is a distinction between holonomy and energy (or scaling dimension) in 2+1 dimensions. While boundary gravitons can contribute to the energy\footnote{It is quite a remarkable fact that all except the global generators of the Virasoro algebra do {\em not} commute with the Hamiltonian, and yet they are a powerful and useful ``symmetry". }, they do not contribute to the holonomy. Holonomy is naturally visible at any radial cut of the geometry, while boundary gravitons are naturally viewed as living at infinity. 

The idea that charges can be visible at any radial cut is very much present in higher dimensions as well. This is a natural property of canonical charges in general relativity. But what is missing, is that there does not seem to be a preferred class of states which are the analogues of boundary gravitons. A related fact is that (bulk) graviton excitations can cause collapse in higher dimensions, but boundary gravitons by themselves can never cause collapse in AdS$_3$. This sharp distinction between heavy microstates and light excitations is a technical feature of AdS$_3$ gravity, which is not present in higher dimensions. We will now present some preliminary arguments however, that this may not be a particularly crucial difference between 2+1 vs higher dimensional AdS/CFT.  

The {\em defining} difference between light and heavy operators is a large-$N$ phenomenon in higher dimensions. however, in 2+1 dimensions, the distinction between boundary gravitons and holonomies can be understood as the distinction between descendants and primaries for {\em any} value of $c$. In that sense, the more useful contrast in 2+1 dimensions is not between primaries and descendants, but between super-threshold primaries and conical defects. This suggests that  conical defects (and not boundary gravitons) may be a better analogue for higher dimensional ``gravitons". Note that conical defects can indeed cause collapse \cite{Sonner}.

Let us also observe that in explicit examples of holographic 2D CFTs, the central charge is expected to be large but finite -- it controls the gap in the spectrum. The (large) value of $c$ is also relevant for controlling the distribution and the dense (but discrete!) spacing of the high-lying spectrum. So it is natural that the largeness of $c$ is actually playing a role in explicit holographic 2D CFTs even though the modular $S$-kernel applies at generic $c$. This is again suggestive that the distinction between conical defects and above-threshold states is the more useful comparison for higher dimensions, than descendants vs primaries. 

We have been able to make some progress in adapting this perspective to higher dimensions and view classes of normalizable modes in Euclidean AdS as being sourced at the core: with some instructive lessons. But we conclude the present paper here without developing it further. It is hard to escape the feeling that AdS$_3$ gravity is ``merely" an extreme version of fundamentally the same holographic principle that is realized in higher dimensions. We hope to come back to these questions in the not-so-distant future.

\section*{Acknowledgments}
We thank Vaibhav Burman for bringing Carlip's Schwarzian paper \cite{Carlip} to our attention and Vishal Gayari for  collaborations. We thank Pallab Basu, Darsh Bhatt, Dmitri Bykov, ChatGPT, Jarah Evslin, Jeremy van der Heijden, Sam van Leuven, Bindusar Sahoo and Konstantinos Zoubos for discussions. We are grateful to Kristan Jensen for a clarification on the status of central charges across orbits and the difficulty with fitting them together in a single CFT. CK thanks the organizers of ICBS-2025 at Beijing for a stimulating conference where this project had its origins. CK also thanks Karapet Mkrtchyan for hospitality at the JME-QFT25 school at Yerevan, where lectures on AdS$_3$/CFT$_2$ from (loosely) this perspective were presented. Thanks are also due to the organizers and participants of Fields \& Strings 2025 (Steklov Mathematical Institute) and the 15th Joburg Workshop on String Theory (Wits Rural Campus) for questions and discussions on talks based on this material. PSP thanks the organizers of Indian Strings Meet 2025 at NISER and IIT Bhubaneswar, and the string group at IISER Bhopal during the Asian Winter School 2026, for their warm hospitality during the final stages of this work. %CK also thanks Hyuk-Jae Park for numerous conversations way back when, on geometric aspects of Chern-Simons theories and beyond. 

\appendix
\section{Contractible Cycles, Thickened Tori and Dehn Surgery}
\label{app:contractible-thermal}

In this appendix we discuss the statement that if we glue a solid torus with a contractible {\em thermal} cycle into the ``drilled hole" in the thickened torus, then the thermal cycle on the \emph{outer} boundary
of the full geometry is contractible in the resulting three-manifold.
The discussion is elementary and standard in 3D topology, but we spell it out: we found this a useful exercise in thinking about contractibility of cycles on a solid torus without the scaffolding provided by the ambient embedding in 3D Euclidean space. 

\subsection{Solid Tori and Cycles}

Recall that a solid torus can be written as
\begin{equation}
  M_0 \cong D^2 \times S^1 ,
\end{equation}
where $D^2$ is a 2-disc and $S^1$ is a circle. Its boundary is a 2-torus
\begin{equation}
  T_{\text{out}} = \partial M_0 \cong S^1 \times S^1 .
\end{equation}
On $T_{\text{out}}$ we pick a basis of 1-cycles
\begin{equation}
  S,\, T \in H_1(T_{\text{out}},\mathbb Z) \cong \mathbb Z \oplus \mathbb Z ,
\end{equation}
which we will call the \emph{spatial} and \emph{thermal} cycles respectively:
\begin{itemize}
  \item $S$ is the boundary of the disc factor at a fixed point on $S^1$:
        it goes once around $\partial D^2$ and is therefore contractible in $M_0$.
        In the main text this is usually\footnote{We do discuss the thermal cycle contractible case when we project the inner boundary annulus state onto a basis state of the conjugate cycle. When this state carries the exceptional orbit label, we have a smoothly contractible thermal cycle. The latter is the BTZ handlebody.} the ``spatial'' cycle.
  \item $T$ wraps the $S^1$ factor at a fixed point on $\partial D^2$:
        it is not contractible in $M_0$. In the main text this is usually the
        ``thermal'' cycle.
\end{itemize}

From a purely topological viewpoint, the labels ``spatial'' and ``thermal''
are just names for two distinguished primitive cycles on $T_{\text{out}}$,
one of which is contractible in $M_0$ and the other is not. A \emph{primitive}
cycle means a 1-cycle which generates a copy of $\mathbb Z$ in homology, i.e. in the
basis $\{S,T\}$ it is of the form $p S + q T$ with coprime integers $p,q$.

We will occasionally refer to the contractible cycle of a solid torus as
a \emph{meridian}, and some chosen non-contractible cycle as a \emph{longitude}.
These words are standard in 3-manifold topology; for our purposes, the meridian
is the unique (up to sign) primitive cycle that bounds a disc in the interior.

\subsection{Drilling out a Tubular Neighbourhood}

We now perform the operation described in the main text: remove a tubular neighbourhood
of the ``thermal core'' of $M_0$. Concretely, consider the core curve
\begin{equation}
  \gamma_{\text{th}} = \{0\} \times S^1 \subset D^2 \times S^1 ,
\end{equation}
and remove a small neighbourhood $N$ of it. Topologically,
\begin{equation}
  N \cong S^1 \times D^2 ,
\end{equation}
and the complement
\begin{equation}
  M := M_0 \setminus N
\end{equation}
is diffeomorphic to a product
\begin{equation}
  M \cong T^2 \times I ,
\end{equation}
where $I$ is an interval. This is the ``thickened torus'' (an annulus in the
radial direction, times a circle).

The boundary of $M$ has two connected components:
\begin{equation}
  \partial M = T_{\text{out}} \,\sqcup\, T_{\text{in}} ,
\end{equation}
where $T_{\text{out}}$ is the original outer boundary, and $T_{\text{in}}$
is the new ``inner'' boundary of the drilled-out tube (the boundary of $N$).

Because $M \cong T^2 \times I$ is literally a cylinder whose cross-section is a torus,
there is a canonical identification between cycles on the outer and inner boundary:
every loop on $T_{\text{out}}$ can be pushed along the interval direction to a
unique loop on $T_{\text{in}}$. In particular, this gives an isomorphism
\begin{equation}
  H_1(T_{\text{out}},\mathbb Z) \xrightarrow{\;\sim\;} H_1(T_{\text{in}},\mathbb Z) .
\end{equation}
We denote the images of $S,T$ under this map by $S_{\text{in}}$ and $T_{\text{in}}$,
and the original cycles by $S_{\text{out}},T_{\text{out}}$ when we need to distinguish
them. Thus
\begin{equation}
  S_{\text{out}} \sim S_{\text{in}},\qquad T_{\text{out}} \sim T_{\text{in}}
\end{equation}
inside $M$. Here $\sim$ just means ``homotopic through the interior of $M$''.

From the physics perspective, this product structure $T^2 \times I$ is expressing
the fact that the thickened torus is locally just a ``radial'' interval times
the thermal/spatial torus, so that it is natural to view the inner torus
as a copy of the boundary torus.

\subsection{Gluing in a New Solid Torus}

Next we glue in a new abstract solid torus $V$ along the inner boundary $T_{\text{in}}$.
Let
\begin{equation}
  V \cong D^2_\mu \times S^1_\lambda
\end{equation}
be a solid torus with boundary $\partial V \cong T^2$. We choose a standard basis
of cycles on $\partial V$:
\begin{itemize}
  \item $\mu$ is the \emph{meridian}: it is the boundary of the disk $D^2_\mu$
        at fixed point on $S^1_\lambda$, hence contractible in $V$.
  \item $\lambda$ is a chosen \emph{longitude}, wrapping the $S^1_\lambda$ factor.
        It is not contractible in $V$.
\end{itemize}

To glue $V$ into $M$ along $T_{\text{in}}$, we choose a diffeomorphism
\begin{equation}
  \phi: \partial V \to T_{\text{in}} .
\end{equation}
This induces a map on homology
\begin{equation}
  \phi_*: H_1(\partial V,\mathbb Z) \to H_1(T_{\text{in}},\mathbb Z)
\end{equation}
and in particular sends the meridian $\mu$ to some primitive cycle on $T_{\text{in}}$:
\begin{equation}
  \phi_*(\mu) = p\, S_{\text{in}} + q\, T_{\text{in}},
  \qquad p,q \in \mathbb Z,\ \gcd(p,q)=1 .
  \label{eq:meridian-slope}
\end{equation}
The pair $(p,q)$ is called the \emph{slope} of the filling: it specifies which linear
combination of the ambient cycles is declared to be contractible once the solid torus
is glued in.

The resulting three-manifold
\begin{equation}
  M' := M \cup_\phi V
\end{equation}
has a single boundary component $T_{\text{out}}$. Topologically $M'$ is again a
solid torus (one can think of $M$ as a genus-2 handlebody and $V$ as capping off
one of the boundary components), but the key point is that the choice of slope
\eqref{eq:meridian-slope} determines \emph{which} cycle on the remaining boundary
$T_{\text{out}}$ is contractible in $M'$.

Two cases are of special interest:

\begin{itemize}
  \item If we choose $\phi_*(\mu)=S_{\text{in}}$ (i.e.\ $(p,q)=(1,0)$), then the spatial
        cycle is contractible in the filling; this is the usual ``thermal AdS''
        filling where the spatial circle pinches off.
  \item If we choose $\phi_*(\mu)=T_{\text{in}}$ (i.e.\ $(p,q)=(0,1)$), then the thermal
        cycle is contractible in the filling; this is the ``BTZ-like'' filling where
        the Euclidean time circle pinches off.
\end{itemize}

In both cases the bulk is topologically a solid torus, but from the boundary point
of view the two saddles are inequivalent, because they declare different primitive
cycles on $T_{\text{out}}$ to be contractible.

\subsection{ Contractibility of the Thermal Cycle}

We now make it plausible that in the second case, where $\phi_*(\mu)=T_{\text{in}}$,
the thermal cycle $T_{\text{out}}$ on the \emph{outer} boundary becomes contractible
in the full manifold $M'$.

By construction, in the solid torus $V$ the meridian $\mu$ bounds a disk, say
$D^2_\mu \subset V$. After gluing, the image $\phi(\mu)$ is identified with
$T_{\text{in}}$ on the inner boundary of $M$, so $T_{\text{in}}$ bounds a disk
embedded in $V \subset M'$.

On the other hand, inside $M \cong T^2 \times I$ the cycles $T_{\text{out}}$ and
$T_{\text{in}}$ are homotopic: we can push the loop $T_{\text{out}}$ along the
interval direction until it lies on $T_{\text{in}}$. This homotopy can be visualized
as a cylinder
\begin{equation}
  C \cong T_{\text{out}} \times I \subset M
\end{equation}
whose boundary is
\begin{equation}
  \partial C = T_{\text{out}} - T_{\text{in}} .
\end{equation}

Now consider the union of this cylinder $C$ in $M$ and the disk $D^2_\mu$ in $V$:
\begin{equation}
  \Sigma := C \cup D^2_\mu \subset M' .
\end{equation}
Its boundary is purely $T_{\text{out}}$:
\begin{equation}
  \partial \Sigma = T_{\text{out}} .
\end{equation}
Thus $T_{\text{out}}$ bounds a 2-dimensional surface in $M'$, and in particular is
null-homotopic. This is exactly what it means for $T_{\text{out}}$ to be
a contractible cycle in the full three-manifold.

In words: by choosing the glued-in solid torus so that its meridian is identified
with the thermal cycle on the inner boundary, we have arranged for the thermal
cycle at the outer boundary to be homotopic to the boundary of a disk in the interior.
Equivalently, the thermal circle at infinity shrinks smoothly somewhere in the bulk.

\subsection{Existence of the Gluing Diffeomorphism}

One might worry\footnote{This worry, is a result of thinking about the torus as being implicitly embedded in 3D Euclidean space. The statement is obvious in terms of the identification picture of the torus, as we show below.} that the previous argument implicitly assumes the existence of
a smooth diffeomorphism
\begin{equation}
  \phi : \partial V \to T_{\text{in}}
\end{equation}
with a prescribed action on cycles, e.g.\ such that $\phi_*(\mu) = T_{\text{in}}$.
Here we re-assert that such a diffeomorphism always exists, and why
choosing the ``slope'' $\phi_*(\mu)=p\,S_{\text{in}} + q\,T_{\text{in}}$ is not
an extra assumption but follows from standard facts in 3-manifold topology.

From a 3-manifold topology point of view, the operation we are performing is a
standard \emph{Dehn filling} along the inner boundary torus. A Dehn filling is
defined exactly by:
\begin{itemize}
  \item choosing a primitive slope $p\,S_{\text{in}} + q\,T_{\text{in}}$ on the
        boundary torus, and
  \item gluing in a solid torus so that its meridian is identified with that slope.
\end{itemize}
It is easy to see (as we do below) that for every primitive slope
there exists such a gluing diffeomorphism.

As a concrete illustration, let us consider the torus
$T^2 = \mathbb R^2 / \mathbb Z^2$ with coordinates $(x,y)$ and identifications
$(x,y) \sim (x+1,y) \sim (x,y+1)$. An integer matrix
\begin{equation}
  A = \begin{pmatrix} p & r \\[2pt] q & s \end{pmatrix} \in \mathrm{SL}(2,\mathbb Z)
\end{equation}
acts linearly on $\mathbb R^2$ by
\begin{equation}
  \begin{pmatrix} x \\[2pt] y \end{pmatrix} \mapsto
  A \begin{pmatrix} x \\[2pt] y \end{pmatrix},
\end{equation}
and this action descends to a smooth diffeomorphism of the quotient torus
$T^2$, because $A$ preserves the integer lattice $\mathbb Z^2$ and has
$\det A = 1$. The map is linear, and the determinant condition guarantees existence of inverse: this therefore constructs our smooth diffeomorohism. In homology, the standard basis cycles correspond to the
vectors $(1,0)$ and $(0,1)$, so the first column $(p,q)$ of $A$ is literally
the image of the ``meridian'' basis vector $(1,0)$. Choosing the slope
$(p,q)$ for the meridian image is therefore equivalent to choosing such a
matrix $A$, and hence choosing a gluing diffeomorphism $\phi$.

\section{Signs and Conventions}\label{AppConventions}

We will largely follow the conventions of \cite{Ammon} in this paper, see their Section 2 (except for one difference which we will emphasize below). This means that we work with the $SL(2)$ generators
\beq
\begin{aligned}
L_0 &= \frac12
\begin{pmatrix}
1 & 0\\
0 & -1
\end{pmatrix},
&
L_{+1} &= 
\begin{pmatrix}
0 & 0\\
-1 & 0
\end{pmatrix},
&
L_{-1} &= 
\begin{pmatrix}
0 & 1\\
0 & 0
\end{pmatrix}
\end{aligned}
\eeq
which obey 
\beq
\left[L_i , L_j\right] = (i - j)L_{i+j}.
\eeq
DS gauge is defined\footnote{The $\rho$ we use below is the one used in \cite{Ammon}. See the concluding paragraph of this Appendix for definitions of various radial coordinates we use in this paper and their relationship to $\rho$.} in holomorphic coordinates $z,\, \bar z$ as \cite{Ammon}:
\beq
\begin{aligned}
   A = \left(e^\rho L_{+1} - \frac{2 \pi \mathcal L}k e^{-\rho} L_{-1}\right)dz + L_0 d\rho\, , \qquad
   \overline A = \left(e^\rho L_{-1} - \frac{2 \pi \overline {\mathcal L}}k e^{-\rho} L_{+1}\right)d \overline {z} - L_0 d\rho. \label{DSBig}
\end{aligned}
\eeq
Writing $A=b^{-1}(d+a)b$ with $b=e^{\rho L_0}$ we find the primitive gauge fields
\begin{equation}
  a(z) = \left( L_{1} - \frac{2\pi}{k}\,L(z)\,L_{-1} \right)\,dz\,,\qquad
  \bar a(\bar z) = \left( L_{-1} - \frac{2\pi}{k}\,\bar L(\bar z)\,L_{1} \right)\,d\bar z\,.\label{holo-a}
\end{equation}
Using the definition of the metric in terms of triads, the Banados solution in Fefferman-Graham form can be reproduced: 
\beq \label{FGBTZ}
ds^2 = d\rho^2 + 8 \pi G_N\left(\mathcal L dz^2 + \overline{\mathcal L} d \overline{z}^2\right) + \left(e^{2\rho} + (8 \pi G_N)^2 \mathcal L \overline{\mathcal L} e^{- 2\rho}\right) dz d \overline{z}.
\eeq

The convenience of the \cite{Ammon} approach is that we can go back and forth between Euclidean and Lorentzian signatures by Wick-rotating the time variable in the definition of $z$ (and $\bar z$, which is the complex conjugate). But there is one difference between our conventions and those of \cite{Ammon}: we will work with $z = \phi - i t_E/\ell$, with $t_E$ being the Euclidean time (we will often set $\ell = 1$). This differs by a sign from the definition in \cite{Ammon}. We have changed the sign because with our choice, and the standard Wick rotation to Lorentzian time $t$ via $t_E=i t$, the chiral asymptotically AdS$_3$ condition simply becomes the standard choice $a_t=a_\phi$ (as follows immediately from \eqref{holo-a}).

Also, with our choice of sign and the constants $\mathcal L$ and $\overline{\mathcal L}$ defined as rescaled versions of the Virasoro zero modes\footnote{For comparison with \cite{Ammon}, in this Appendix we will absorb the Casimir energy shift on the cylinder into the generators. But this is less important, and we will always have the explicit $-c/24$ elsewhere in this paper. We have also called the anti-chiral gauge field by $\overline{A}$.} 
\beq 
\mathcal L = \frac1{2\pi} L_0 \, , \quad \quad \overline{\mathcal L} = \frac1{2\pi} \overline{L}_0,
\eeq
we find the following relations in terms of $r_\pm$, 
\beq
L_0 = \frac{(r_+ + r_-)^2}{16 G_N} = \frac{M + J}2 \, , \quad \quad \overline{L}_0 = \frac{(r_+ - r_-)^2}{16 G_N} = \frac{M - J}2,
\eeq
where $M$ and $J$ are the physical mass and angular momentum of the Black hole, and the Euclidean BTZ metric takes the from \eqref{FGBTZ}:  
\beq
ds^2 = \frac{(r^2 - r^2_+)(r^2 - r^2_-)}{r^2} dt^2_E + \frac{r^2}{(r^2 - r^2_+)(r^2 - r^2_-)}dr^2 + r^2 \left(d\phi - \frac{i r_+ r_-}{r^2}dt_E\right)^2. \label{rotBTZ}
\eeq
These relations differ by some signs from those in \cite{Ammon}, e.g., the $L_0$ and $\overline{L}_0$ are interchanged (which we believe is a typo even within the conventions of \cite{Ammon}) and the black hole is rotating in the opposite direction. We call the above metric the Euclidean BTZ metric: it is complex, because we do not analytically continue $r_-$ when defining it. This is to be contrasted to the {\em real} Euclidean metric, which we define below.

When thinking about the torus with a non-trivial complex structure as the asymptotic boundary, we keep metric real and define a complex modulus $\tau\equiv\tau_1-i \tau_2$ (with $\tau_2>0$). It is convenient to pass to a real Euclidean saddle by analytically continuing the rotation parameter:
\begin{equation}
r_-^E \;\equiv\; -\,i\,r_- \qquad\Longleftrightarrow\qquad r_- = i\,r_-^E,
\label{eq:rminusE_def}
\end{equation}
with $r_-^E\in\mathbb R$, which leads to a manifestly real Euclidean metric:
\begin{equation}
ds^2=\frac{(r^2-r_+^2)(r^2+(r_-^E)^2)}{\ell^2 r^2}\,dt_E^2
+\frac{\ell^2 r^2}{(r^2-r_+^2)(r^2+(r_-^E)^2)}\,dr^2
+r^2\!\left(d\phi+\frac{r_+\,r_-^E}{\ell\,r^2}\,dt_E\right)^{\!2}.
\label{eq:BTZ_E_real}
\end{equation}
To define the boundary complex structure, we use the complex coordinate
\begin{equation}
w \;\equiv\; \phi-\frac{i}{\ell}\,t_E,
\qquad
w\sim w+2\pi \sim w+2\pi\tau,
\qquad
\tau\equiv\tau_1-i\tau_2,
\label{eq:w_tau_def}
\end{equation}
equivalently
\begin{equation}
(\phi,t_E)\sim(\phi+2\pi,t_E)\sim(\phi+2\pi\tau_1,\;t_E+2\pi\ell\,\tau_2).
\label{eq:torus_idents_tau}
\end{equation}
For the smooth Euclidean BTZ filling, regularity at $r=r_+$ is easiest to see in the co-rotating angle
\begin{equation}
\tilde\phi \;\equiv\; \phi+\frac{r_-^E}{\ell\,r_+}\,t_E,
\label{eq:corot_angle}
\end{equation}
so that the shrinking (contractible) cycle is traversed at fixed $\tilde\phi$. This implies the thermal identification
\begin{equation}
(t_E,\phi)\sim\Bigl(t_E+\beta,\;\phi-\beta\,\frac{r_-^E}{\ell\,r_+}\Bigr),
\qquad
\beta=\frac{2\pi \ell^2 r_+}{r_+^2+(r_-^E)^2}.
\label{eq:thermal_ident_beta}
\end{equation}
Comparing \eqref{eq:thermal_ident_beta} with \eqref{eq:torus_idents_tau} we read off
\begin{equation}
\tau_2=\frac{\beta}{2\pi\ell}=\frac{\ell\,r_+}{r_+^2+(r_-^E)^2},
\qquad
\tau_1=-\frac{\beta}{2\pi}\frac{r_-^E}{\ell\,r_+}
=-\frac{\ell\,r_-^E}{r_+^2+(r_-^E)^2},
\label{eq:tau_from_rplus_rminusE}
\end{equation}
and hence
\begin{equation}
\tau=\tau_1-i\tau_2
=-\,\frac{\ell}{r_+^2+(r_-^E)^2}\,\Bigl(r_-^E+i r_+\Bigr).
\label{eq:tau_compact}
\end{equation}
Note that $\tau_2$ is manifestly positive (semi-)definite, but the choice of sign in \eqref{eq:rminusE_def} is a convention degenerate with the choice of orientation of $\phi$. 

We now briefly comment on how our radial coordinates are related to each other and to the radial coordinates $\rho$ and $r$ in \cite{Ammon} (that we have used in this Appendix). The radial coordinate $\rho$ in \eqref{DSBig}, \eqref{FGBTZ} and in \cite{Ammon} is what we have called $r$ in our discussions of DS and Banados gauges in this paper. But in discussions of Chern-Simons quantization we sometimes work in radial gauge, where we have $A_r=0$ instead of $A_r=L_0$ or $a_r=0$. This is unlikely to cause any confusion. (In writing the metrics \eqref{rotBTZ} and \eqref{eq:BTZ_E_real}, there is a standard variable change from $\rho$ in \eqref{FGBTZ} to a new radial coordinate $r$, which should not be confused with our $r$. We never use the $r$-coordinate of the BTZ metric \eqref{rotBTZ}, except for a somewhat standalone comment in and around \eqref{non-rotBTZ}. The $\rho$ coordinate is standard when working with Fefferman-Graham metrics, but in discussions of radial-type gauges for Chern-Simons, $r$ seems standard. Since the paper is mostly in the Chern-Simons language, we have decided to stick with the latter and never use $\rho$.)

\section{Chemical Potential and Residual Freedom in DS Gauge}\label{ResidualDS}

In this Appendix, we will illustrate the kind of calculations that are standard in proving claims about DS gauge made in the text.  

Let us briefly recap our conventions. We work with $SL(2,\mathbb R)$ Chern-Simons in radial gauge $A=b^{-1}(a+d)b$ with $b(r)=e^{r L_0}$ and $[L_m,L_n]=(m-n)L_{m+n}$ for $m,n\in\{-1,0,1\}$. We denote $\partial_\phi$ by a prime and $\partial_t$ by a dot. The angular component is fixed to Drinfel'd-Sokolov (highest-weight) gauge
\begin{equation}
a_\phi \;=\; L_{+1}\;-\;\alpha\,L_{-1}, 
\qquad 
\alpha(t,\phi):=\frac{2\pi}{k}\,\mathcal L(t,\phi).
\label{eq:DS-aph}
\end{equation}
The most general time component that preserves this form while also satisfying the flatness condition $F_{t\phi}=0$ is parametrized by a source (chemical potential) $\mu(t,\phi)$:
\begin{equation}
a_t \;=\; \mu\,L_{+1}\;-\;\mu'\,L_0\;+\;\Big(\tfrac{1}{2}\mu''-\alpha\,\mu\Big)L_{-1}.
\label{eq:DS-at}
\end{equation}
The straightforward way to check these claims is to assume a general ansatz with three independent coefficients for the three generators, and then let the constraints constrain these coefficients. In the above instance, all three coefficients get determined in terms of one independent function $\mu(t,\phi)$. We will repeatedly use this approach below. 

\subsection{DS-preserving Gauge Transformations}

Consider the most general $sl(2)$-valued parameter $\lambda=\varepsilon L_{+1}+\sigma L_0+\tau L_{-1}$ and demand that $\delta a_\phi=\partial_\phi\lambda+[a_\phi,\lambda]$ keeps $a_\phi$ in the DS form: no $L_0$ term and unit $L_{+1}$ coefficient. Using $
[L_0,L_{\pm1}]=\mp L_{\pm1}, [L_{+1},L_{-1}]=2L_0$,
%\label{eq:sl2-alg}
one finds:
\begin{align}
\delta a_\phi
&= (\varepsilon'+\sigma)\,L_{+1} + (\sigma'+2\tau+2\alpha\varepsilon)\,L_0 + (\tau'+\alpha\sigma)\,L_{-1}.
\end{align}
Imposing $\varepsilon'+\sigma=0$ and $\sigma'+2\tau+2\alpha\varepsilon=0$ gives
\begin{equation}
\sigma=-\varepsilon',\qquad
\tau=\tfrac{1}{2}\varepsilon''-\alpha\,\varepsilon.
\label{eq:sigma-tau}
\end{equation}
Thus the unique DS-preserving transformation is
\begin{equation}
\lambda(\varepsilon)
=\varepsilon\,L_{+1}-\varepsilon'\,L_0+\Big(\tfrac{1}{2}\varepsilon''-\alpha\,\varepsilon\Big)L_{-1}\;.
\label{eq:lambda-DS}
\end{equation}
Using \eqref{eq:lambda-DS} back in $\delta a_\phi$, we find 
\begin{equation}
\delta a_\phi
=\Big(\tfrac{1}{2}\varepsilon'''-\alpha'\varepsilon-2\alpha\,\varepsilon'\Big)L_{-1}
= -\,\delta_\varepsilon\alpha\,L_{-1},
\end{equation}
so the induced variation of $\alpha$ (equivalently of $\mathcal L$) is the Virasoro coadjoint action
\begin{equation}
\delta_\varepsilon \alpha \;=\; \varepsilon\,\alpha'+2\alpha\,\varepsilon'-\tfrac{1}{2}\,\varepsilon'''\;,
\ \ {\rm or} \ \ \
\delta_\varepsilon \mathcal L \;=\; \varepsilon\,\mathcal L'+2\mathcal L\,\varepsilon'-\frac{k}{4\pi}\,\varepsilon'''.
\label{eq:virasoro-coadjoint}
\end{equation}

\subsection{Chemical Potential Compatibility}
We now require that \emph{both} $a_\phi$ and $a_t$ retain the DS form \eqref{eq:DS-aph}--\eqref{eq:DS-at} under the gauge transformation generated by $\lambda(\varepsilon)$, while holding the source $\mu$ fixed. Compute
\begin{equation}
\delta_\varepsilon a_t \;=\; \partial_t \lambda + [a_t,\lambda],
\qquad
\lambda=\lambda(\varepsilon)\ \text{of \eqref{eq:lambda-DS}}.
\end{equation}
Projecting on to the $L_{+1}$ component and using $[L_{+1},L_0]=L_{+1}$ one finds
\begin{equation}
\big(\delta_\varepsilon a_t\big)\big|_{L_{+1}}
= \dot\varepsilon\;-\;\mu\,\varepsilon'\;+\;\mu'\,\varepsilon
=:\delta\mu.
\label{eq:delta-mu}
\end{equation}
In DS gauge the $L_{+1}$ coefficient of $a_t$ \emph{is} the source $\mu$, hence \eqref{eq:delta-mu} is precisely its variation. Demanding that the boundary condition keeps $\mu$ fixed, $\delta\mu=0$, yields the sought condition:
\begin{equation}
\dot\varepsilon\;=\;\mu\,\varepsilon'\;-\;\mu'\,\varepsilon\;.
\label{eq:eps-transport}
\end{equation}
Intuitively, \eqref{eq:eps-transport} states that $\varepsilon$ is Lie-dragged along the time-dependent vector field $v=\mu(t,\phi)\,\partial_\phi$ on the boundary circle. Using \eqref{eq:eps-transport} one immediately gets
\begin{equation}
\big(\delta_\varepsilon a_t\big)\big|_{L_0}
= -\partial_\phi(\delta\mu)=0\qquad(\text{when }\delta\mu=0),
\end{equation}
so the $L_0$ structure of $a_t$ is automatically preserved.

For the $L_{-1}$ slot, a straightforward expansion with \eqref{eq:DS-at} and \eqref{eq:lambda-DS} gives
\begin{align}
\big(\delta_\varepsilon a_t\big)\big|_{L_{-1}}
&=\tfrac{1}{2}\dot\varepsilon''-\alpha\,\dot\varepsilon-\dot\alpha\,\varepsilon
-\tfrac{1}{2}\mu'\varepsilon''+\mu'\alpha\,\varepsilon+\tfrac{1}{2}\mu''\varepsilon'
-\alpha\,\mu\,\varepsilon'.
\label{eq:Ltominusraw}
\end{align}
Inserting \eqref{eq:eps-transport} and the source-driven evolution of $\alpha$ (or equivalently of $\mathcal L$) obtained from flatness $F_{t\phi}=0$,
\begin{equation}
\dot\alpha=\mu\,\alpha'+2\alpha\,\mu'-\tfrac{1}{2}\,\mu'''\qquad
\Big(\text{equivalently } \dot{\mathcal L}
=\mu\,\mathcal L'+2\mathcal L\,\mu'-\tfrac{k}{4\pi}\,\mu'''\Big),
\label{eq:alpha-eom}
\end{equation}
and simplifying, one finds
\begin{equation}
\big(\delta_\varepsilon a_t\big)\big|_{L_{-1}}
= -\,\mu\Big(\varepsilon\,\alpha'+2\alpha\,\varepsilon'-\tfrac{1}{2}\varepsilon'''\Big)
= -\,\mu\,\delta_\varepsilon\alpha.
\label{eq:Ltominusfinal}
\end{equation}
Equation \eqref{eq:Ltominusfinal} is exactly what is required for $a_t$ to retain the DS form \eqref{eq:DS-at} with $\mu$ held fixed: only the field $\alpha$ shifts by $\delta_\varepsilon\alpha$, consistent with \eqref{eq:virasoro-coadjoint}.

\section{Cylinder Cobordism $\,\Sigma\times I\,$ as an Identity Map}
\label{app:cylinder_identity}

This appendix reviews the standard TQFT statement that the cylinder cobordism $\Sigma\times I$ acts as the identity operator on the Chern-Simons Hilbert space $\mathcal H_\Sigma$. As often in Chern-Simons theory, the argument can be presented in a succinct way at the expense of being somewhat abstract: but as elsewhere in the paper, we have tried to be fairly explicit and computational in our presentation. Our discussion is largely agnostic about the choice of the Chern-Simons gauge group. 

\subsection{The Interpretation of Boundary Terms}
\label{app:cyl_operator}

We first discuss the role of boundary terms and explain that their role is to prepare suitable wave functionals. We also outline the general idea that the bulk works as a trivial propagator of the gauge invariant data, namely holonomies. 

Let $M$ be a three-manifold whose boundary is the disjoint union of two closed oriented surfaces,
$\partial M = \overline{\Sigma_-}\sqcup \Sigma_+$.
The Chern-Simons path integral on $M$, with boundary conditions specified by states
$|\psi_-\rangle\in \mathcal H_{\Sigma_-}$ and $|\psi_+\rangle\in \mathcal H_{\Sigma_+}$, defines the transition amplitude
\bea\label{eq:Z_as_matrix_element}
\langle \psi_+|\,U(M)\,|\psi_-\rangle
 \;&\equiv&\; \int D A_+ DA_-\psi_+^*[A_+]\;\psi_-[A_-]\int_{A|_{\Sigma_-}=A_-}^{A|_{\Sigma_+}=A_+ }
\!\!\!\mathcal D A\;
e^{\,i S_{\rm D}[A]}\\
\;&\equiv&\; \int D A_+ DA_-\psi_+^*[A_+]\;\psi_-[A_-]\;Z_M[A_+,A_-]\,.
\eea
Here $S_{\rm D}[A]$ is the Chern-Simons action plus boundary terms needed to implement Dirichlet boundary 
conditions. The functionals $\psi_{\pm}[A_{\pm}] = \langle A_{\pm}|\psi_{\pm}\rangle$ are the boundary wave
functionals in the Dirichlet polarization, and in the second line we have defined the corresponding Dirichlet kernel
$Z_M[A_+,A_-]$. The resulting $U(M):\mathcal H_{\Sigma_-}\to\mathcal H_{\Sigma_+}$ is the linear map assigned to
the cobordism $M$. The functional integral with Dirichlet boundary conditions is an intermediate object we have introduced for illustration. In principle we could have introduced boundary terms that prepared the states $\psi_{\pm}$ directly and done away with $A_{\pm}$ entirely. In our presentation above on the other hand, the objects $\psi_{\pm}[A_{\pm}]$ are to be viewed as boundary terms, which when absorbed into the action, turn the Dirichlet boundary condition into our chosen boundary condition.

Depending on the chosen boundary condition, some would-be gauge degrees of freedom at $\Sigma_\pm$ can become
physical ``edge'' degrees of freedom (for example, chiral/WZW or Alekseev-Shatashvili boundary modes). This is what we
mean by ``boundary dynamics''. In radial quantization, this boundary dynamics is entirely encoded in the endpoint wave functional $|\psi_{\pm}\rangle$: it is dynamics intrinsic to $\Sigma_\pm$ implied by the boundary condition, not propagation
of bulk local degrees of freedom along the cobordism.

In the present discussion, $\mathcal H_\Sigma$ is the gauge-invariant Chern-Simons
Hilbert space on  $\Sigma$, i.e.\ the quantization of the reduced phase space obtained after imposing
Gauss law (equivalently, the quantization of the moduli space of flat connections on $\Sigma$). A boundary-theory
description (WZW/AS, etc.) is best understood as a convenient way to \emph{construct and represent} particular vectors in
this same $\mathcal H_\Sigma$: its path integral yields a gauge-invariant wavefunctional (and hence overlaps
$\langle P|\psi\rangle$ in a holonomy basis for  $\mathcal H_\Sigma$), rather than defining kets outside the physical Hilbert space. Even if one
temporarily keeps additional boundary variables (or gauge parameters) in an ``extended'' description, the final state vector that it constructs should live in the gauge invariant $\mathcal H_\Sigma$.  So only flat-connection/holonomy sector label is transmitted through the bulk. This is analogous to using a path integral with extra fields to compute a wavefunction in ordinary QM: the extra fields are not enlarging the physical Hilbert space -- they are a computational representation of a vector in it. (Let us emphasize that this discussion is in radial quantization.)

Since Chern-Simons theory has no local propagating bulk degrees of freedom, the bulk path integral on $\Sigma\times I$
can only identify the gauge-invariant labels between the two ends (e.g.\ the flat-connection/holonomy moduli on $\Sigma$).
Any nontrivial dependence associated with boundary excitations is carried by the choice of states $|\psi_\pm\rangle$
themselves. With this separation in mind, the statement below that the cylinder implements the identity map should be read as a
statement about the \emph{bulk cobordism map} for $\Sigma\times I$ (i.e.\ no operator insertions in the interior of the
cylinder), while allowing that the chosen boundary condition may endow $|\psi_\pm\rangle$ with nontrivial
boundary-theory content.

The claim we wish to justify then, is that for the cylinder cobordism $M=\Sigma\times I$ (with no extra insertions, so that
any boundary degrees of freedom implied by the chosen boundary condition are entirely contained in the endpoint states
$|\psi_\pm\rangle$),
\begin{equation}\label{eq:cyl_claim}
U(\Sigma\times I) \;=\; \mathbf 1_{\mathcal H_\Sigma}\,.
\end{equation} 

\subsection{Hamiltonian Chern-Simons on $\Sigma\times I$}
\label{app:cyl_hamiltonian_form}

We start from the Chern-Simons action
\begin{equation}\label{eq:CS_action_app}
S_{\rm CS}[A] \;=\; \frac{k}{4\pi}\int_{M}\Tr\!\left(A\wedge dA+\frac{2}{3}A\wedge A\wedge A\right),
\qquad M=\Sigma\times I\,.
\end{equation}
Choose a coordinate $r\in I=[0,L]$ along the interval and local coordinates $x^a$ on $\Sigma$ ($a=1,2$).  Decompose
\begin{equation}
A \;=\; A_r\,dr + A_a\,dx^a \;\equiv\; A_r\,dr + A_\Sigma\,,
\qquad
F_\Sigma \;\equiv\; d_\Sigma A_\Sigma + A_\Sigma\wedge A_\Sigma\,.
\end{equation}
A standard computation gives 
\begin{equation}\label{eq:CS_hamiltonian_form_app}
S_{\rm 1}[A]
\;=\;
\frac{k}{4\pi}\int_0^L\!dr\,
\left[
\int_{\Sigma}\Tr\!\big(A_\Sigma\wedge \partial_r A_\Sigma\big)
+2\int_{\Sigma}\Tr\!\big(A_r\,F_\Sigma\big)
\right]
\;+\; S_{\partial M}\,,
\end{equation}
where $S_{\partial M}$ collects the boundary terms implementing the chosen boundary conditions/polarization. It encodes the choice of which wavefunction(al) the path integral prepares, as discussed in Appendix \ref{app:cyl_operator}.

Two immediate consequences follow from \eqref{eq:CS_hamiltonian_form_app}:

\begin{itemize}
\item The component $A_r$ appears \emph{linearly} and plays the role of a Lagrange multiplier.  Integrating over $A_r$ in
the path integral imposes the constraint
\begin{equation}\label{eq:F_constraint_app}
F_\Sigma(r,x)=0 \qquad \text{for all $r\in I$.}
\end{equation}

\item The remaining term $\frac{k}{4\pi}\int dr \int_\Sigma \Tr(A_\Sigma\wedge \partial_r A_\Sigma)$ is of the universal
phase-space form $\int dr\,\theta_i(X)\,\dot X^i$ with \emph{no} ordinary Hamiltonian; all ``evolution'' is pure gauge and
constraint-generated.
\end{itemize}

The constraint \eqref{eq:F_constraint_app} says that, for each $r$ slice, the spatial connection $A_\Sigma(r)$ is flat.
Gauge transformations act as $A_\Sigma\mapsto g^{-1}(d_\Sigma+A_\Sigma)g$, so the gauge-invariant configuration space on
a slice is the moduli space of flat connections,
\begin{equation}
\mathcal M_{\rm flat}(\Sigma,G)
\;\equiv\;
\big\{A_\Sigma\ \big|\ F_\Sigma=0\big\}\big/\mathcal G_\Sigma\,,
\end{equation}
where $\mathcal G_\Sigma$ is the group of gauge transformations on $\Sigma$.

For a closed surface $\Sigma$ there is no gauge-invariant local Hamiltonian in
the bulk.  The ``Hamiltonian'' associated with $r$-translations is a linear combination of constraints.
Indeed, reading off \eqref{eq:CS_hamiltonian_form_app}, the generator is
\begin{equation}\label{eq:H_is_constraint_app}
H[A_r] \;=\; -\frac{k}{2\pi}\int_{\Sigma}\Tr\!\big(A_r\,F_\Sigma\big).
\end{equation}
Note that in the case that is most interesting to us, $\Sigma=T^2$.
On the physical phase space $F_\Sigma=0$, so $H$ vanishes. Equivalently, in the quantum theory physical states obey the
constraints (Gauss law/flatness) and are annihilated by them. Therefore the $r$-evolution operator on the physical Hilbert space is trivial:
\begin{equation}\label{eq:U_is_identity_app}
U(\Sigma\times I) \;=\; \mathcal N_{\rm cyl} \mathbf 1_{\mathcal H_\Sigma}.
\end{equation}
We will argue below that $\mathcal N_{\rm cyl}$ can be set to 1.

\subsection{Cylinder Kernel as a Delta Function}
\label{app:cyl_kernel}

It is often useful to translate \eqref{eq:U_is_identity_app} into a statement about kernels.
Let $\{|X\rangle\}$ be a basis that diagonalizes a complete set of gauge-invariant coordinates on
$\mathcal M_{\rm flat}$ (for $\Sigma=T^2$ these can be taken to be holonomy data).
Because $U \sim 1$, its kernel in this basis is the delta function on the physical configuration space:
\begin{equation}\label{eq:cyl_kernel_delta_app}
\langle X_+|\,U(\Sigma\times I)\,|X_-\rangle
\;=\;
\mathcal N_{\rm cyl}\,\delta_{\mathcal M_{\rm flat}}(X_+, X_-).
\end{equation}
The appearance of the delta function (rather than a delta functional on the redundant field $A_\Sigma$)
reflects the fact that $A_\Sigma$ contains gauge redundancy, while $X$ parametrizes gauge-invariant data.

Equation \eqref{eq:cyl_kernel_delta_app} is the kernel form of \eqref{eq:cyl_claim}.  In words: the cylinder does not mix
superselection sectors of $\mathcal H_\Sigma$, it only identifies the physical labels at its two ends.

\subsection{Normalization $\mathcal N_{\rm cyl}$}
\label{app:cyl_normalization}

Two general properties can be used to argue that the $\mathcal N_{\rm cyl}$-factor can be taken to be unity:

\begin{enumerate}
\item \textbf{Length independence.}
The $L$ in the cylinder $\Sigma\times[0,L]$ is a choice of coordinate chart. Chern-Simons theory integrates a form over a manifold and is independent of chart choice. So the $U$ map we have obtained cannot depend on $L$.  Denoting the operator by $U(L)$, this implies $U(L)=U(L')$ for
all $L,L'$.

\item \textbf{Gluing.}
Gluing two cylinders gives a cylinder:
$\Sigma\times[0,L_1+L_2]\simeq (\Sigma\times[0,L_2])\circ(\Sigma\times[0,L_1])$.
Functoriality of the path integral therefore gives
\begin{equation}\label{eq:gluing_cyl_app}
U(L_1+L_2) \;=\; U(L_2)\,U(L_1).
\end{equation}
Now $U(L)$ is independent of $L$. Call it $U$.  Then \eqref{eq:gluing_cyl_app} says $U=U^2$. In terms of the normalization, this translates to $\mathcal N_{\rm cyl}^2 =\mathcal N_{\rm cyl}$, which has the non-trivial solution $\mathcal N_{\rm cyl}=1$. 
\end{enumerate}
Combining these facts, we arrive at \eqref{eq:cyl_claim}:
\begin{equation}
U(\Sigma\times I) \;=\; \mathbf 1_{\mathcal H_\Sigma}.
\end{equation}

\subsection{Specialization to $\Sigma=T^2$ and the Holonomy Basis}
\label{app:cyl_torus_holonomy}

For $\Sigma=T^2$, the moduli space $\mathcal M_{\rm flat}(T^2,G)$ is described (classically) by commuting holonomies
$(H_C,H_{\widetilde C})$ modulo conjugation, reflecting $\pi_1(T^2)=\mathbb Z\times\mathbb Z$.
Choosing a basis that diagonalizes the conjugacy class of the spatial-cycle holonomy $[H_C]$,
we can label states by the corresponding continuous (and possibly discrete) data, schematically denoted by $P$ and $\mu$.
Then \eqref{eq:cyl_kernel_delta_app} becomes
\begin{equation}\label{eq:cyl_kernel_holonomy_app}
\langle P;C|\,U(T^2\times I)\,|P';C\rangle \;=\; \delta(P-P'),
\qquad
\langle \mu;C|\,U(T^2\times I)\,|\mu';C\rangle \;=\; \delta_{\mu\mu'}.
\end{equation}
This is the gauge-invariant statement that ``bulk propagation through the cylinder is trivial'' in the holonomy
basis: the cylinder identifies holonomy sectors and produces no nontrivial mixing or dynamics.

\section{Inner Symplectic Structure and Gauge Edge Modes}\label{app:symp}

With the canonical bulk action with \emph{no} additional inner-boundary term, (i) the variational principle at $\partial M_-$ is Dirichlet in the configuration variable
$A_\phi$ and (ii) the resulting inner-boundary canonical charges for gauge transformations are
trivial. We derive the boundary-charge variation formula from the bulk canonical symplectic structure of the theory, in this Appendix. Since we eventually mod out by {\em all} periodic gauge transformations at the inner boundary, this discussion is of somewhat ``academic" interest. But it is interesting as an exercise in seeing that unlike the outer boundary, Dirichlet leads to no interesting dynamics at the inner boundary. 

We recall the canonical bulk action (for one chiral copy)
\begin{equation}
S[A]\equiv \frac{k\ell}{4\pi}\int_M d^3x\,\Tr\!\left(A_\phi \dot A_r - A_r \dot A_\phi + 2 A_t F_{r\phi}\right),
\qquad
F_{r\phi}=\partial_r A_\phi-\partial_\phi A_r+[A_r,A_\phi],
\label{eq:S-canonical-5.34}
\end{equation}
and its combination with the outer boundary term
\begin{equation}
S_1 = S[A]-S[ \overline A]+S_{\rm bdry}[A, \overline A],
\label{eq:S1-5.35}
\end{equation}
with $S_{\rm bdry}$ given in \eqref{Sbdry}. Since our focus here is the \emph{inner} boundary, we will
ignore the details of the outer boundary term and the corresponding outer boundary conditions
in what follows.

When $M$ has two boundaries, an outer $\partial M_+$ and an inner $\partial M_-$, the variation
of $S_1$ takes the form
\begin{equation}
\delta S_1
=
\int_M (\text{EOM})
-\frac{k\ell}{2\pi}\int_{\partial M_+} dt\,d\phi\,\Tr\!\left(A_-\,\delta A_\phi+ \overline A_+\,\delta\ \overline A_\phi\right)
-\frac{k\ell}{2\pi}\int_{\partial M_-} dt\,d\phi\,\Tr\!\left(A_t\,\delta A_\phi- \overline A_t\,\delta \overline A_\phi\right).
\label{eq:deltaS1-5.36}
\end{equation}
At the inner boundary we impose Dirichlet boundary conditions
\begin{equation}
\delta A_\phi\big|_{\partial M_-}=\delta \overline A_\phi\big|_{\partial M_-}=0.
\label{eq:Dirichlet-inner-5.37}
\end{equation}
With this choice, the inner-boundary contribution in \eqref{eq:deltaS1-5.36} vanishes without
adding any extra boundary term at $\partial M_-$. Equivalently, the field-space one-form
(‘‘symplectic potential’’) coming from the inner torus,
\begin{equation}
\Theta_- \;=\; -\frac{k\ell}{2\pi}\int_{\partial M_-} dt\,d\phi\,
\Tr\!\left(A_t\,\delta A_\phi- \overline A_t\,\delta \overline A_\phi\right),
\label{eq:Theta-minus-5.38}
\end{equation}
vanishes on the allowed variations. This already signals that the inner boundary does not
support dynamical edge modes of WZW/AS type: there is no nontrivial symplectic pairing for
fluctuations that live purely at $\partial M_-$.

To see how this implies \emph{trivial} inner charges, it is useful to phrase the discussion in
canonical language. Fix a time slice $\Sigma$ (an annulus)
with coordinates $(r,\phi)$ and boundary $\partial\Sigma=\partial\Sigma_+\sqcup\partial\Sigma_-$.
From the first-order kinetic term in \eqref{eq:S-canonical-5.34}, the canonical symplectic
potential on $\Sigma$ is
\[
\Theta_\Sigma(\delta A)
=
\frac{k\ell}{4\pi}\int_{\Sigma} dr\,d\phi\,
\Tr\!\left(A_\phi\,\delta A_r - A_r\,\delta A_\phi\right),
\]
so the associated (pre-)symplectic form is
\[
\Omega_\Sigma(\delta_1 A,\delta_2 A)
=
\delta\Theta_\Sigma(\delta_1 A,\delta_2 A)
=
\frac{k\ell}{2\pi}\int_{\Sigma} dr\,d\phi\,
\Tr\!\left(\delta_1 A_\phi\,\delta_2 A_r-\delta_2 A_\phi\,\delta_1 A_r\right).
\]
This is the standard canonical structure for Chern-Simons in temporal slicing: $(A_r,A_\phi)$
are the canonical pair, while $A_t$ is a Lagrange multiplier imposing the constraint
$F_{r\phi}=0$ on each time slice.

Now let $\lambda(r,\phi)$ be a Lie-algebra-valued gauge parameter. The infinitesimal gauge
transformation is $\delta_\lambda A_\mu = D_\mu\lambda$ with
$D_\mu\lambda=\partial_\mu\lambda+[A_\mu,\lambda]$. On phase space we need only\footnote{Note that $A_t$ is a Lagrange multiplier and therefore has trivial contraction with the pre-sympletic form defined above.}
\[
\delta_\lambda A_r = D_r\lambda,\qquad \delta_\lambda A_\phi = D_\phi\lambda.
\]
In the canonical/covariant phase-space approach \cite{ReggeTeitelboim,LeeWald,IyerWald,BarnichBrandt,BarnichCompere},
the Hamiltonian generator $H[\lambda]$ (when it exists) is defined by
\[
\delta H[\lambda] \;=\; \Omega_\Sigma(\delta A,\delta_\lambda A).
\]
We now compute the right-hand side explicitly and isolate the boundary term that defines the
surface charge.

Substituting $\delta_\lambda A_r=D_r\lambda$ and $\delta_\lambda A_\phi=D_\phi\lambda$ into the
symplectic form gives
\begin{align*}
\Omega_\Sigma(\delta A,\delta_\lambda A)
&=
\frac{k\ell}{2\pi}\int_{\Sigma} dr\,d\phi\,
\Tr\!\left(\delta A_\phi\,D_r\lambda-\delta A_r\,D_\phi\lambda\right).
\end{align*}
The key manipulation is a covariant integration by parts, using cyclicity of the trace. Inserting the identities
\begin{align*}
\Tr(\delta A_\phi\,D_r\lambda)
&=
\partial_r\,\Tr(\lambda\,\delta A_\phi)
-\Tr\!\left(\lambda\,D_r\delta A_\phi\right),
\\
\Tr(\delta A_r\,D_\phi\lambda)
&=
\partial_\phi\,\Tr(\lambda\,\delta A_r)
-\Tr\!\left(\lambda\,D_\phi\delta A_r\right).
\end{align*}
into the expression gives
\begin{align*}
\Omega_\Sigma(\delta A,\delta_\lambda A)
&=
\frac{k\ell}{2\pi}\int_{\Sigma} dr\,d\phi\,
\Big[\partial_r\Tr(\lambda\delta A_\phi)-\partial_\phi\Tr(\lambda\delta A_r)\Big]
-\frac{k\ell}{2\pi}\int_{\Sigma} dr\,d\phi\,
\Tr\!\left(\lambda\,(D_r\delta A_\phi-D_\phi\delta A_r)\right).
\end{align*}
The combination $D_r\delta A_\phi-D_\phi\delta A_r$ is precisely $\delta F_{r\phi}$ (standard
consequence of $F_{r\phi}=\partial_rA_\phi-\partial_\phi A_r+[A_r,A_\phi]$). Therefore
\[
\Omega_\Sigma(\delta A,\delta_\lambda A)
=
\frac{k\ell}{2\pi}\int_{\Sigma} dr\,d\phi\,
\Big[\partial_r\Tr(\lambda\delta A_\phi)-\partial_\phi\Tr(\lambda\delta A_r)\Big]
-\frac{k\ell}{2\pi}\int_{\Sigma} dr\,d\phi\,
\Tr(\lambda\,\delta F_{r\phi}).
\]
The $\partial_\phi$-total-derivative integrates to zero because $\phi$ is periodic on each
boundary circle, leaving only the $r$-boundary contribution:
\[
\Omega_\Sigma(\delta A,\delta_\lambda A)
=
\frac{k\ell}{2\pi}\int_{\partial\Sigma} d\phi\,\Tr(\lambda\,\delta A_\phi)
-\frac{k\ell}{2\pi}\int_{\Sigma} dr\,d\phi\,\Tr(\lambda\,\delta F_{r\phi}).
\]
(Here $\int_{\partial\Sigma}$ is understood with the induced orientation; e.g., if one keeps both boundaries
$\int_{\partial\Sigma}=\int_{\partial\Sigma_+}-\int_{\partial\Sigma_-}$.)

The second term is the variation of the bulk constraint functional
$\frac{k\ell}{2\pi}\int_\Sigma \Tr(\lambda F_{r\phi})$ (we will assume $\lambda$ is field-independent),
so we can rewrite the result as
\[
\Omega_\Sigma(\delta A,\delta_\lambda A)
=
-\delta\!\left(\frac{k\ell}{2\pi}\int_{\Sigma} dr\,d\phi\,\Tr(\lambda\,F_{r\phi})\right)
+\frac{k\ell}{2\pi}\int_{\partial\Sigma} d\phi\,\Tr(\lambda\,\delta A_\phi).
\]
This is a standard structure: the generator of gauge transformations is “constraint plus
surface charge’’. If we set
\[
H[\lambda]
=
-\frac{k\ell}{2\pi}\int_{\Sigma} dr\,d\phi\,\Tr(\lambda\,F_{r\phi}) \;+\; Q[\lambda],
\]
then the defining relation $\delta H[\lambda]=\Omega_\Sigma(\delta A,\delta_\lambda A)$
implies that the boundary functional $Q[\lambda]$ must satisfy
\[
\delta Q[\lambda]=\frac{k\ell}{2\pi}\int_{\partial\Sigma} d\phi\,\Tr(\lambda\,\delta A_\phi).
\]
Restricting attention to the inner boundary component $\partial\Sigma_-\cong S^1$ gives the
inner surface-charge variation. Writing it directly as an integral over $\partial M_-$ at fixed
time (so only $d\phi$ appears), we obtain
\begin{equation}
\delta Q[\lambda]
=
\frac{k\ell}{2\pi}\int_{\partial M_-} d\phi\,\Tr\!\left(\lambda\,\delta A_\phi\right).
\label{eq:deltaQ-inner-5.39}
\end{equation}

With our inner Dirichlet boundary condition \eqref{eq:Dirichlet-inner-5.37}, we have
$\delta A_\phi|_{\partial M_-}=0$, and therefore \eqref{eq:deltaQ-inner-5.39} implies
$\delta Q[\lambda]=0$ for all $\lambda$. In other words, all gauge transformations at the inner
boundary carry vanishing canonical charge: they remain pure gauge and do \emph{not} generate
physical edge modes at $\partial M_-$. 

Note that this is a specific fact about Chern-Simons theory: for Yang-Mills, this is {\em not} true \cite{Donnelly}. Even in Chern-Simons theories, the oft-studied chiral boundary conditions lead to boundary gravitons as we noted earlier. There the corresponding $\delta Q[\lambda]$ is nonzero and integrates to a nontrivial
charge algebra, leading to boundary dynamics (and, in the AdS$_3$ setting,
eventually to WZW/AS degrees of freedom). When the inner cut is
a Dirichlet problem in $A_\phi$ on the other hand, the only inner dynamics that survives is not
a local edge field but the global holonomy zero-mode.

\section{Boundary Terms in Lorentzian and Euclidean Path Integrals}
\label{app:pq-boundary}

In this subsection we use a simple first--order mechanical system to show explicitly
that the boundary term which converts Dirichlet to Neumann boundary conditions has the
same functional form in Lorentzian and Euclidean signature, and that its contribution
to the path integral is the same phase factor $e^{-i p q}$ in both cases. This is the
sense in which the Fourier kernel and its $i$ are insensitive to Wick rotation.

\subsection{Lorentzian First-Order Action and Neumann Boundary}

Consider a single degree of freedom with phase--space variables $(q(t),p(t))$ and
Hamiltonian $H(p,q)$. The standard first--order Lorentzian action is
\begin{equation}
  S_L[q,p]
  \;=\;
  \int_{t_i}^{t_f} dt\;\big(p\,\dot q - H(p,q)\big)\,,
  \label{eq:SL-pq}
\end{equation}
where $\dot q \equiv dq/dt$. Varying $q$ and $p$ gives
\begin{align}
  \delta S_L
  &= \int_{t_i}^{t_f} dt\;\Big(\delta p\,\dot q + p\,\delta\dot q
        - \partial_q H\,\delta q - \partial_p H\,\delta p\Big)
  \nonumber\\
  &= \Big[p\,\delta q\Big]_{t_i}^{t_f}
   + \int_{t_i}^{t_f} dt\;\Big[(\dot q - \partial_p H)\,\delta p
          +(-\dot p - \partial_q H)\,\delta q\Big]\,.
  \label{eq:deltaSL}
\end{align}
The bulk term gives Hamilton's equations, while the boundary term is
\begin{equation}
  \delta S_L\big|_{\text{bdy}} \;=\; \Big[p\,\delta q\Big]_{t_i}^{t_f}\,.
\end{equation}
With Dirichlet boundary conditions $\delta q(t_i)=\delta q(t_f)=0$ the variational
problem is well--posed.

Suppose instead that we want $p$ fixed at the endpoints, $\delta p(t_i)=\delta p(t_f)=0$,
with $q$ allowed to fluctuate. We can achieve this by adding the boundary term
\begin{equation}
  S_{L,\text{bdy}}[q,p]
  \;:=\;
  -\,\Big[p(t)\,q(t)\Big]_{t_i}^{t_f}\,.
  \label{eq:SL-bdy}
\end{equation}
Its variation is
\begin{equation}
  \delta S_{L,\text{bdy}}
  \;=\;
  -\,\Big[\delta p\,q + p\,\delta q\Big]_{t_i}^{t_f}\,,
\end{equation}
so the total boundary variation becomes
\begin{align}
  \delta\big(S_L + S_{L,\text{bdy}}\big)\big|_{\text{bdy}}
  &= \Big[p\,\delta q\Big]_{t_i}^{t_f}
     - \Big[\delta p\,q + p\,\delta q\Big]_{t_i}^{t_f}
  \nonumber\\[2pt]
  &= -\,\Big[q\,\delta p\Big]_{t_i}^{t_f}\,.
\end{align}
Thus, for $\delta p(t_i)=\delta p(t_f)=0$, the Lorentzian Neumann action
\begin{equation}
  S_L^{(N)} \;:=\; S_L + S_{L,\text{bdy}}
\end{equation}
has a well--posed variational principle with $p$ fixed at the endpoints.

In the Lorentzian path integral the weight is $e^{i S_L^{(N)}}$. The extra factor from the
boundary term \eqref{eq:SL-bdy} is
\begin{equation}
  e^{i S_{L,\text{bdy}}}
  \;=\;
  \exp\!\Big(i\,\big[-p q\big]_{t_i}^{t_f}\Big)
  \;=\;
  \exp\!\Big(-\,i\,p q\Big)
\end{equation}
if we focus on a single endpoint (say, the final one). For the overlap with $q$ fixed
at the initial boundary and $p$ fixed at the final boundary, this factor reproduces the
usual plane--wave kernel $\langle p_f|q_f\rangle \propto e^{-i p_f q_f}$.

\subsection{Euclidean First-Order Action}

The Euclidean evolution operator is $e^{-\beta H}$, and its phase--space path integral
representation has the standard form
\begin{equation}
  \langle q_f|e^{-\beta H}|q_i\rangle
  \;=\;
  \int \mathcal{D}q\,\mathcal{D}p\; e^{-S_E[q,p]}\,.
\end{equation}
Time--slicing the matrix element
$\langle q_f|e^{-\beta H}|q_i\rangle$ and inserting resolutions of the
identity in the momentum basis with $\langle q|p\rangle \propto e^{i p q}$ gives, in the
continuum limit,
\begin{equation}
  -\,S_E[q,p]
  \;=\;
  \int_0^\beta d\tau\;\big(i\,p\,\dot q - H(p,q)\big)\,,
\end{equation}
where now $\dot q \equiv \partial_\tau q$. Thus the Euclidean first--order action is
\begin{equation}
  S_E[q,p]
  \;=\;
  \int_{\tau_i}^{\tau_f} d\tau\;\big(-i\,p\,\dot q + H(p,q)\big)\,.
  \label{eq:SE-pq}
\end{equation}

The same result follows from standard Wick rotation. Starting from \eqref{eq:SL-pq}
and setting $t = -i\tau$, $dt = -i\,d\tau$, one finds
$\dot q_L \equiv dq/dt = i\,\dot q$ in terms of the Euclidean $\tau$--derivative.
Then
\begin{equation}
  S_L
  =
  \int dt\,(p\,\dot q_L - H)
  =
  \int_{\tau_i}^{\tau_f} d\tau\;\big(p\,\dot q + i\,H\big)\,,
\end{equation}
so
\begin{equation}
  i S_L
  =
  \int_{\tau_i}^{\tau_f} d\tau\;\big(i\,p\,\dot q - H\big)
  =
  -\,S_E\,,
\end{equation}
i.e.\ $e^{i S_L} = e^{-S_E}$ with $S_E$ given by \eqref{eq:SE-pq}. Thus the
time--slicing derivation and the standard Wick rotation give the same Euclidean
first--order action: 
\bea
S_E=-i S_L|_{t=-i \tau}\, .
\eea

\subsection{Euclidean Neumann Boundary and Comparison}

Now vary the Euclidean action \eqref{eq:SE-pq}:
\begin{align}
  \delta S_E
  &= \int_{\tau_i}^{\tau_f} d\tau\;\Big(-i\,\delta p\,\dot q
         - i\,p\,\delta\dot q + \partial_p H\,\delta p
         + \partial_q H\,\delta q\Big)
  \nonumber\\
  &= \Big[-i\,p\,\delta q\Big]_{\tau_i}^{\tau_f}
   + \int_{\tau_i}^{\tau_f} d\tau\;\Big[(-i\,\dot q + \partial_p H)\,\delta p
         + (i\,\dot p + \partial_q H)\,\delta q\Big]\,.
\end{align}
The bulk terms again give the (imaginary--time) Hamilton equations. The boundary term is
\begin{equation}
  \delta S_E\big|_{\text{bdy}} \;=\; \Big[-i\,p\,\delta q\Big]_{\tau_i}^{\tau_f}\,.
\end{equation}
Therefore $S_E$ defines a Dirichlet problem with $q$ fixed at the Euclidean endpoints.

To obtain a Neumann problem (fixing $p$ at the endpoints) we add a boundary term
\begin{equation}
  S_{E,\text{bdy}}[q,p]
  \;:=\;
  \alpha\,\Big[p(\tau)\,q(\tau)\Big]_{\tau_i}^{\tau_f}
\end{equation}
with a (possibly complex) constant $\alpha$. Its variation is
\begin{equation}
  \delta S_{E,\text{bdy}}
  \;=\;
  \alpha\,\Big[\delta p\,q + p\,\delta q\Big]_{\tau_i}^{\tau_f}\,,
\end{equation}
so the total boundary variation becomes
\begin{align}
  \delta\big(S_E + S_{E,\text{bdy}}\big)\big|_{\text{bdy}}
  &= \Big[-i\,p\,\delta q
       + \alpha\,p\,\delta q + \alpha\,q\,\delta p\Big]_{\tau_i}^{\tau_f}
  \nonumber\\[2pt]
  &= \Big[(-i+\alpha)\,p\,\delta q + \alpha\,q\,\delta p\Big]_{\tau_i}^{\tau_f}\,.
\end{align}
Requiring the coefficient of $\delta q$ to vanish fixes $\alpha = i$, so that
\begin{equation}
  S_{E,\text{bdy}}
  \;=\;
  i\,\Big[p(\tau)\,q(\tau)\Big]_{\tau_i}^{\tau_f}\,,
\end{equation}
and
\begin{equation}
  \delta\big(S_E + S_{E,\text{bdy}}\big)\big|_{\text{bdy}}
  \;=\;
  \Big[i\,q\,\delta p\Big]_{\tau_i}^{\tau_f}\,.
\end{equation}
Thus for $\delta p(\tau_i)=\delta p(\tau_f)=0$ the Euclidean Neumann action
\begin{equation}
  S_E^{(N)} \;:=\; S_E + S_{E,\text{bdy}}
\end{equation}
has a well--posed variational principle with $p$ fixed at the Euclidean endpoints.

In the Euclidean path integral the weight is $e^{-S_E^{(N)}}$, so the extra boundary
factor from $S_{E,\text{bdy}}$ is
\begin{equation}
  e^{-S_{E,\text{bdy}}}
  \;=\;
  \exp\!\Big(-\,i\,\big[p(\tau)\,q(\tau)\big]_{\tau_i}^{\tau_f}\Big)\,.
\end{equation}
Again, focusing on a single endpoint, this gives a factor $e^{-i p q}$.

Putting the two signatures side by side, we have
\begin{align}
  &\text{Lorentzian Neumann:} &&
  S_{L,\text{bdy}} = -[p q]_{t_i}^{t_f},
  & e^{i S_{L,\text{bdy}}} &= \exp\!\big(-i [p q]_{t_i}^{t_f}\big), \\[2pt]
  &\text{Euclidean Neumann:} &&
  S_{E,\text{bdy}} = i [p q]_{\tau_i}^{\tau_f},
  & e^{-S_{E,\text{bdy}}} &= \exp\!\big(-i [p q]_{\tau_i}^{\tau_f}\big).
\end{align}
Thus the exponentiated boundary contribution is \emph{literally the same} phase
$\exp(-i p q)$ in the two formulations. In particular, for an endpoint where $p$ is held
fixed and $q$ is fluctuating, the boundary term produces exactly the Fourier kernel
relating the $q$-- and $p$--polarizations.

\section{A Compact $SU(2)_k$ Analogue}
\label{sec:compact_vs_noncompact}

The picture we presented in section~\ref{sec:CFT-bra} and section \ref{sec:mod-bootstrap} builds a basic structure: from the bulk Chern-Simons point of view, the holonomy phase space $M_{\rm flat}(T^2)$ is naturally {\em continuous}, whereas a microscopic holographic CFT on the cylinder has
a \emph{discrete} set of primaries $\{h_n\}$. This tension is (at least partly) what modular bootstrap exploits. It is useful to compare this picture to the finite-dimensional version of a similar story that arises in compact Chern-Simons theory. There, the quantized space of holonomies is {\em finite} dimensional and the set of characters appearing in a full CFT constructed from them is also {\em finite}. This is the setting of rational CFTs. To keep things concrete, we will focus on $SU(2)$ Chern-Simons.

\subsection{Compact $SU(2)_k$ Chern-Simons: ``Modular Functor'' }

For compact gauge group $SU(2)$ at integer level $k$, Chern-Simons theory is a unitary
$(2+1)$-dimensional TQFT. Canonical quantization on $\Sigma\times\mathbb{R}$ reduces the
classical phase space to the moduli space of flat connections on $\Sigma$,
$M_{\rm flat}(\Sigma,SU(2))=\{F_\Sigma=0\}/\mathcal{G}_\Sigma$, equipped with the
Atiyah--Bott symplectic form. Quantization produces a finite-dimensional Hilbert space
$\mathcal{H}_\Sigma^{SU(2)_k}$ and assigns linear maps to cobordisms $M:\Sigma_{\rm in}\to\Sigma_{\rm out}$,
compatible with gluing. 

On $\Sigma=T^2$, flat connections are commuting holonomies along the two primitive cycles,
modulo conjugation. The reduced phase space is compact, so quantization yields a \emph{finite}
basis of states. A standard basis is labeled by integrable highest weights (spins)
$j=0,\frac12,\dots,\frac{k}{2}$, so $\dim\mathcal{H}_{T^2}=k+1$. The mapping class group
$\mathrm{MCG}(T^2)=SL(2,\mathbb{Z})$ acts on $\mathcal{H}_{T^2}$ by finite matrices $S$ and $T$.
For $SU(2)_k$ one may take
\bea
S_{jj'}=\sqrt{\frac{2}{k+2}}\,
\sin\!\Big(\frac{(2j+1)(2j'+1)\pi}{k+2}\Big), 
\qquad
T_{jj}=e^{2\pi i\,(h_j-\frac{c}{24})}, \\
\qquad
h_j=\frac{j(j+1)}{k+2},\ \qquad \ c=\frac{3k}{k+2}. \hspace{2.1cm}
\eea
Conceptually, this is the compact analogue of the bulk Fourier kernel between $(P,\widetilde P)$:
it is a change-of-polarization (equivalently, change of primitive cycle) on the torus holonomy phase space.
In particular, the cylinder $T^2\times I$ implements the identity operator in a fixed polarization,
while expressing the same cobordism between conjugate polarizations produces the overlap kernel,
which on $T^2$ is precisely the modular $S$-matrix.

\subsection{Chiral Conformal Blocks vs. Characters}

With chiral boundary conditions\footnote{Here we are describing a single chiral sector. When we build a full local CFT out of it, we will require
an independent anti-chiral copy so that torus amplitudes pair $\chi_i(\tau)$ with $\bar\chi_{\bar j}(\bar\tau)$ as in \eqref{Nijbar}.}, compact Chern-Simons reduces to a chiral
$SU(2)_k$ WZW theory on $\Sigma$. In this setting one can identify
$\mathcal{H}^{SU(2)_k}_\Sigma$ with the space of chiral conformal blocks on $\Sigma$
for the affine algebra $\widehat{\mathfrak{su}}(2)_k$ (equivalently, the state space of the
associated modular tensor category \cite{Fuchs}). A useful point of clarification is that a ``conformal block''
is a holomorphic building block of correlators determined by chiral Ward identities. It depends
holomorphically on the complex structure moduli of $\Sigma$ (and insertion points and channels when present).
On the torus without insertions, this space is spanned by the affine characters $\chi_j(\tau)$ (we set the $J^3$ fugacities of the $SU(2)$ to unity for simplicity).
Thus, characters are a special case of chiral conformal blocks.

We can view our Virasoro characters $\chi_P(\tau)$ in the $SL(2,\mathbb{R})$ CS case, similarly: they are the genus-one no-insertion chiral blocks of Virasoro, written in a basis
labelled by the holonomy parameter $P$.

\subsection{From Chiral Data to a Full 2D CFT}

A key structural lesson from compact Chern-Simons is that the chiral/topological data
$(\mathcal{H}_\Sigma,\ S,\ T,\ \text{fusion},\ \text{sewing of blocks})$ does \emph{not} by itself pick a unique
 (nonchiral, local) 2D CFT. A full RCFT torus partition function takes the form
\begin{equation}
Z(\tau,\bar\tau)=\sum_{i,\bar j} N_{i\bar j}\,\chi_i(\tau)\,\overline{\chi_{\bar j}(\tau)},
\qquad
N_{i\bar j}\in\mathbb{Z}_{\ge 0}, \label{Nijbar}
\end{equation}
and modular invariance imposes $NS=SN$ and $NT=TN$. But fixing the chiral algebra (hence fixing
$S,T$ and the characters) does not uniquely fix $N_{i\bar j}$. For $SU(2)_k$ there are multiple modular
invariants, associated to the ADE classification \cite{EberhardtReview}. The quantization of the {\em non-chiral} WZW model (for example) corresponds to the diagonal case where $N_{i \bar j}=\delta_{i\bar j}$ and the sum is over the full basis of $k+1$ states of $\mathcal{H}_{T^2}$. This is just the $A$-series. Moreover, modular invariance at genus one is only the first constraint:
a full CFT must satisfy factorization, sewing, crossing, OPE associativity, higher-genus consistency, ...
In the 3D TQFT construction of RCFT correlators, this extra input is packaged as additional algebraic data
(e.g.\ a suitable Frobenius algebra object) selecting a consistent pairing of left and right blocks \cite{Fuchs}.

The key point however, is that in the compact CS case the moduli space is compact and Hilbert space is finite dimensional, so a systematic classification is possible. In the character language this leads to the statement that there is only a finite number of them present in the theory: the CFT is rational. 

\subsection{Non-Compact $SL(2,\mathbb{R})$: the RCFT vs. Non-RCFT Divide}

AdS$_3$ gravity provides a non-compact analogue of the above structure.
Quantization of the moduli space $M_{\rm flat}(T^2,SL(2,\mathbb{R}))$ yields a torus Hilbert space with a continuum of
hyperbolic labels $P\ge 0$ (and possibly additional discrete sectors depending on quantization choices),
and the change of polarization between the two primitive cycles is governed by an integral kernel
$S(P,P')$ rather than a finite matrix. In our setting, the outer Brown-Henneaux 
boundary conditions implement a Virasoro (rather than affine) chiral theory at infinity. The resulting
boundary path integral produces the universal descendant factor and therefore the Virasoro character
$\chi_P(\tau)$ as the coefficient of $|P;C\rangle$ in the annulus state.

From this viewpoint, the universal ``bulk toolkit'' produced by the $SL(2,\mathbb{R})$ Chern-Simons
quantization is the \emph{continuum} family of  blocks/characters $\chi_P(\tau)$ together with the
mapping-class-group kernels implementing change of polarization on $T^2$. But a specific holographic CFT is not obtained by the chiral bulk data alone. In our radial-quantization
language, specifying the microscopic CFT corresponds to choosing an ``inner bra''
$\langle\!\langle \Psi_{\rm CFT}||$ (or, equivalently, a spectral measure $\rho(P,\bar P)$) so that
$ Z_{\rm CFT}(\tau,\bar\tau)=\int dP\,d\bar P\;\rho(P,\bar P)\,\chi_P(\tau)\,\chi_{\bar P}(\bar\tau)$, as discussed in the main text. This provides the analogue of $N_{i \bar j}$ in \eqref{Nijbar}. For a compact, unitary CFT on the cylinder, $\rho(P,\bar P)$
is expected to be a delta-function comb (a discrete spectrum with nonnegative integer degeneracies),
whereas the bulk $S$-kernel has continuum support. \\

\noindent
Thus the compact and non-compact stories have some similarities:
\begin{itemize}
\item compact $SU(2)_k$: finite label set $j$, finite matrices $S,T$, and the \emph{full} CFT is specified by
an additional modular invariant $N_{i\bar j}$ plus sewing/factorization data;
\item non-compact $SL(2,\mathbb{R})$: continuous label $P$, integral kernels $S(P,P')$, and the \emph{full}
holographic CFT is specified by additional UV data $\rho(P,\bar P)$ (expected to be discrete) plus the
non-rational analogues of sewing/factorization.
\end{itemize}
In both cases, Chern-Simons provides the universal chiral or topological ``skeleton'' (characters and
mapping-class-group kernels), while the choice of a specific unitary CFT is an extra input.

The key difference however is that in the non-compact, infinite dimensional case, solving these constraints explicitly to obtain a discrete spectrum, has not been done\footnote{But note that while we do not know how to systematically solve the constraints to get such a CFT, we do know how to construct simple examples that satisfy the constraints: toroidal Narain type CFTs are an explicit example.}. If such a solution, with further constraints like large-$c$, sparse low-lying spectrum, etc. exists, then we have a candidate holographic CFT. 

\subsection{Philosophical Differences from Virasoro TQFT}
\label{app:G5:VirTQFT}

Virasoro TQFT is a consistent quantization/completion of the gravitationally relevant (Teichmuller) component of the $PSL(2,\IR)$ flat-connection moduli space. Its Hilbert spaces do contain the same kind of hyperbolic continuum labels that appear in our torus-holonomy analysis \cite{Eberhardt}. So let us emphasize some points which we view are crucial distinctions between the two approaches.

The goal of the Virasoro TQFT program is to define a bulk topological TQFT that can be viewed as the quantization of 3D pure Einstein gravity. Our goal is to take hints from the quantization of the moduli space of flat $SL(2,R)$ connections on the thickened torus, to get a better understanding of the states of a holographic CFT on the cylinder -- especially from the bulk.

The RCFT discussion above serves as a prototype for what we have in mind. In rational theories the chiral/topological data (modular functor, $S$ and $T$, ...) do not uniquely fix a local non-chiral CFT.
In the $3$d TQFT construction of RCFT correlators, the extra input needed to obtain a full CFT is packaged as a
choice of additional algebraic data, e.g.\ a (symmetric special) Frobenius algebra object $A$ in the
relevant modular tensor category \cite{Fuchs}.
For $SU(2)_k$ this organizing principle connects with the existence of multiple modular invariants
(the ADE classification): one is not merely given a single ``canonical'' left-right pairing by the chiral theory.

The Virasoro TQFT proposal has a different goal.
Rather than using modular invariance and sewing as classification tools for 
boundary CFTs, it aims to define a \emph{bulk} topological theory for $3$d pure gravity amplitudes by
cutting and sewing with the Virasoro modular-functor data.
To make such a TQFT well-defined one must specify a gluing-compatible pairing on state spaces.
In the Virasoro TQFT framework this left-right pairing is taken to be the canonical
Hermitian inner product\footnote{Schematically, this is the ``absolute square" pairing, reminiscent of the pairing in non-chiral $SU(2)_k$ WZW CFT. Since there is a lot of terminological looseness in the literature, let us be precise: what we mean by non-chiral $SU(2)_k$ WZW CFT here, is the full CFT defined by the standard WZW action on the group $SU(2)$, to be distinguished from a generic CFT with chiral algebra $su(2)_k$.}. The bulk ``gravity'' answer is then obtained by the appropriate mapping-class-group sums. The canonical pairing is part of the definition of the gravitational TQFT.

An important question for our approach (which is less central for the Virasoro TQFT program) is whether a consistent holographic CFT that solves the constraints, and has a {\em discrete} spectrum, {\em exists}! The general consensus seems to be that holographic conformal field theories exist at least when there is supersymmetry: the $\mathcal{N}=(4,4)$ SCFT on the cylinder is believed to be holographic and it should have a discrete spectrum. Note that unlike Virasoro TQFT, our stated goals do not involve pure Einstein gravity. The D1D5 gravity system is simply one example of matter-coupled Einstein gravity for us. See our discussion in Section \ref{WilsonSpool} for related comments.

\section{Maloney-Witten Poincar\'e Series and $S$-kernels}\label{App:Maloney-Witten}

In this appendix we review some known material to show that the spectrum of ``pure 3D Einstein gravity" as defined by the Maloney-Witten Poincar\'e series can be written in terms of Virasoro modular kernels. This essentially {\em defines} the spectral density of the putative holographic dual  of ``pure 3D Einstein gravity". The ingredients are due to Maloney-Witten \cite{MaloneyWitten, MaloneyKeller} and Benjamin-Collier-Maloney \cite{Benjamin}: some of our notation follows Pal-Qiao \cite{Pal}.

\subsection{General Spectral Representation}

We work with $q = e^{2\pi i\tau}$, $\bar q = e^{-2\pi i\bar\tau}$ and factor out the universal descendant piece
\begin{equation}
  F(\tau) := \prod_{n=1}^\infty \frac{1}{1-q^n}
  = q^{-1/24}\,\frac{1}{\eta(\tau)},
  \label{eq:F-def-appendix}
\end{equation}
so that
\begin{equation}
  \chi_h(\tau) = F(\tau)\,q^{h-c/24},
  \qquad
  \chi_{\rm vac}(\tau)
  = F(\tau)\,q^{-c/24}(1-q).
\end{equation}
As before, we parametrize normal orbits by a continuous label
$P\ge0$, with
\begin{equation}
  h(P) = \frac{c-1}{24} + P^2,
  \qquad
  \chi_P(\tau) := \chi_{h(P)}(\tau)
  = F(\tau)\,q^{P^2-1/24}
  = \frac{q^{P^2-1/24}}{\eta(\tau)}.
  \label{eq:F-chiP-def}
\end{equation}
In particular there is a gap at $h_{\rm gap} = \frac{c-1}{24}$ between the
vacuum and the onset of the continuum.

For any modular invariant CFT with a “twist gap’’ at
$\min(h,\bar h)\ge (c-1)/24$, the reduced
partition function can be written as a vacuum piece plus an integral over
a non-negative spectral density $\rho(h,\bar h)$ (see e.g., \cite{Pal} eq.\ (2.8)).  In our notation this can be expressed  as
\begin{equation}
  Z(\tau,\bar\tau)
  = F(\tau)\,\bar F(\bar\tau)\Big[
      (1-q)(1-\bar q)
      + \int_{h,\bar h\ge (c-1)/24} \!\! dh\,d\bar h\;
        \rho(h,\bar h)\,
        q^{h-c/24}\,\bar q^{\bar h-c/24}
    \Big].
  \label{eq:F-PQ-structure}
\end{equation}
Changing variables to the normal-orbit parameters
$h = \frac{c-1}{24} + P^2$, $\bar h = \frac{c-1}{24} + \bar P^2$, and
absorbing Jacobians into $\rho$, one can rewrite
\eqref{eq:F-PQ-structure} as
\begin{equation}
  Z(\tau,\bar\tau)
  = F(\tau)\,\bar F(\bar\tau)\Big[
      (1-q)(1-\bar q)
      + \int_0^\infty dP\int_0^\infty d\bar P\;
        \rho(P,\bar P)\,
        q^{P^2-1/24}\,\bar q^{\bar P^2-1/24}
    \Big],
  \label{eq:F-PQ-P}
\end{equation}
i.e.\ as a “vacuum + continuum’’ decomposition with a gap at
$h=(c-1)/24$.

\subsection{Maloney-Witten Spectral Density from Crossing Kernels}

Maloney and Witten define a “pure gravity’’ torus partition function as a
Poincar\'e series seeded by the vacuum amplitude $\chi_{\rm vac}$
\cite{MaloneyWitten}:
\begin{equation}
  Z_{\rm MW}(\tau,\bar\tau)
  \;\sim\;
  \sum_{\gamma\in\Gamma_\infty\backslash PSL(2,\mathbb Z)}
  \chi_{\rm vac}(\gamma\tau)\,\overline{\chi_{\rm vac}(\gamma\tau)},
  \label{eq:F-MW-Poincare}
\end{equation}
where ``$\sim$'' indicates that regularization factors in
$\Im(\gamma\tau)$ are omitted here for brevity.
Benjamin, Collier and Maloney \cite{Benjamin} note that
each modular image in \eqref{eq:F-MW-Poincare} can be expanded in the
continuous character basis using a Virasoro modular crossing kernel (we try to follow their notations instead of our $S_{PP'}$ in this appendix)
$k^{(\gamma)}_{h_0 h}$:
\begin{equation}
  \chi_h(\gamma\tau)
  = \int_{(c-1)/24}^\infty\! dh_0\;
    k^{(\gamma)}_{h_0 h}\,\chi_{h_0}(\tau),
  \label{eq:F-crossing}
\end{equation}
with explicit $k^{(\gamma)}_{h_0 h}$ given in eqs.\ (2.1)-(2.3) of
\cite{Benjamin}.  The MW seed is the usual vacuum
combination
\[
  \chi_{\rm vac}(\tau,\bar\tau)
  = (\chi_0(\tau)-\chi_1(\tau))(\bar\chi_0(\bar\tau)-\bar\chi_1(\bar\tau)),
\]
so that one can define a “vacuum kernel’’
\[
  K_\gamma(h_0)
  := k^{(\gamma)}_{h_0\,0} - k^{(\gamma)}_{h_0\,1},
\]
and write
\[
  \chi_{\rm vac}(\gamma\tau)
  = \int_{(c-1)/24}^\infty\! dh_0\;
    K_\gamma(h_0)\,\chi_{h_0}(\tau).
\]
When $\gamma$ is the $S$-operation, $K_\gamma(h_0)$ takes the vacuum row form that we have used in the main body of the paper. The general form is well-understood and straightforward to make explicit.
Inserting this into \eqref{eq:F-MW-Poincare} and comparing with the
generic spectral form \eqref{eq:F-PQ-structure}, one finds that the MW
spectral density can be written schematically as
\begin{equation}
  \rho_{\rm MW}(h,\bar h)
  \sim \sum_{\gamma}
    K_\gamma(h)\,K_\gamma(\bar h),
  \label{eq:F-rho-h}
\end{equation}
up to the same regularization of the Poincar\'e sum that is already
present in \cite{MaloneyWitten,Benjamin}.

Passing to the normal-orbit labels via
$h=(c-1)/24+P^2$, $\bar h=(c-1)/24+\bar P^2$, and absorbing
Jacobians into the kernels (but we still call them $K$), one may equally well write
\begin{equation}
  \rho_{\rm MW}(P,\bar P)
  = \sum_\gamma K_\gamma(P)\,K_\gamma(\bar P),
  \label{eq:F-rho-P}
\end{equation}
so that the continuous part of the MW partition function takes the form
\begin{equation}
  Z_{\rm MW}(\tau,\bar\tau)
  \;\sim\;
  F(\tau)\,\bar F(\bar\tau)\Big[
      (1-q)(1-\bar q)
      + \int_0^\infty dP\int_0^\infty d\bar P\;
        \rho_{\rm MW}(P,\bar P)\,
        q^{P^2-1/24}\,\bar q^{\bar P^2-1/24}
    \Big].
  \label{eq:F-MW-final}
\end{equation}
This is in the Pal-Qiao form \eqref{eq:F-PQ-P}, with the
specific “pure gravity’’ spectral density
$\rho(P,\bar P)=\rho_{\rm MW}(P,\bar P)$ determined by products of
Virasoro modular kernels. A more explicit version of this result can be found in \cite{Benjamin}.

\end{document}